%% file: paperC2UnconventionalSCinScars_v2.tex
\documentclass[aps,prx,groupedaddress,twocolumn,showpacs,longbibliography,10pt]{revtex4-2}

\usepackage{amsmath}
\usepackage{amsthm,amssymb,bbold}
\usepackage{bm}
\usepackage{braket}
\usepackage{graphicx}
\usepackage{color}
\usepackage{mathtools,tikz}
\usepackage{natbib}
\usepackage{hyperref}
\usepackage{lipsum}

\hypersetup{colorlinks=true,
}

\newcommand{\CO}{{\cal O}}

\newcommand{\tSp}{{\widetilde{\text{Sp}}}}

\newcommand{\ttU}{{\widetilde{\text U}}}

\newcommand{\SU}{{\text{SU}}}

\newcommand{\CM}{{\cal M}}
\usepackage{amsthm}

\makeatletter
\newcommand*{\rom}[1]{\expandafter\@slowromancap\romannumeral #1@}
\makeatother

\usepackage{verbatim}
\usepackage{prettyref}

\usepackage{etoolbox}

 \usepackage{slashed}
\usepackage{esvect}
\usepackage{dsfont}

\definecolor{darkgreen}{rgb}{0,0.5,0}
\definecolor{darkblue}{rgb}{0,0,0.6}
\definecolor{purple}{rgb}{0.4,.2,0.7}

\numberwithin{equation}{section}
\numberwithin{figure}{section}
\numberwithin{table}{section}

\def\CM{{\cal M}}
\def\CN{{\cal N}}

\DeclareFontShape{OT1}{cmr}{mx}{n}{<->cmr10}{}

\date{\today}

\begin{document}

\fontseries{mx}\selectfont

\title{ Unconventional superconducting correlations in fermionic many-body scars}

\author{Kiryl Pakrouski$^{1}$ and K. V. Samokhin$^{2}$ }
\affiliation{$^{1}$Institute for Theoretical Physics, ETH Zurich, 8093 Zurich, Switzerland}
\affiliation{$^{2}$Department of Physics, Brock University, St. Catharines, Ontario, Canada L2S 3A1}

\begin{abstract}

Weak ergodicity breaking in interacting quantum systems may occur due to the existence of a subspace dynamically decoupled from the rest of the Hilbert space. In two-orbital spinful lattice systems, we construct such subspaces that are in addition distinguished by strongest inter-orbital and spin-singlet or spin-triplet, long-range superconducting pairing correlations. All unconventional pairing types we consider are local in space and unitary. Alternatively to orbitals, the additional degree of freedom could originate from the presence of two layers or through any other mechanism. Required Hamiltonians are rather non-exotic and include chemical potential, Hubbard, and spin-orbit interactions typically used for two-orbital superconducting materials. Each subspace is spanned by a family of group-invariant quantum many-body scars combining both $2e$ and $4e$ pairing/clustering contributions. One of the basis states has the form of a BCS wavefunction and can always be made the ground state by adding a mean-field pairing potential. Analytical results in this
work are lattice-, dimension- and (mostly) system size-independent. We confirm them by exact numerical diagonalization in small systems.

\end{abstract}

\maketitle

\section{Introduction}

Understanding microscopic physics of unconventional superconductivity remains one of the most interesting problems of modern science even after many years of active research. A novel approach to it may thus be helpful to make further progress. An interesting development in that regard is the emerging evidence for possible interconnection between superconductivity and weak ergodicity breaking \cite{imai2024quantummanybodyscarsunconventional,paperC1}. In this work we extend this evidence by constructing ergodicity breaking subspace that exhibits unconventional Cooper pairing significantly stronger than the rest of the Hilbert space.   

Weak ergodicity breaking occurs when a subspace of special states called many-body scars (MBS) \cite{RydbergExperimentRevivals,Turner_2018} is dynamically decoupled from the rest of the Hilbert space in the absence of a Hamiltonian symmetry that would isolate them in a separate sector \cite{Serbyn:2020wys,Moudgalya:2021xlu,Papic2022,Chandran:2022jtd}. Found in a variety of systems \cite{PhysRevA.86.041601,AbaninScarsSU2Dynamics,Moudgalya:2018,Khemani:2019vor,Sala_2020,Prem:2018,Schecter:2019oej,BucaNature2019,SciPostPhys.3.6.043,IadecolaHubbardAlmostNUPRL2019,Shibata:2020yek,michailidis2020stabilizing,2020MarkMotrEtaPairHub,PRLPapicClockModels,VedikaScarsVsIntegr,Pal2020ScarsFromFrustration,mark2020unified,iadecola2020quantum,PhysRevB.101.220305,Nielsen2020ChiralScars,Hsieh2020PXP2D,Regnault2020MPStoFindSc,FloquetPXPScars,PapicWeaklyBrokenAlgebra,kuno2021multiple,banerjee2020quantum,Pakrouski:2021jon,chaoticDickeAllScarred2021,maskara2021DrivenScars,langlett2021rainbow,2021arXiv210807817R,2021arXiv211011448T,Schindler:2021lma,Dodelson:2022eiz,Liska:2022vrd,scarsInSchwingerModel2023,TruncatedSchwingerModel2023,budde2024quantummanybodyscarsarbitrary,NielsenTowerInLocalized,shen2024enhancedmanybodyquantumscars,osborne2024quantummanybodyscarring21d} it can be understood in terms of several generalized frameworks \cite{Shiraishi2017ScarsConstruction,moudgalya2020etapairing,pakrouski2020GroupInvariantScars,Moudgalya:2022nll,PhysRevResearch.2.043305,PhysRevLett.126.120604,PhysRevResearch.6.023041,PhysRevB.112.125165}, including the group-invariant approach \cite{pakrouski2020GroupInvariantScars}. In the latter, the MBS subspace is formed by the states invariant under a continuous group $G$ with generators $T_l$ for any Hamiltonian of the form
\begin{gather}
\label{eq:H0PlusOT}
H=H_0+\sum_l O_l T_l.
 \end{gather}
Here $H_0$ respects $G$ (see more detailed conditions in Ref. \cite{pakrouski2020GroupInvariantScars}) and $O_l$ is any operator such that the product $O_lT_l$ is Hermitian. $G$-invariant states are eigenstates of $H$, while $O_l$ can be chosen such that $H$ has no symmetries. 
 
The first hint at the possible connection between superconductivity and weak ergodicity breaking is provided by the fact that the $\eta$-pairing states \cite{etaPairingYang89}, which are known to have long-range superconducting correlations, are also \cite{SciPostPhys.3.6.043,2020MarkMotrEtaPairHub} MBS in the Hilbert space of single-flavour spinful fermions, where they are stable to certain disorder types \cite{KolbStabilityPRXQuantum2023}. Experimentally, $\eta$-pairing correlations have been observed in pump-probe experiments \cite{Kaneko2019pumpProbeEta,Werner2020RevivalsOfZetaStatesMultiband}, while theoretical generalizations of the $\eta$-pairing states to multi-flavour systems have been developed \cite{japaneseGeneralizedEtaPairingSep2021,Nakagawa:2022jsg,paper3Case1Majorana,imai2024quantummanybodyscarsunconventional,ZhaiGeneralizedEtaPairing2005,pwaveScarsInSpinlessFermMazza2022}. Connection between superconductivity and $\eta$-pairing in flat-band systems was made in recent works \cite{scFlatBand1,scFlatBand2}.

\subsection{Unconventional superconductivity \label{sec:scIntro}}

According to the Bardeen-Cooper-Schrieffer (BCS) theory, superconductivity is attributed to the formation of the Cooper pairs of electrons, which condense into the same state whose wave function plays the role of the order parameter. The superconducting state exhibits an off-diagonal long-range order (ODLRO) \cite{yang1962concept}, which means that the two-point correlation function $\braket{\CO^\dagger_i \CO_j}$, where $\CO^\dagger_i$ are the pair creation operators, does not exponentially fall off (is actually constant) at large spatial separations between the sites $i$ and $j$.    
As shown in Ref. \cite{etaPairingYang89}, single-flavour $\eta$-pairing states do possess superconducting correlations. For these states, the pair creation operator $\CO^\dagger_j = c^\dagger_{j\uparrow} c^\dagger_{j\downarrow}$ corresponds to a one-band, or one-orbital, spin-singlet electron pairing which is local in space. However, in many superconductors of current interest, the pairing is ``unconventional'', e.g. nonlocal, spin-triplet, and/or involves several orbital species of electrons. 

Multiband, or multi-orbital, superconductors have been actively studied since the discovery of two-gap superconductivity in MgB$_2$ \cite{Naga01,Budko15}. The list of materials in which multiband effects may play an important role also includes Sr$_2$RuO$_4$ \cite{MM03}, 
various heavy-fermion compounds \cite{Bauer04,Kohori01}, iron-based superconductors \cite{Hirsch11}, doped topological insulators \cite{Wray11,Fu10}, superconducting oxide interfaces \cite{Trevi18,Singh22}, Moir\'e materials \cite{experimentSyntheticTwoBandHubbard}, and many others. 

The BCS theory can be straightforwardly extended to multiband systems if the Cooper pairs are formed by the quasiparticles from the same band \cite{Suhl59}.
The pair scattering from one band to another produces a ``Josephson coupling'' between the bands, which depends on the relative phase of the two pair condensates. This coupling gives rise to a number of features peculiar to multiband superconductors, such as the Leggett modes \cite{Legg66}, phase solitons \cite{Tanaka01}, and fractional vortices \cite{Baba02,Iguchi23}.

More recently, theory of multiband superconductivity has been further extended by taking into account the pairing of quasiparticles from different bands, i.e. interband pairing \cite{Moreo09,Fisch13,Ram16,Nomoto16,Sam20,Sam24}. Interband terms in the Hamiltonian naturally appear in the BCS framework if the energy scale of the pairing interaction exceeds the energy splitting between the bands. 
Alternatively, one can start with some model of real-space pairing between electrons in different atomic orbitals and then transform the pairing interaction into the band representation, which in general produces both intraband and interband pairing terms. 

Regarding the spin structure of the superconducting order parameter, the pairs can be spin-singlet (total spin $S=0$) or spin-triplet ($S=1$) \cite{SU-review,TheBook}. While superconductors described by the standard BCS model exhibit singlet pairing, there is a growing number of materials, e.g. UPt$_3$ \cite{UPt302}, Sr$_2$RuO$_4$ \cite{SRO12}, UTe$_2$ \cite{Ran19}, and ferromagnetic superconductors (UGe$_2$, URhGe, UCoGe) \cite{Aoki-FMSC19}, that are viewed as strong candidates for triplet pairing. 

In order to understand the physics of novel superconductors that exhibit unconventional multiband or multi-orbital pairing, we propose to use a two-orbital, spin-$1/2$ Hilbert space as the simplest playground to allow for the minimal required complexity. We aim to answer the following question: Could many-body scar states have an unconventional pairing ODLRO? Recent Ref. \cite{imai2024quantummanybodyscarsunconventional} showed that single-band generalizations of the $\eta$-pairing states with unconventional pairing are MBS for certain, somewhat artificial, multi-body interactions. Ref. \cite{paperC1} considered multi-orbital spinful fermions and rather general (both ``superconducting'' and ``magnetic'') pairing and demonstrated that the resulting states are group-invariant scars that include certain BCS wavefunctions.
In this work, we specialize the approach of Ref. \cite{paperC1} to construct local \textit{non-exotic} Hamiltonians that host MBS with an unconventional pairing in the two-orbital spinful Hilbert space relevant for the materials reviewed above.

We will focus (see Appendix \ref{sec:motivatePairing} for motivation) on two unconventional pairing types, both local in space. One of them is the space-even, spin-singlet, and orbital-triplet pairing described by the following pair creation operator at the $j$th site:
\begin{gather}
\label{eq:OdagEBTSS}
\CO^{\dagger}_{0\nu,j} = \frac{1}{2}\sum_{pq,\alpha\beta} c^{p,\dagger}_{j\alpha} c^{q,\dagger}_{j\beta} (i\hat\sigma_2)_{\alpha\beta}(i\hat\tau_\nu\hat\tau_2)_{pq},
\end{gather}
where the $\sigma$ Pauli matrices act on the spin ($\alpha,\beta=\uparrow,\downarrow$) and the $\tau$ Pauli matrices -- on the orbital ($p,q=x,y$) degrees of freedom. We denote the two orbitals $x,y$ motivated by the relevant orbitals $d_{xz}$ and $d_{yz}$ in ${\mathrm{Sr}}_{2}\mathrm{Ru}{\mathrm{O}}_{4}$ \cite{typicalUSCHamZinkl2021}, although our results do not depend on the specific nature of the orbitals. The intra-orbital pairing is described by the $\nu=1,2$ components, while the $\nu=3$ component describes the inter-orbital pairing. Note that $\nu=0$ is not included, because it would not satisfy the Pauli exclusion principle.
The other pairing operator we consider is space-even, spin-triplet, and orbital-singlet:
\begin{gather}
\label{eq:evenBSST}
\CO^{\dagger}_{\mu0,j} = \frac{1}{2}\sum_{pq,\alpha\beta} c^{p,\dagger}_{j\alpha} c^{q,\dagger}_{j\beta}  (i\hat\sigma_\mu\hat\sigma_2)_{\alpha\beta}(i\hat\tau_2)_{p q}.
\end{gather}
This describes a purely inter-orbital pairing, with $\mu=1,2,3$ labeling the three components of spin-triplet pairing \cite{SU-review,TheBook}.
Any space-odd pairing operators would be incompatible with the specific choices of the group $G$ we will make but could be considered in future work. 

The MBS in this work are to a great extent lattice- and dimension-independent. However, multi-site pairing operators would depend on specific lattice choice and we defer their full classification to future studies.

\subsection{Recap results of Ref. \cite{paperC1}}
\label{sec: recap}

Two out of five families of scars considered in this work are special cases of the construction developed in Ref. \cite{paperC1}, which is briefly reviewed below for the specific case of spinful fermions and two orbitals.

Consider an arbitrary lattice with $N$ sites (the Hilbert space dimension equals $2^{4N}$) and define the pair annihilation operator at each site $j$ as
\begin{align}
\label{eq:type1O}
\CO_j = \frac{1}{2} \sum_{\rho,\rho'=1}^{4}c_{j\rho} A_{\rho\rho'} c_{j\rho'} ,
\end{align}
where $\rho$ and $\rho'$ are multi-indices that include both the spin and orbital degrees of freedom and $A$ is an anti-symmetric unitary matrix. 

The operators $\CO^{\dagger}_j$ for a fixed $A$ belong to a realization of SU$(2)$ algebra and can be interpreted as ``spin''-lowering operators \cite{paperC1} where the allowed ``spin'' values are 0,$\frac{1}{2}$,1 and $(\CO^{\dagger}_j)^2\ket{1}=\mathrm{const}\ket{0}$. All algebras share the same, unique product states with maximum $\ket{1}$ (fully empty) and minimum $\ket{0}$ (fully filled) ``spins". Therefore all the operators $(\CO^{\dagger}_j)^2$ represent the same transformation and are identical up to a constant factor. For the operators in Eqs. \eqref{eq:OdagEBTSS} and \eqref{eq:evenBSST} we have
\begin{align}
\label{eq:squaredRelations}
(\CO^{\dagger}_{0\nu,j})^2 = - (\CO^{\dagger}_{\mu0,j})^2,
\end{align}
which holds for any $\mu$ and $\nu$ and will be relevant in our discussion of 4e clustering.

The O$(N)$-invariant states 
\begin{align}
\label{eq:genTowerWf}
\ket{\phi_n} = \frac{(\sum_j\CO^{\dagger}_j)^n \ket{0}}{P_N(n)}, \quad P_N(n) = \sqrt{ \frac{ (2N)!n!}{ (2N-n)! } } ,
\end{align}
with $0\le n\le 2N$, are MBS for any Hamiltonian of the form \eqref{eq:H0PlusOT}, for the group $G=\text{O}(N)$. A simple though inexact way to understand O$(N)$-invariance is as independence on arbitrary site relabeling. Generators of O$(N)$ will be discussed in Sec. \ref{sec:HbHoppingTerms}. The form of the states \eqref{eq:genTowerWf} is inspired by Yang's $\eta$-pairing states \cite{etaPairingYang89}.

The first part of the relevant local Hamiltonians, see Eq. \eqref{eq:H0PlusOT}, can be represented in the form
\begin{align}
\label{eq:H0generalSU2}
    &H_0 = - \sum_j \mu_{\rm eff}  (2-n_j) + H_\Delta + \mathrm{const},\\ 
    &H_\Delta  = \sum_j \left( \Delta e^{i\theta} \CO^\dagger_j+\Delta e^{-i\theta} \CO_j \right),
\label{eq:dH0general}
\end{align}
where $\mu_{\rm eff}$, $\Delta>0$ and $\theta\in[0,2\pi)$ are real numbers and $n_j=\sum_\rho c_{j\rho}^\dagger c_{j\rho}$ is the particle number operator on the $j$th site. The first term in Eq. \eqref{eq:H0generalSU2} is the site-independent effective chemical potential. The second term can be interpreted as the pairing potential that could arise in the mean-field approximation, with $\Delta e^{i\theta}$ being a complex superconducting gap function. Such a term could also be a model for proximity-induced superconductivity. As expected, the specific value of the phase $\theta$ does not have a material influence on any observables.
We also note that the substitution of Eq. \eqref{eq:type1O} in $H_\Delta$ brings the latter to the form
\begin{equation}
    H_\Delta=\frac{1}{2}\sum_{j,\rho\rho'} \Delta_{\rho\rho'} c^\dagger_{j\rho}c^\dagger_{j\rho'} + \text{H.c.},
\end{equation}
where $\hat\Delta=\Delta e^{i\theta} A^\dagger$ is the gap function matrix in the orbital and spin space. This matrix has the property $\hat\Delta\hat\Delta^\dagger=\Delta^2$ and therefore describes a ``unitary'' superconducting pairing. 
In the single-orbital case, the Cooper pairs in the unitary superconducting states have zero spin magnetic moment \cite{SU-review,TheBook}. 

For any Hamiltonian \eqref{eq:H0PlusOT} with $H_0$ given by Eq. \eqref{eq:H0generalSU2}, the states $\ket{\phi_n}$ \eqref{eq:genTowerWf} are eigenstates and MBS for $\Delta=0$. 
For $\Delta\ne0$ the basis rotation occurs in the scar subspace. The lowest-energy state is given by the BCS-like wavefunction

\begin{eqnarray}
\label{eq:spgs}
 \ket{z_0} & \propto & \prod_j \exp\left( -\frac{v}{u} \CO_j^\dagger\right)\ket{0} \nonumber\\
 & = & \prod_j \left( 1 - \frac{v}{u} \CO^{\dagger}_{j} + \frac{v^2}{2u^2}\CO^{\dagger,2}_{j} \right)\ket{0},
\end{eqnarray}
where 
\begin{align}
\label{eq:uvXdef}
 u=\sqrt{\frac{1}{2}\left(1+\frac{\mu_{\rm eff}}{E}\right)}, \quad 
 v=e^{i\theta}\sqrt{\frac{1}{2}\left(1-\frac{\mu_{\rm eff}}{E}\right)},
\end{align}
and $E = \sqrt{\Delta^2+\mu_{\rm eff}^2}$.
Compared to the standard BCS state, the wavefunction \eqref{eq:spgs} also includes the quadratic term $\CO^{\dagger,2}_{j}$, whereas orders up to ${\cal N}$ would appear in a system with ${\cal N}$ orbitals/flavours per site \cite{paperC1}. For $\CO_j$ from \eqref{eq:OdagEBTSS} or \eqref{eq:evenBSST} the $\CO^{\dagger}_{j}\ket{0}$ is a half-filled state on site $j$ while $(\CO^{\dagger}_{j})^2\ket{0}$ is a fully filled state and therefore $(\CO^{\dagger}_{j})^m\ket{0}$ vanishes for $m>2$.

All other states in the scar subspace have the form analogous to Eq. \eqref{eq:genTowerWf}:
\begin{align}
\label{eq:scarsInZnBasis}
\ket{z_n} = \frac{(\sum_j \CO^{\gamma,\dagger}_{j})^n \ket{z_0} }{P_N(n)},
\end{align}
where $0<n\le 2N$ and
\begin{align}
\label{eq:OgammaGeneral}
\CO^{\gamma,\dagger}_{j}  = \frac{\Delta}{2E}\left(2 - n_j + \frac{u}{v}e^{2i\theta}\CO^{\dagger}_{j} - \frac{v}{u}e^{-2i\theta} \CO_{j} \right).
\end{align}
Alternatively to the derivation presented in Ref.  \cite{paperC1} this expression can be obtained by replacing the original fermions in $\CO_{j}$ with the Bogoliubov-transformed fermions. Therefore, at $\Delta\ne0$ the scar tower in the rotated basis $\ket{z_n}$ consists of the same scar states \eqref{eq:genTowerWf} but written in terms of the Bogoliubov fermions instead of the original fermionic operators $c^p_{j\alpha}$. The highest state in the tower $\ket{z_{2N}}$ coincides with the BCS state \eqref{eq:spgs} but with the flipped sign in the exponent. 

The states in the scar subspace (spanned by \eqref{eq:genTowerWf} and \eqref{eq:scarsInZnBasis}) are endowed with strong pairing correlations of the type $\CO$ specified by the matrix $A$ in Eq. \eqref{eq:type1O}. In particular, the BCS state \eqref{eq:spgs} maximizes the absolute value of the one-point function
\begin{align}
\label{eq:1ptFunInSPGSK1}
\braket{z_n|\CO^{\dagger}_{j}|z_n}  = -\frac{\Delta e^{-i\theta} }{E} \left( 1-\frac{n}{N} \right)
\end{align}
over the entire Hilbert space. It can therefore always be made the ground state by adding a sufficiently strong pairing potential $H_\Delta$. Comparing Eqs. \eqref{eq:1ptFunInSPGSK1} and \eqref{eq:H0generalSU2} we note that the expectation value $\braket{H_\Delta}$ and thus the energy of scars do not depend on the phase $\theta$.

The two-point function
\begin{eqnarray}
\label{eq:2ptFunInZInTermsOfMu}
\braket{z_n|\CO^{\dagger}_{i} \CO_{j}|z_n}  = \frac{1}{N(2N-1)}\biggl\{ 2n(2N-n) \nonumber \\ 
     + \frac{\Delta^2}{E^2}\left[3n^2 - 6nN + N(2N-1)\right] \biggr\}
\end{eqnarray}
is also high in the states in the scar subspace and its average over the scar subspace,
\begin{align}
\label{eq:2ptFunAvgOverScarSubsp}
\overline{\braket{z_n|\CO^{\dagger}_{i} \CO_{j}|z_n}} = \frac{\sum_{n=0}^{2N} \braket{z_n|\CO^{\dagger}_{j} \CO_{i}|z_n}}{2N+1} = \frac{2}{3},
\end{align}
which is basis-independent, exceeds its value in almost every state outside the scar subspace for typical Hamiltonians. Furthermore, when measured in any O$(N)$-invariant scar state the two-point function \eqref{eq:2ptFunInZInTermsOfMu} does not depend on the coordinates of the two sites, which by definition amounts to the presence of the ODLRO of type $\CO_{j}$ in any scar state.

For $\Delta=0$, the states $\ket{z_n}$ turn into $\ket{\phi_n}$ but the analytical expressions \eqref{eq:1ptFunInSPGSK1} and \eqref{eq:2ptFunInZInTermsOfMu} remain valid in that basis too.
Both one- and two-point functions remain finite in the thermodynamic limit of large $N$.

\subsection{Dynamical probes}

As long as the full Hamiltonian has the form \eqref{eq:H0PlusOT}, infinite and exact revivals are guaranteed for any initial state from the $G$-invariant subspace. Any such state is a linear combination of $\ket{\phi_n}$ or $\ket{z_n}$.

One possibility to observe revivals is by implementing the Hamiltonian
\begin{align}
\label{eq:HToPrepRevivals}
    H(\lambda) = - \sum_j \mu_{\rm eff}  (2-n_j) + \lambda H_\Delta + (1-\lambda)OT,
\end{align}
where $H_\Delta$ is given by Eq. \eqref{eq:dH0general}. For $\lambda=1$, the above expression coincides with Eq. \eqref{eq:H0generalSU2}, so that all eigenstates are given by $\ket{z_n}$ and the ground state is the BCS-like state $\ket{z_0}$.  
Tuning $\lambda$ to $0$ will turn on the $OT$ terms leading to an ergodic full Hamiltonian. It will also change the basis in the invariant subspace to $\ket{\phi_n}$, with the initial $\ket{z_0}$ state being a suitable linear combination exhibiting revivals.

There is also a possibility that the required linear combination may be created by a short external influence such as in optical pump-probe experiments. Further investigations for specific materials and scar families are needed to carefully study and either confirm or refute this possibility.

\subsection{Outline}

\begin{figure}[htp!]
				\includegraphics[width=0.58\columnwidth]{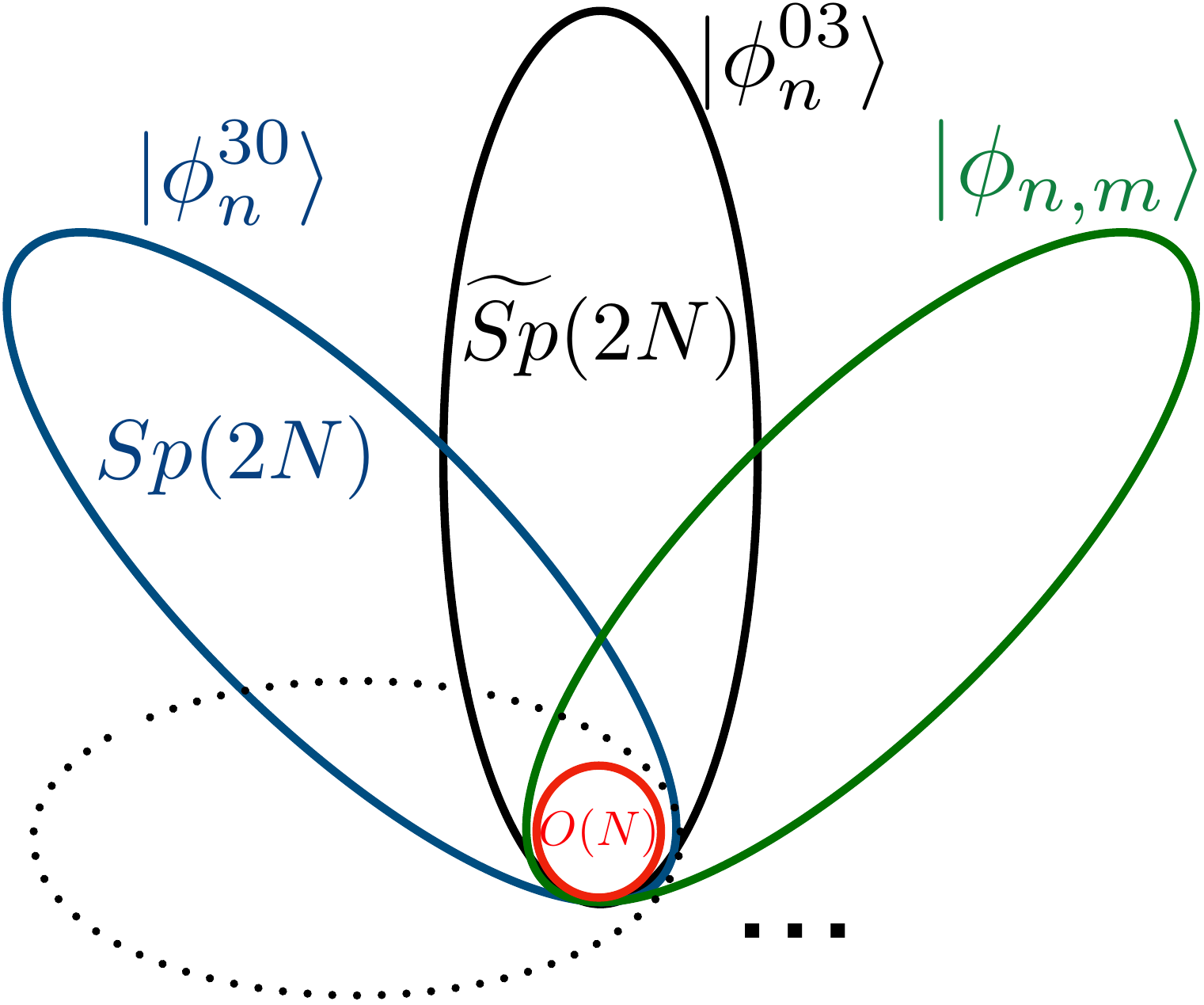}
				\includegraphics[width=0.4\columnwidth]{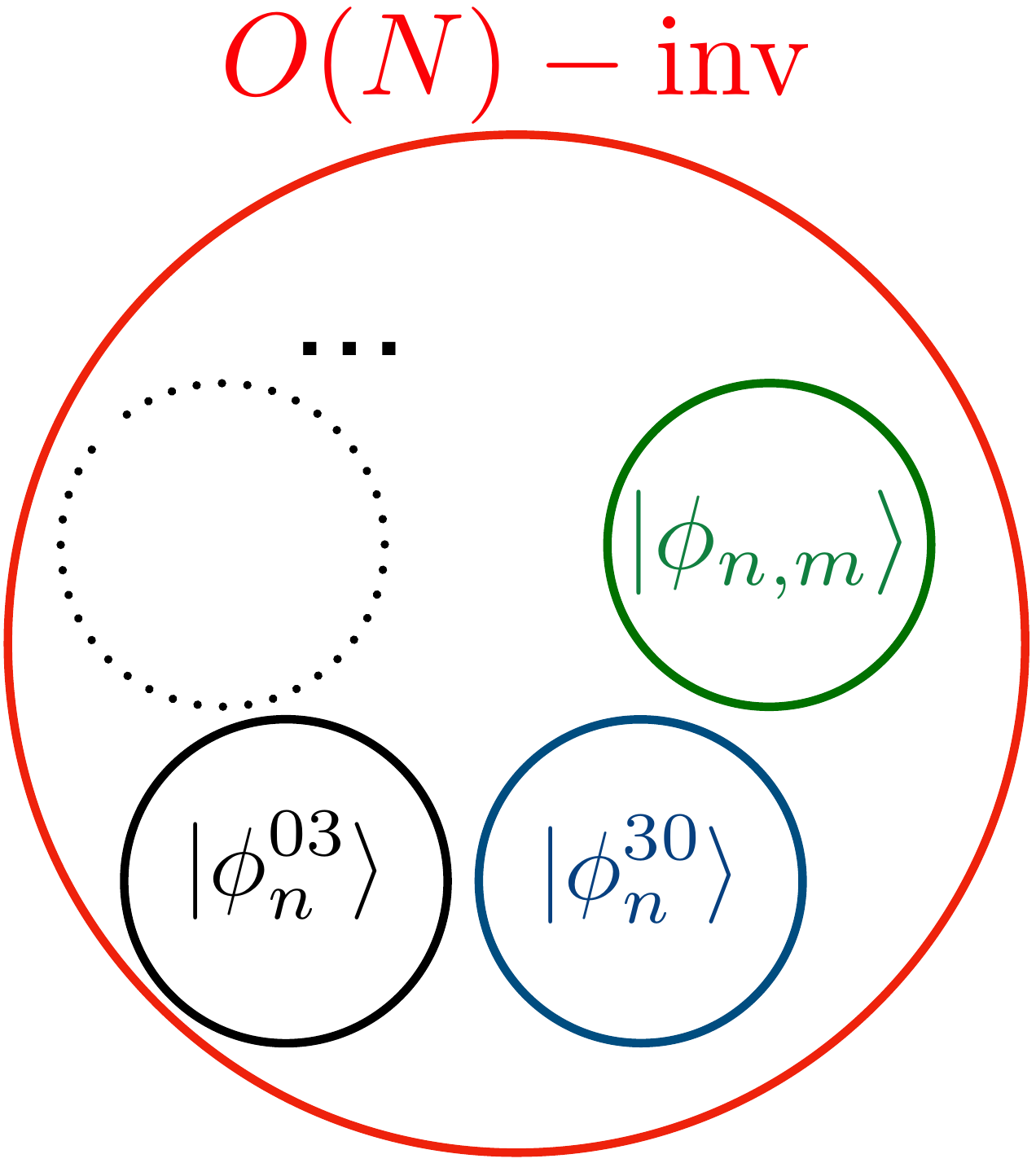}
\caption{a) Some of the groups $G$ appearing in this work and their inter-dependencies. b) Scar subspaces corresponding to these groups. }
\label{fig:schemesoutline}
\end{figure}

The different symmetry groups $G$ (see Fig. \ref{fig:schemesoutline}a) of the scar families we consider all share one common subgroup O$(N)$. The full symmetry group of any particular scar family is higher and has more generators annihilating a specific scar subspace. For the group-invariant scar subspaces schematically depicted in Fig. \ref{fig:schemesoutline}b, the relation is inverse: subspaces with higher symmetry are all sub-sets of the larger subspace of the O$(N)$-invariant states.

Sections \ref{sec:ioPairingSpecialEta}, \ref{sec:triplType1} and \ref{sec:triplType2} each handle the scars corresponding to one of the ``petals'' in Fig. \ref{fig:schemesoutline}a.
In Sec. \ref{sec:ioPairingSpecialEta} we discuss the states $\ket{\phi^{03}_n}$ that are built according to the prescription from the previous subsection using $\CO^{\dagger}=
\CO^{\dagger}_{03}$ \eqref{eq:OdagEBTSS} and show that they have $\widetilde{Sp}(2N)$ symmetry and inter-orbital spin-singlet pairing. Closely related states $\ket{\phi^\pm_n}$, which do not have the general structure of Eq. \eqref{eq:genTowerWf} and are not shown in Fig. \ref{fig:schemesoutline}, arise as a by-product of this discussion. Applying the prescription of \cite{paperC1}, Eq. \eqref{eq:genTowerWf} with $\CO^{\dagger}=\CO^{\dagger}_{30}$ \eqref{eq:evenBSST}, in Sec. \ref{sec:triplType1} we discuss the scars $\ket{\phi_n^{30}}$ with Sp$(2N)$ symmetry and inter-orbital spin-triplet pairing. In Sec. \ref{sec:triplType2} we discuss one further family of scars $\ket{\phi_{n,m}}$ with spin-triplet pairing that goes beyond the construction of Ref. \cite{paperC1}.


\section{Relevant interaction terms \label{sec:relevantInteractions}}

All the scar symmetry groups we will consider share a common O$(N)$ subgroup as illustrated in Fig. \ref{fig:schemesoutline}. For this reason, the Hamiltonian terms $H_0$ and the generators $T$, in the sense of Eq. \eqref{eq:H0PlusOT}, which are suitable for $G=\text{O}(N)$ can also be used as $H_0$ and $T$ for any scar family with a higher symmetry and we list such Hamiltonian terms here. We also discuss the spin-orbit (SO) coupling that happens to be a generator $T$ for several symmetry groups of interest. 

\subsection{$H_0$}

The suitable $H_0$ terms are the O$(N)$-invariant orbital-dependent chemical potential
\begin{align}
\label{eq:H0eta}
H_\mu = \sum_{j,p} \mu_p (n_{jp\uparrow}+ n_{jp\downarrow}),
\end{align}
with $n_{jp\alpha}=c_{j\alpha}^{p,\dagger}c_{j\alpha}^p$, and a specific generalization of the Hubbard interaction
\begin{align}
\label{eq:HhubShort}
H_{\mathrm{Hub}} = \frac{U}{2} \sum_j \left(n_j - M^2_j\right),
\end{align}
where $M_j=\sum_p(n_{jp\uparrow}-n_{jp\downarrow})$ is the on-site magnetization. In the single-orbital case, the latter interaction coincides with the standard Hubbard interaction $n_{j\uparrow}n_{j\downarrow}$.
It can also be obtained from the Hubbard Hamiltonian often considered in two-orbital systems (e.g. in $\rm{Sr_2RuO_4}$, see Ref. \cite{typicalUSCHamZinkl2021}), which includes intraorbital ($V$), interorbital ($K$), and Hund’s rule ($J$) couplings:
\begin{eqnarray}
\label{eq:Hhub}
 H_{\mathrm{Hub}} &=& (K-J)\sum_{ j,\alpha} n_{j x\alpha} n_{jy\alpha}+ V \sum_{j,p} n_{j p\uparrow}n_{j p\downarrow} \nonumber \\
 && +K\sum_{j,pq} n_{j p\uparrow}\tau^1_{pq}n_{j q\downarrow} ,
\end{eqnarray}
with a specific choice $V=K=U$, $J=2U$.
Two other generalizations of the Hubbard interaction are compatible with one of the scar families and will be discussed in Sec. \ref{sec:type2tripletH}.

\subsection{Generators $T$}

\subsubsection{Hopping terms \label{sec:HbHoppingTerms}}

The orbital- and spin-blind imaginary-amplitude hopping between the sites $i$ and $j$,
\begin{eqnarray}
\label{eq:regularImAmplHopping}
    T_{ij} = i\sum_{p,\alpha} \left(c^{p,\dagger}_{i\alpha}\, c^{p}_{j\alpha}-\text{H.c.}\right),
\end{eqnarray}
is a generator of O$(N)$ and thus annihilates all the MBS in this work. 
The group generated by the hopping terms depends on its amplitude being imaginary, real, or complex, as was first discussed in the single-orbital case \cite{pakrouski2020GroupInvariantScars,Pakrouski:2021jon}. While the corresponding groups are closely related, their singlet subspaces \eqref{eq:genTowerWf} are distinct. To keep this work concise we only focus on the above hopping with purely imaginary amplitude. We expect that the results for purely real hopping can be obtained by a mapping in the same fashion as in the single-orbital case \cite{Pakrouski:2021jon}.

\subsubsection{SO coupling}

The SO coupling terms in solids depend on the type of orbitals present, their orientation and dimensionality of the system. We leave for the future work a systematic study of more general SO couplings that support MBS. As an example, here we consider a particular type of SO coupling that is relevant under certain approximations in ${\mathrm{Sr}}_{2}\mathrm{Ru}{\mathrm{O}}_{4}$ and more generally in 2D systems with a square lattice and two degenerate orbitals per site with in-plane dispersion. 

We generalize the SO coupling on site $j$ from Ref. \cite{typicalUSCHamZinkl2021} as follows:
\begin{eqnarray}
\label{eq:Hsooriginal}
T_{\mathrm{SO},j}^A = -i \sum_{\alpha\beta} c^{x,\dagger}_{j\alpha} \sigma^A_{\alpha\beta }c^y_{j\beta} + \mathrm{H.c.},
\end{eqnarray}
by allowing the spin direction $A$ to take values 1 and 2 in addition to $A=3$. The corresponding Hamiltonian term may in general have site-dependent strength:
\begin{equation}
\label{eq:TsoA-r}
    T_{\mathrm{SO}}^A = \sum_j r^{\mathrm{SO}}_j T_{\mathrm{SO},j}^A.
\end{equation}
The operators (\ref{eq:Hsooriginal}) annihilate several scar families in this work and are generators of their full symmetry group.

\subsection{Common Hamiltonian \label{sec:bareH}}

The minimal Hamiltonian shared by all the scar families in this work is given by
\begin{align}
\label{eq:bareH}
 H_{\mathrm{O}(N)} =  H^{c}_0 + T_h,
\end{align}
where 
\begin{align}
\label{eq:bareH0}
H^{c}_0 =  H_\mu +H_{\mathrm{Hub}},
\end{align}
and
\begin{align}
\label{eq:bareOT}
T_h=  \sum_{\braket{i,j}} T_{ij},
\end{align}
with $T_{ij}$ defined in Eq. \ref{eq:regularImAmplHopping}. For the sake of simplicity, we will focus on the nearest-neighbour hopping, while longer-range terms with $i,j$-dependent strength could be used just as well.
Strength of the hopping term is set to 1 and therefore all the other energy scales in this work are measured in units of hopping.

Note that the chemical potentials in Eq. \eqref{eq:H0eta} are orbital-dependent, while Ref. \cite{paperC1} parametrizes the corresponding $H_0$ term \eqref{eq:H0PlusOT} as $\mu_{\rm eff} \sum_j [\CO_{j},\CO^{\dagger}_{j}] = \mu_{\rm eff}\sum_j(2-n_j)$, with an orbital-independent $\mu_{\rm eff}$. In $H_0^c$, there is another contribution containing $n_j$, which comes from the Hubbard term. Both pairing operators we consider produce exactly equal occupation in each of the two orbitals, the property inherited by the MBS states. For these reasons, the effective chemical potential to be used in the expressions in Sec. \ref{sec: recap} is
\begin{align}
\label{eq:bareH0effectiveMu}
\mu_{\rm eff} = \frac{\mu_x + \mu_y}{2}+\frac{U}{2}.
\end{align}

The Hamiltonian \eqref{eq:bareH}, example Hamiltonians in the sections below and all the analytical results in this work are lattice-, dimension- and (mostly) system size-independent. For numerical tests of our results we will implement exact diagonalization of these Hamiltonians on a 1D chain with $N=4$ sites and periodic boundary conditions for hopping.

\subsection{The auxiliary $OT$ term}

The low entanglement and periodic revivals exhibited by many-body scars are truly non-trivial in a chaotic Hamiltonian without symmetries. While the many-body Hamiltonians are in general expected to be of this type, one sometimes can only clearly confirm this numerically (by studying the level statistics for example) for relatively large systems. To ensure that most symmetries are broken and that the bulk of the spectrum (excluding scars) is chaotic already for small systems, we will add to the Hamiltonian an $OT$ term specified below in the numerical simulations.

As an auxiliary term it is not meant to have a clear physical origin. We design it to be short-range and it could be interpreted as a quite drastic perturbation that by the virtue of being of the $OT$ form has no influence on scars but strongly mixes all other states, which is achieved by building in some randomness. 
 
Specifically, we will use
\begin{gather}
\label{eq:TOT2}
OT =  \sum_{\braket{i,j},p} T_{ij} V_{i}^p T_{ij},  \\ \nonumber
V_{i}^p =  r^{(1),p}_{i,\alpha\beta}c^{p,\dagger}_{i\alpha} c^{p+1}_{i\beta} + r_{i,\alpha\beta}^{(2),p}c^{p,\dagger}_{i\alpha} c^{p+1,\dagger}_{i\beta} + {\rm H.c.},
\end{gather}
where $r^{(1),p}_{i,\alpha\beta}$ and $r^{(2),p}_{i,\alpha\beta}$ are real random numbers, uniformly distributed between -0.5 and 0.5, and $T_{ij}$ are the hopping terms from Eq. \eqref{eq:regularImAmplHopping}, which are the generators of the common symmetry subgroup O$(N)$ for all the scars in this work. Additional generators $T$ will be specified for each scar family according to its full symmetry group.


\section{Scar families with inter-orbital spin-singlet pairing \label{sec:ioPairingSpecialEta}}

This section is devoted mainly to the family of scars $\ket{\phi_n^{03}}$ with inter-orbital pairing. However, another family $\ket{\phi_n^\pm}$ naturally accompanies it and will be briefly discussed as well.

\subsection{Inter-orbital $\eta$ states and their full symmetry group}
\label{sec:fullSymmGroupForInterOrbital}

The inter-orbital $\eta$ states 
\begin{equation}
\label{eq:phi-03-definition}
    \ket{\phi_n^{03}} = \frac{(\sum_j\CO^{\dagger}_{03,j})^n \ket{0}}{P_N(n)}
\end{equation}
are obtained using the general form \eqref{eq:genTowerWf} and the spin-singlet inter-orbital pairing operator $\CO^{\dagger}_{03,j}=c^{x,\dagger}_{j\uparrow}c^{y,\dagger}_{j\downarrow}+c^{y,\dagger}_{j\uparrow}c^{x,\dagger}_{j\downarrow}$. For $H^{c}_0$ \eqref{eq:bareH0}, the energy of these states is 
\begin{gather}
\label{eq:ioetaEnergies}
    E_n = n(\mu_x+\mu_y+U),
\end{gather}
and they become degenerate when $\mu_x+\mu_y=-U$.
The states \eqref{eq:phi-03-definition} have the same particle number $n$ for both orbitals $x$ and $y$. Since $n$ cannot be larger than $2N$, there are $2N+1$ such states in total and $\ket{\phi_{2N}^{03}}$ is fully occupied. 

In the single-orbital models, the $\eta$-pairing states are created by the $\eta$-pairing operator $\eta^\dagger = \sum_j c_{j\uparrow}^\dagger c_{j\downarrow}^\dagger$.
These states are invariant under a symplectic group, denoted by $\tSp(N)$, which is the semi-direct product of $\ttU(N)$ and the spin SU$(2)$ group \cite{Pakrouski:2021jon}. The $\ttU(N)$ group is defined such that $c_{j\uparrow}$ and $c_{j\downarrow}^\dagger$ transform in its fundamental representation. Precisely, the (complexified) generators of $\tSp(N)$ are 
\begin{align}\label{Spbasis1}
&\widetilde{\cal T}_{i,j}=c^\dagger_{i\uparrow}c_{j\uparrow}-c^\dagger_{j\downarrow}c_{i\downarrow}, \nonumber\\
&\widetilde{\cal T}^+_{i,j}=c^\dagger_{i\uparrow}c_{j\downarrow}+c^\dagger_{j\uparrow}c_{i\downarrow}, \\
&\widetilde{\cal T}^-_{i,j}=c^\dagger_{i\downarrow}c_{j\uparrow}+c^\dagger_{j\downarrow}c_{i\uparrow}, \nonumber
\end{align}
where $\widetilde{\cal T}_{i, j}$ generate $\ttU(N)$.
We will show that in the two-orbital models, the inter-orbital $\eta$ states also admit a similar symplectic structure.

At each site $j$,  define fermionic operators 
\begin{align}\label{uvspinor}
    w_{j\alpha} \equiv  \frac{c_{j\alpha}^x-i c^y_{j\alpha}}{\sqrt{2}}, \quad \bar w_{j\alpha} \equiv  \frac{c_{j\alpha}^x+i c^y_{j\alpha}}{\sqrt{2}} ,
\end{align}
and their corresponding creation operators. They satisfy the standard anti-commutation relations $\{w, w^\dagger\} =\{\bar w, \bar w ^\dagger\} = 1$ and $\{w, \bar w^\dagger\} =\{\bar w, w^\dagger\} = 0$. Using $w$ and $\bar w$, we define a new single-orbital spinor $\psi_{I\alpha}$ ($1\le I\le 2N$): 
 \begin{align}\label{psispinor}
 \psi_{I\alpha} = \begin{cases} e^{i\pi/4} w_{I\alpha} ,   &1\le I\le N, \smallskip \\ e^{-i\pi/4} \bar w_{I-N, \alpha} ,   &N+1\le I\le 2N, \end{cases}
 \end{align}
 which can be thought of as a Dirac fermion defined on a lattice of $2N$ sites with $I$ being the site label. 
 
A crucial property of this ``extended'' lattice model is that the operator 
$$
 \widetilde{\mathfrak{t}}^+ = \frac{1}{2} \sum_j (c^{x,\dagger}_{j\uparrow}c^{y,\dagger}_{j\downarrow}+c^{y,\dagger}_{j\uparrow}c^{x,\dagger}_{j\downarrow}) = \frac{1}{2}\sum_j \CO^{\dagger}_{03,j},
$$
which generates the inter-orbital $\eta$ states, coincides with the $\eta$-pairing operator $\eta^+_\psi = \sum_I \psi^\dagger_{I\uparrow}\psi^\dagger_{I\downarrow}$ of the $\psi_{I\alpha}$ fermions.
Because of this hidden one-orbital $\eta$-pairing structure, the $|\phi_n^{03}\rangle$ states are invariant under an $\tSp(2N)$ group, whose {\bf $2N(4N+1)$} generators can be obtained by  substituting $c_{j\alpha}\to \psi_{I\alpha}$ in \eqref{Spbasis1}:
 \begin{align}\label{LargedSpbasis1}
&\widetilde{\mathcal{T}}_{I, J}=\psi_{I\uparrow}^\dagger\psi_{J\uparrow}-\psi^\dagger_{J\downarrow}\psi_{I\downarrow} ,\nonumber\\
& \widetilde{\mathcal{T}}^+_{I, J}=\psi^\dagger_{I\uparrow}\psi_{J\downarrow}+\psi^\dagger_{J\uparrow}\psi_{I\downarrow} ,\\
&\widetilde{\mathcal{T}}^-_{I,J}=\psi^\dagger_{I\downarrow}\psi_{J\uparrow}+\psi^\dagger_{J\downarrow}\psi_{I\uparrow} .\nonumber
\end{align}

Next, we discuss some important subgroups of $\tSp(2N)$. By restricting $I, J$ to $\{1, 2, \cdots, N\}$ or $\{N+1, N+2, \cdots, 2N\}$, we obtain two commuting symplectic groups $\tSp_w(N) = \ttU_w(N)\rtimes \SU_w(2)$ and $\tSp_{\bar w}(N) = \ttU_{\bar w}(N)\rtimes \SU_{\bar w}(2)$ respectively. Here $\SU_w(2)$ denotes the spin group of the $w$ spinors and $\SU_{\bar w}(2)$ denotes the spin group of the $\bar w$ spinors. They form an SO$(4)$ group, whose generators include the spin generators of the original $c^p_{j\alpha}$ fermions $S_\mu =\sum_j S_{\mu, j}$, with
\begin{align}
\label{eq:spinGens}
    S_{\mu, j}= \frac{1}{2}\sum_{p,\alpha\beta} c^{p,\dagger}_{j\alpha}\sigma^\mu_{\alpha\beta}c^p_{j\beta},
\end{align}
as well as $K_\mu = \sum_j K_{\mu,j}$, with
\begin{eqnarray}
\label{eq:gensKSU2main}
K_{\mu,j} = -\frac{i}{2} \sum_{\alpha\beta} \sigma^\mu_{\alpha\beta} \left(c^{x,\dagger}_{j\alpha}
c^y_{j\beta} -c^{y,\dagger}_{j\alpha} c^x_{j\beta}\right),
\end{eqnarray}
where $\mu=1, 2, 3$. In particular, the SO coupling $T_{\mathrm{SO}}^3$, see Eq. (\ref{eq:TsoA-r}) with all $r^{\mathrm{SO}}_j=1$, equals $2K_3$ and hence annihilates all the $|\phi_n^{03}\rangle$ states.

The $K_\mu$ operators transform as a vector under spin $\SU(2)$, i.e. $[S_\mu, K_\nu] = i\sum_\rho\epsilon_{\mu\nu\rho} K_\rho $, and their commutators yield the spin generators, i.e. $[K_\mu, K_\nu] = i\sum_\rho\epsilon_{\mu\nu\rho} S_\rho $. The inter-orbital $\eta$ states $\ket{\phi_n^{03}}$ (and some further scar families in this work) are annihilated not only by $K_\mu$ and $S_{\mu}$ but also by every on-site generator $K_{\mu,j}$ and $S_{\mu, j}$ which means that these terms, including the SO coupling $T_{\mathrm{SO}}^3$ \eqref{eq:Hsooriginal}, may appear in the Hamiltonian \eqref{eq:H0PlusOT} with a site-dependent strength.

Spin generators \eqref{eq:spinGens} are a subset of the algebra corresponding to the full symmetry group of the $\ket{\phi^{03}_n}$ scars. As a consequence these states have exactly zero magnetic moment. This property holds for all superconducting states with unitary pairing in the single-orbital case \cite{SU-review,TheBook}. In Sec. \ref{sec:triplType1} we will present an explicit two-orbital example of a unitary pairing with non-zero total magnetic moment. 

Under the Shiba transformation $c_{j\downarrow}^p\to c_{j\downarrow}^{p,\dagger}$, the inter-orbital $\eta$ states are mapped to inter-orbital $\zeta$ states \cite{paperC1}, which is the obvious generalization of the single-orbital $\zeta$ states \cite{Pakrouski:2021jon}. Since $K_3$ is invariant under the Shiba transformation, the  inter-orbital $\zeta$ states are also annihilated by the SO coupling $T_{\mathrm{SO}}^3$.

\subsection{Triplet inter-orbital $\eta$ states}

Another family of scar states, the spin-triplet inter-orbital $\eta$ states $\ket{\phi^{\pm}_n }$, arises naturally as a by-product for some of the Hamiltonians hosting $\ket{\phi_n^{03}}$. These additional states are constructed analogously to $\ket{\phi_n^{03}}$ \eqref{eq:phi-03-definition}, but the very first ($n=1$) pair that is placed into vacuum is replaced with the spin-triplet orbital-singlet pair:
\begin{align}
\label{eq:triplSpecEtaStates}
\ket{\phi^{\pm}_n } \equiv \frac{(\sum_j \CO^{\dagger}_{03,j})^n \ket{0^{\pm}_t}}{ P_{N-1}(n)}, \quad 0\le n\le 2N-2,
\end{align}
where
\begin{align}
\ket{0^{+}_t} =  \frac{-i}{\sqrt{2N}} \sum_j \CO^{\dagger}_{20,j} \ket{0}, \
\ket{0^{-}_t} =  \frac{-1}{\sqrt{2N}} \sum_j \CO^{\dagger}_{10,j} \ket{0}.
\end{align}
The energies of these states for $H_0^c$ \eqref{eq:bareH0} are
\begin{align}
\label{eq:ietaTriplEnergies}
E^{\text{tripl}}_n = (n+1)(\mu_x+\mu_y+U)-2U.
\end{align}
The triplet inter-orbital $\eta$ states $\ket{\phi^{\pm}_n }$ are O$(N)$-invariant and belong to the  (1,1,0,...,0) representation of the flattened $\tSp(2N)$ with Casimir $4N$.

\subsection{Compatible Hamiltonian terms \label{sec:specEtaCompHterms}}

The orbital-dependent chemical potential term $H_\mu$ \eqref{eq:H0eta} and the Hubbard interaction $H_{\mathrm{Hub}}$ \eqref{eq:HhubShort} are both valid $H_0$ terms in the sense of Eq. \ref{eq:H0PlusOT}.

\subsubsection{$OT$ \label{sec:OTinH}}

Besides the generators of the full symmetry group provided in Sec. \ref{sec:fullSymmGroupForInterOrbital}, valid $OT$ terms include the site-dependent Zeeman coupling with the magnetic field strength absorbed into the real coefficients $r^{\text{Z}}_j$:
\begin{align}
\label{eq:siteDepZeeman}
    H_{\text{Z}}^{A} = \sum_j r^{\text{Z}}_j H_{\text{Z},j}^{A} = \sum_{j,p,\alpha\beta} r^{\text{Z}}_j c^{p,\dagger}_{j\alpha} \sigma^A_{\alpha\beta} c^{p}_{j\beta},
\end{align}
which annihilates the inter-orbital $\eta$ states $\ket{\phi^{03}_n}$ (because they are spin-singlets) and the following triplet inter-orbital $\eta$ states: $H_{\text{Z},j}^{1}\ket{\phi^{-}_n}=H_{\text{Z},j}^{2}\ket{\phi^{+}_n}=0$. All other combinations of $A$ and triplet inter-orbital $\eta$ states are incompatible.

The orbital-blind spin-dependent hopping terms 
\begin{align}
\label{eq:soHopping}
T^A_{\text{s},ij} = \sum_{p,\alpha\beta} c^{p,\dagger}_{i\alpha} \sigma^A_{\alpha\beta} c^{p}_{j,\beta}+{\rm H.c.}
\end{align}
are the generators of the diagonal subgroup of $\tSp_w(N)\times \tSp_{\bar w}(N)\subset \tSp(2N)$ and therefore annihilate $\ket{\phi_n^{03}}$. Their action on the triplet inter-orbital $\eta$ states is analogous to the Zeeman terms: the states get mixed with the thermal bulk for most combinations, but are annihilated in two cases $T^{1}_{\text{s},ij} \ket{\phi^{-}_n}=T^{2}_{\text{s},ij} \ket{\phi^{+}_n}=0$.

Expression \eqref{eq:soHopping} is the simplest special case of terms of this type. Hopping in 2D with direction-dependent strength or long-range hopping or other variations may be used instead, as long as the spin- and orbital dependence remains the same as in \eqref{eq:soHopping}.

\subsection{Example Hamiltonian }

Our example Hamiltonian supporting the inter-orbital $\eta$ scars $\ket{\phi_n^{03}}$,
\begin{align}
\label{eq:Hminimal}
H^{A}_{\text{i}\eta} =H_{\mathrm{O}(N)} + H_{\text{Z}}^{A} + T_{\mathrm{SO}}^3,
\end{align}
is obtained by combining the Zeeman term with the common part $H_{\mathrm{O}(N)}$ \eqref{eq:bareH}, which includes the chemical potential, hopping, and the Hubbard interaction and with the SO coupling terms $T_{\mathrm{SO}}^3$ \eqref{eq:Hsooriginal} that mix all the O$(N)$-invariant states except for $\ket{\phi_n^{03}}$ with thermal bulk. The scar subspace of the Hamiltonian \eqref{eq:Hminimal} has the full $\tSp(2N)$ symmetry. While in the absence of the SO coupling the states $\ket{\phi_n^{03}}$ would also be eigenstates, they would typically be degenerate (and of mixed character) with other O$(N)$-invariant states. 
The hopping $T_h$ is natural in a Hamiltonian such as \eqref{eq:Hminimal} but is optional from the scars point of view.

\begin{figure}[htp!]
	\begin{center}
				\includegraphics[width=0.91\columnwidth]{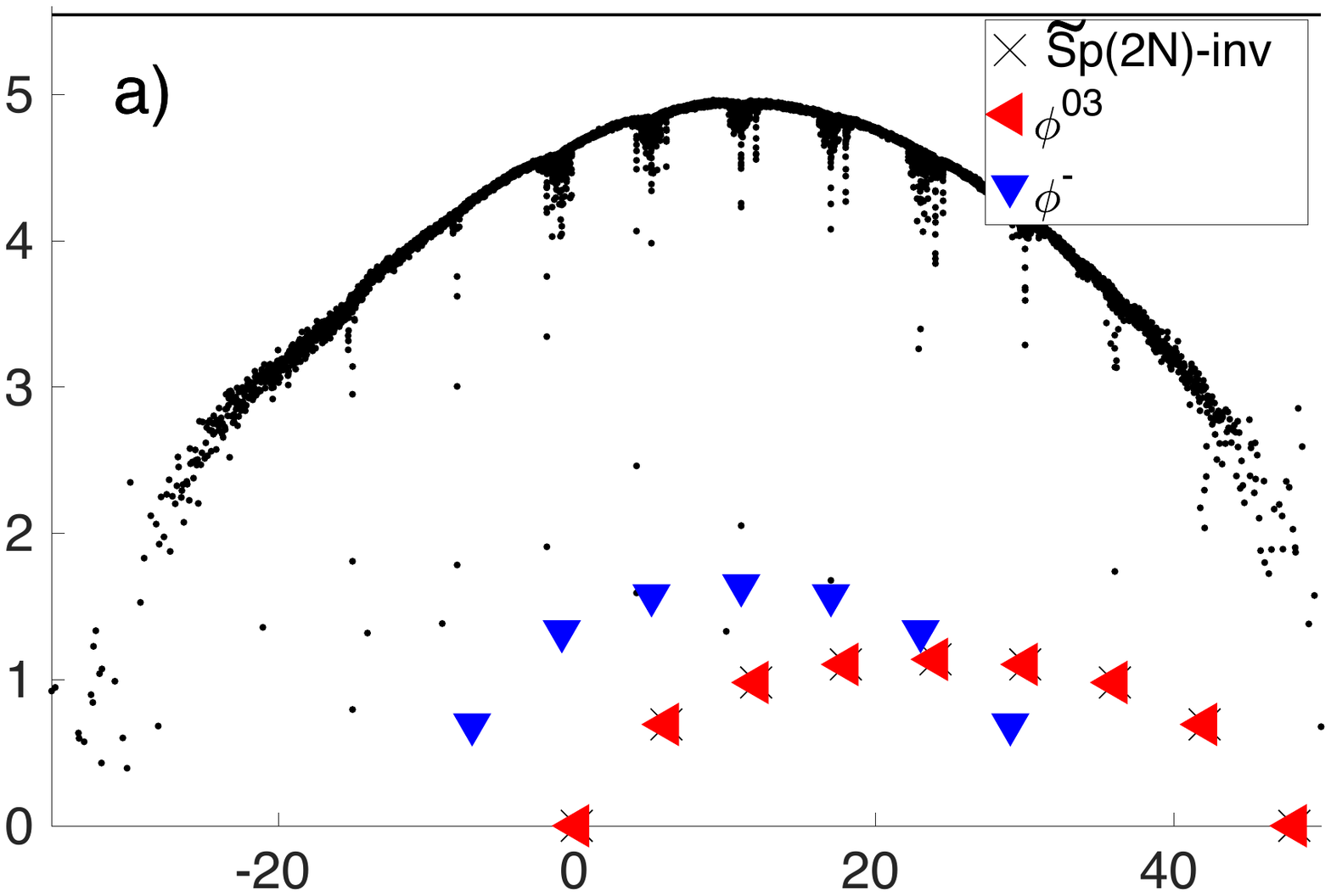}
				\includegraphics[width=0.98\columnwidth]{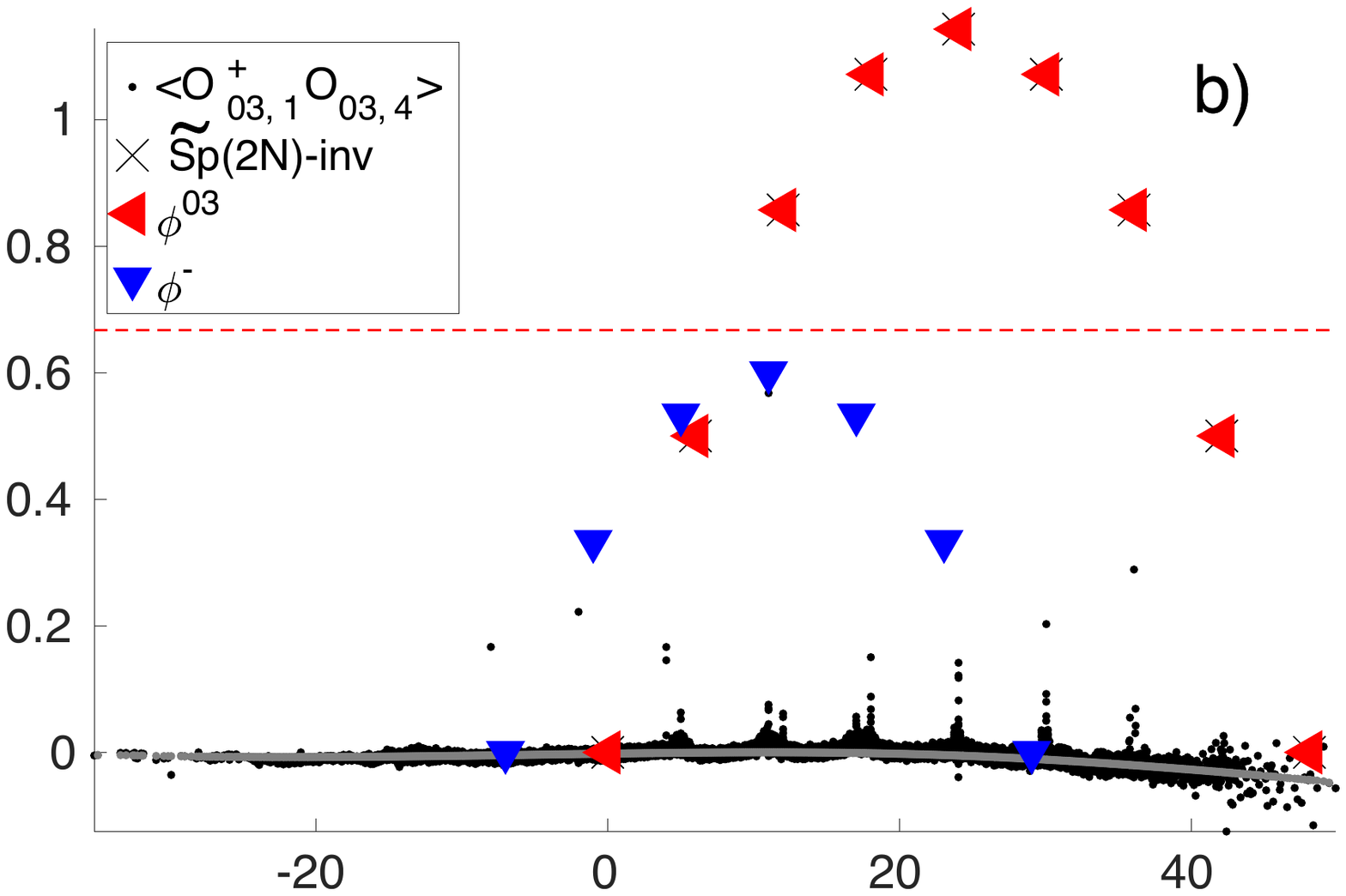}
				\includegraphics[width=\columnwidth]{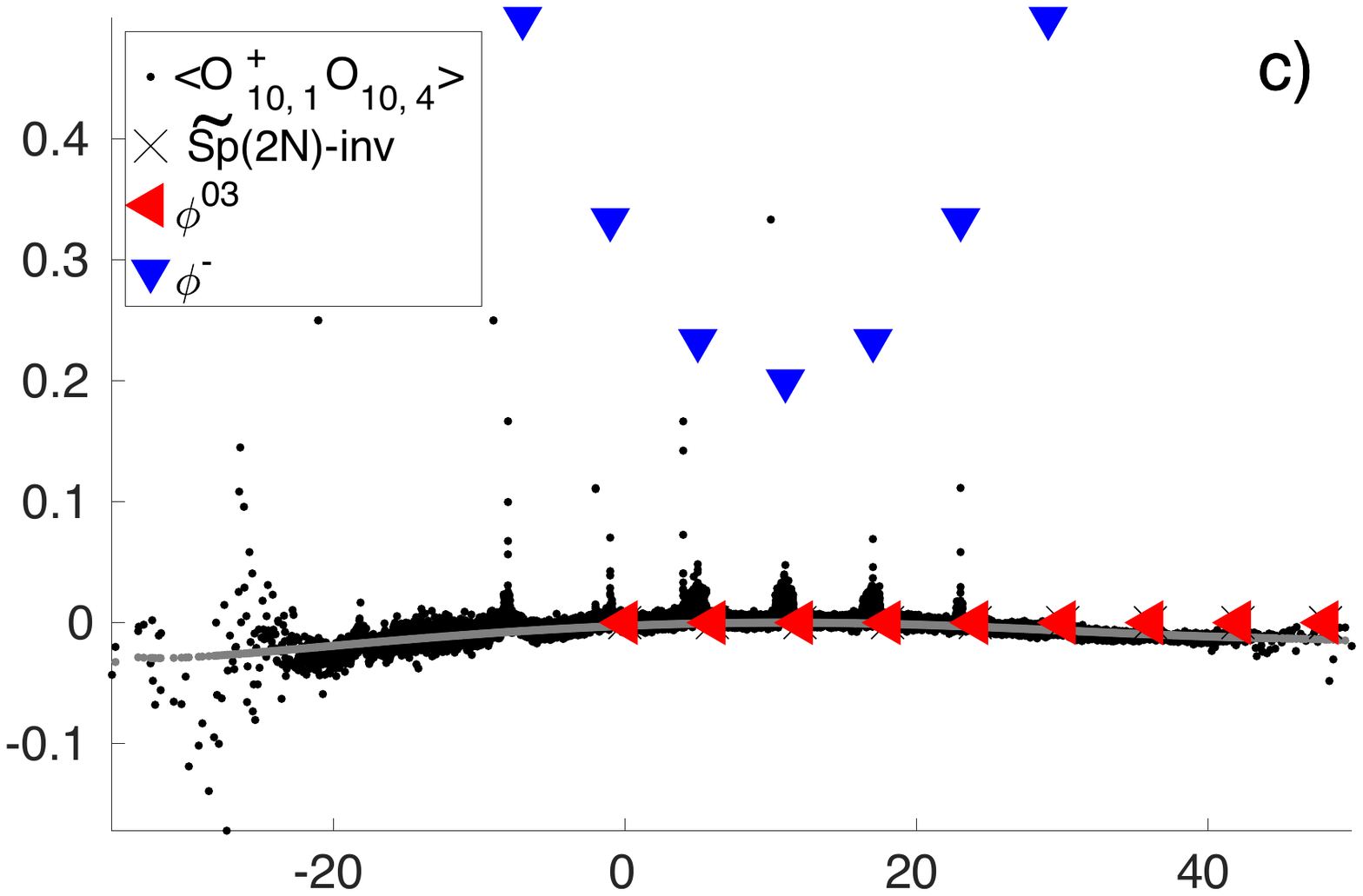}
	\end{center}
\caption{Numerical results for the $H^{1}_{\text{i}\eta}$ \eqref{eq:Hminimal} and $N=4$. Horizontal axis in this and all other figures is the energy. a) Entanglement entropy b) Spin-singlet orbital-triplet ($\mu=0$, $\nu=3$) pairing ODLRO. Dashed red line shows the average over the $\ket{\phi_n^{03}}$ states. c) Spin-triplet orbital-singlet ($\mu=1$, $\nu=0$) pairing ODLRO. In both cases, the two-point function is measured between the most distant sites ($i=1$ and $j=N$).}
\label{fig:minH1Plots}
\end{figure}

The states $\ket{\phi_n^{03}}$ are scars for any $A$, with or without the Zeeman term. The Hamiltonian $H^{1}_{\text{i}\eta}$, due to the Zeeman term, also singles out $\ket{\phi^{-}_n}$ scars and $H^{2}_{\text{i}\eta} $ - the $\ket{\phi^{+}_n}$ scars, as can be seen in Fig. \ref{fig:minH1Plots}, where we show numerical results for $A=1$, $\mu_x=-0.23$, $\mu_y=-0.27$, $r_j^Z=0.1$, $r_j^{\mathrm{SO}}=0.2$, and $U=6.51$. Note that one could choose site-dependent Zeeman and/or SO coupling without significant changes to the results. In the absence of the Zeeman term, $\ket{\phi^{-}_n}$ can no longer be identified by overlap as these states likely become degenerate with some other states. 

In our simulation the $OT$ term given by Eq. \eqref{eq:TOT2} with strength 1.252 is extended by adding the term $0.5008\sum_j T_{\mathrm{SO},j}^3 O_j T_{\mathrm{SO},j}^3$ with the generators \eqref{eq:Hsooriginal}. The generic low-entropy states seen in Fig. \ref{fig:minH1Plots}a could be eliminated (better mixed with thermal states) by increasing the strength of the $OT$ term or by adding further eligible $T$ terms such as $T^A_{s}$ (with $A=1$ or $2$).
The $OT$ term Eq. \eqref{eq:TOT2} is the only particle number non-conserving term in the model used for Fig. \ref{fig:minH1Plots} and it was so chosen in order to have all the particle number sectors in one figure. However, an expression preserving the particle number would be equally valid.

The absolute value of the ODLRO expectation value (i.e. the two-point correlation function) corresponding to the unconventional inter-band pairing \eqref{eq:OdagEBTSS} in the $\ket{\phi_n^{03}}$ scar family is on average a factor of 87 (43 for the $\ket{\phi^{-}_n}$ scars) higher than in thermal states. These large factors result by dividing the constant average within the scar subspace, see Eq. \eqref{eq:2ptFunAvgOverScarSubsp}, by small, parameter-sensitive averages over the generic, thermal states. The values for the individual scar states are in agreement with the analytical expression \eqref{eq:2ptFunInZInTermsOfMu}. 

The states $\ket{\phi^{-}_n}$ also possess a significant unconventional spin-triplet ODLRO of the type given by Eq. \eqref{eq:evenBSST} that in absolute value is on average a factor of 60 higher than in thermal states.
Analytical expressions for the ODLRO expectations values in the spin-triplet inter-orbital states are derived analogously to those in $\ket{\phi_n^{03}}$ states \cite{paperC1}:
\begin{equation}
    \langle\phi^{\pm}_n|\CO^{\dagger}_{03,i}\CO_{03,j}|\phi^{\pm}_n\rangle  = \frac{2 n (N-2) (2N-2-n)}{N(N-1) (2N-3)} ,
\end{equation}
and
\begin{eqnarray}
    \langle \phi^{-}_n|&&\CO^{\dagger}_{1 0, i} \CO_{1 0, j} |\phi^{-}_n\rangle \nonumber\\
    && = 2\frac{(N-n-1)^2+(N-1)(N-2)}{N (N-1)(2N-3)},
\end{eqnarray}
and are in agreement with the numerical results.

Both scar families form equally-spaced towers, with the energies given by Eqs. \eqref{eq:ioetaEnergies} and \eqref{eq:ietaTriplEnergies} in the middle of the spectrum and have low entanglement, as analytically expected for group-invariant scars \cite{paper3Case1Majorana}.

In Appendix \ref{sec:appMoreGeneralHForIoEta} we discuss how only the inter-orbital $\eta$ states $\ket{\phi_n^{03}}$ remain scars for a more general Hamiltonian \eqref{eq:Hfull} that includes all the components of the spin-dependent hopping, while the triplet states $\ket{\phi^{-}_n}$ are no longer scars in that case. We also discuss there how the same families of states remain scars for the Hamiltonian \eqref{eq:Hminimal} and small Hubbard $U$, which can then be considered a small perturbation added to the mean-field without meaningfully affecting the scar subspace.

\subsection{Making inter-orbital $\eta$ the ground state \label{sec:ws521g}}

\begin{figure}[htp!]
	\begin{center}
	\includegraphics[width=0.8\columnwidth]{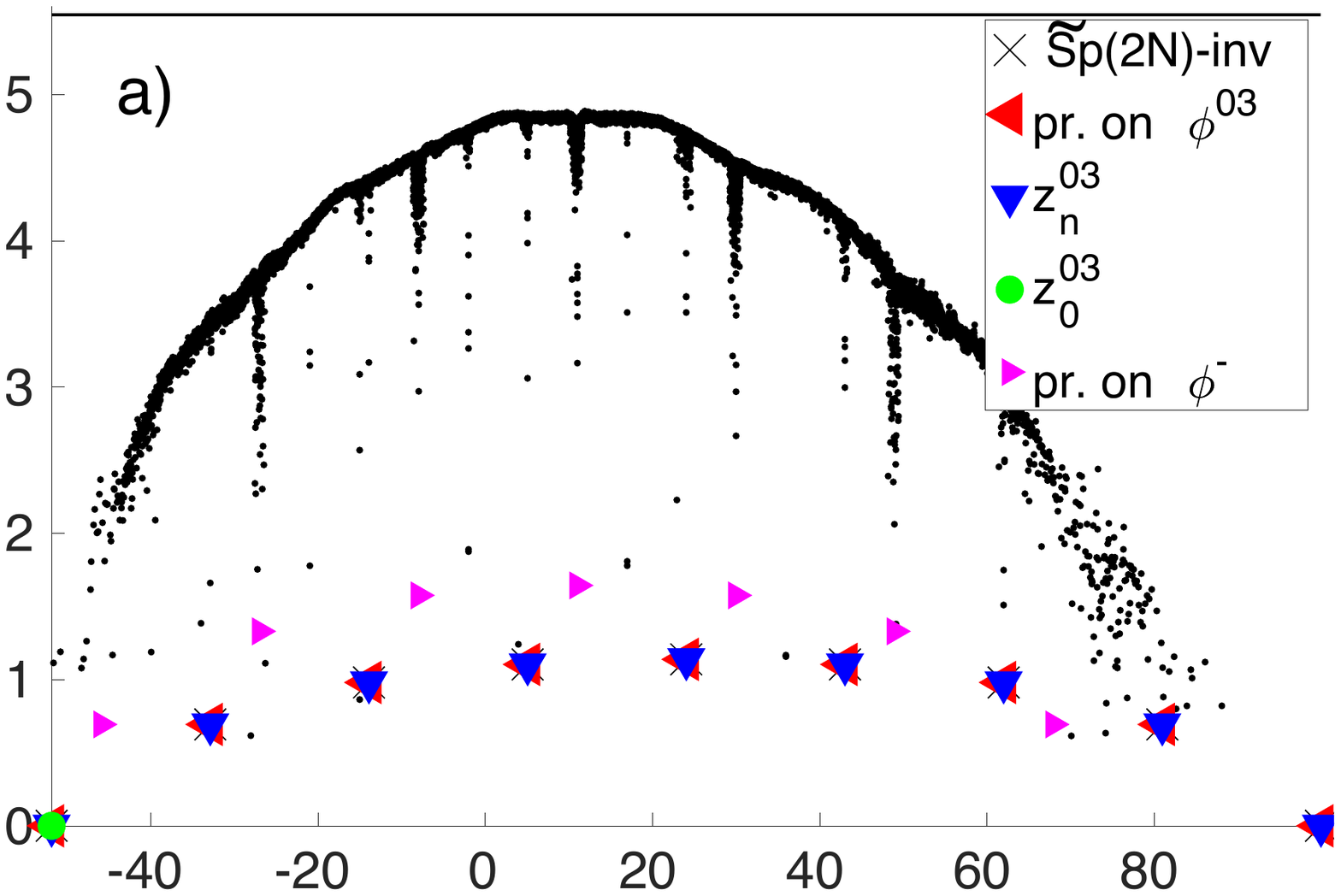}
	\includegraphics[width=0.88\columnwidth]{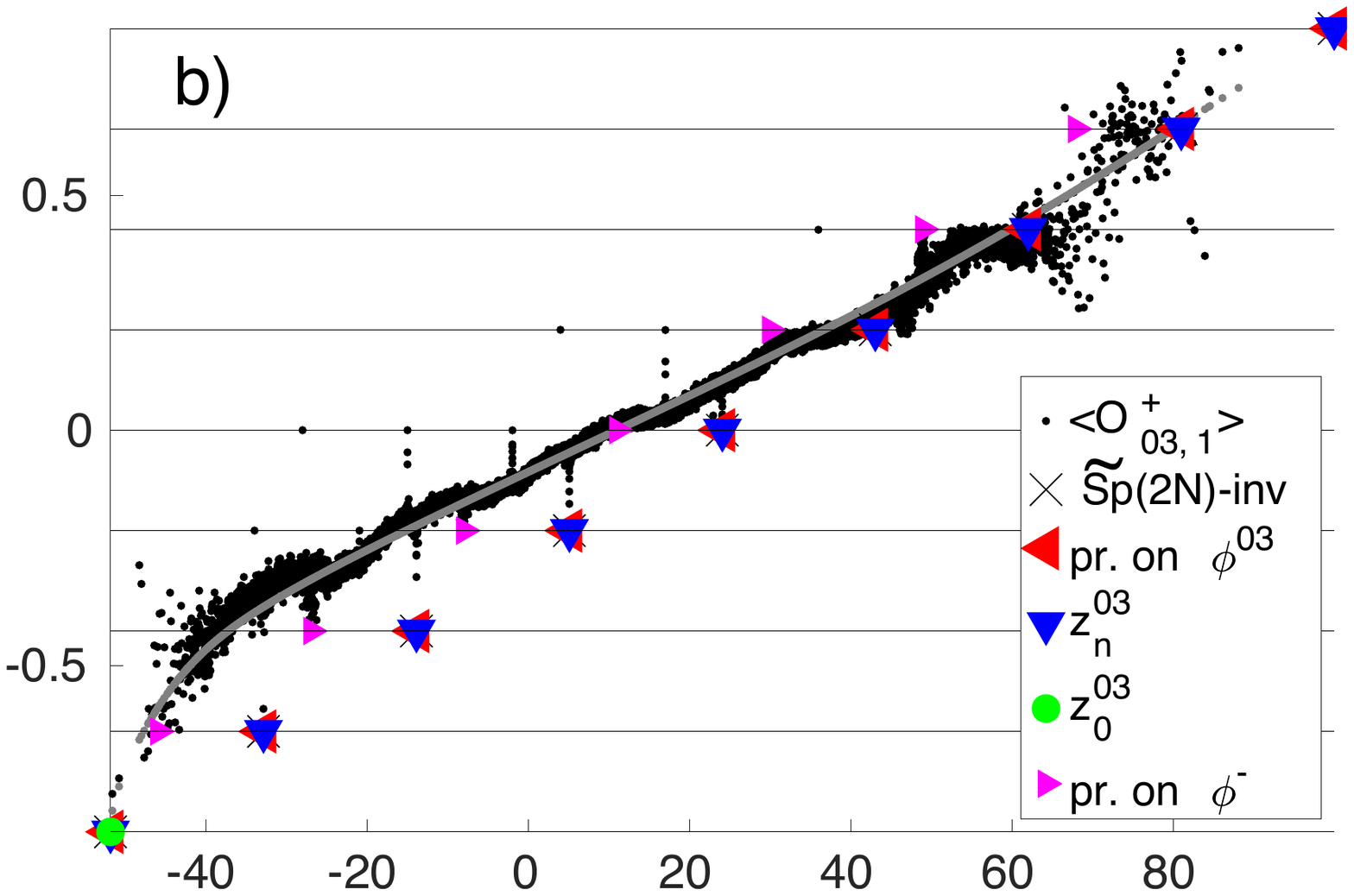}	
		\includegraphics[width=0.85\columnwidth]{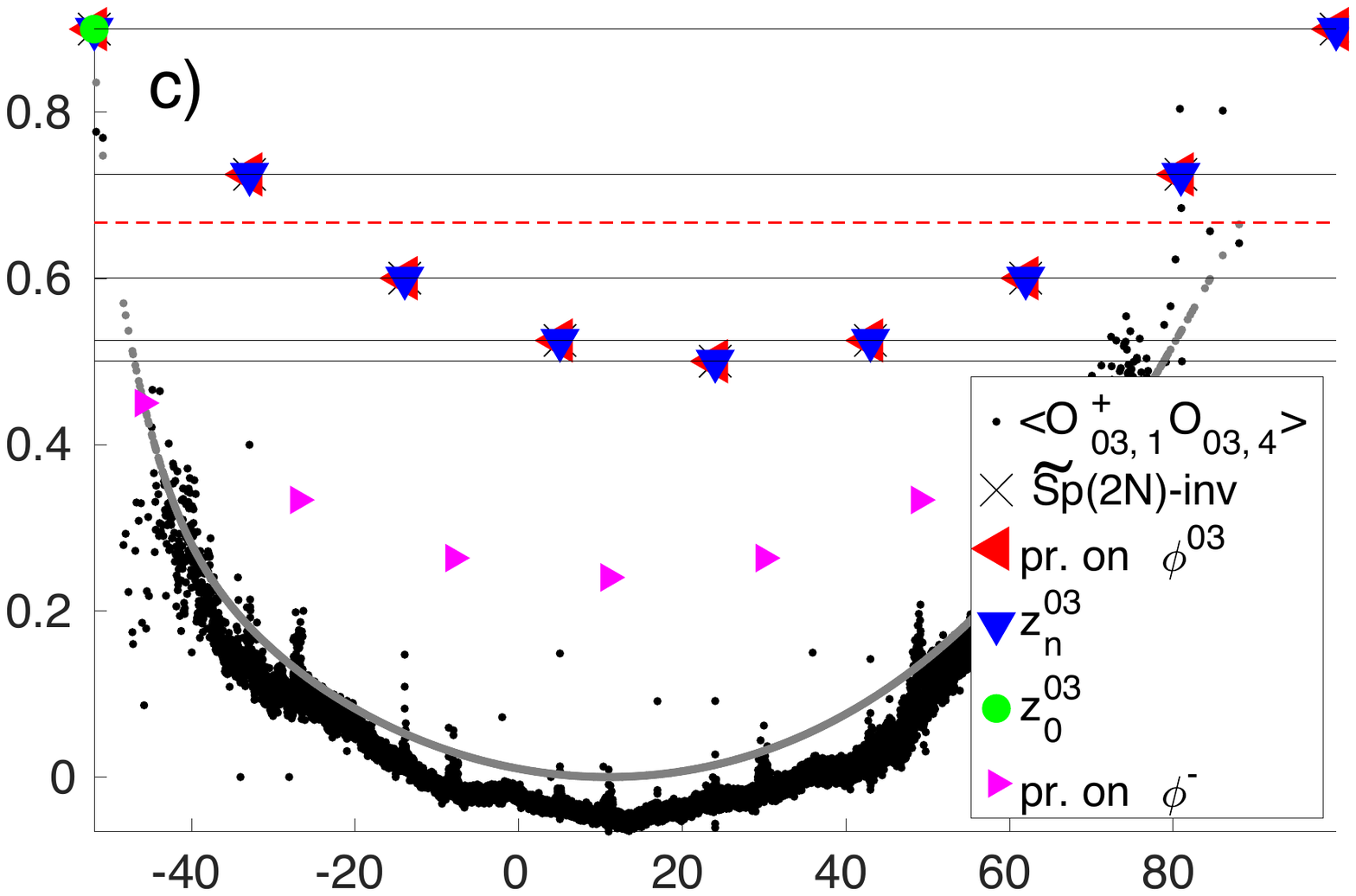}	
			\includegraphics[width=0.88\columnwidth]{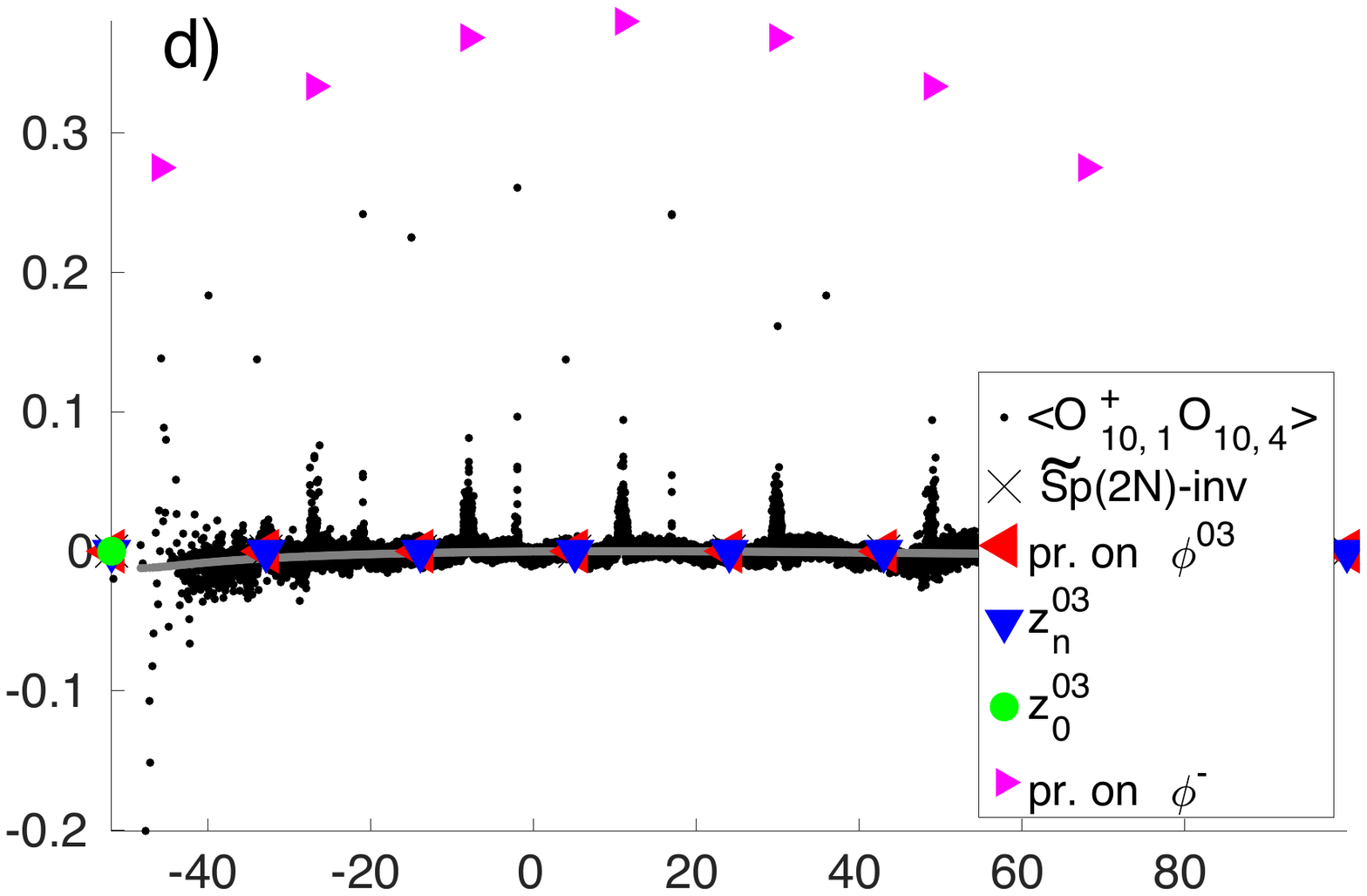}	
	\end{center}
\caption{Numerical results for the inter-orbital ($\CO_{03}$) pairing potential \eqref{eq:dH0general} added to the Hamiltonian \eqref{eq:Hminimal}. a) Entanglement entropy. b) Real part of the 1-point function $\braket{\CO^{\dagger}_{03,j}}$. Horizontal lines indicate analytical values. c) 2-point function $\braket{\CO^{\dagger}_{03,1} \CO_{03,N}}$. Dashed line indicates average over $\ket{z_n^{03}}$ scar subspace. d) Two-point function $\braket{\CO^{\dagger}_{10,1} \CO_{10,N}}$.}
\label{fig:addOdagGetPsi0sp}
\end{figure}

The inter-orbital $\eta$ states $\ket{\phi^{03}_n}$ are a special case of the states considered in Ref. \cite{paperC1} and a superposition of these states can be made the ground state by adding a mean-field pairing potential of the $\CO_{03}$ type to the Hamiltonian \eqref{eq:Hminimal}:
\begin{gather}
    H = H^A_{i\eta} + H_\Delta,
\end{gather}
where $H_\Delta$ is given by Eq. (\ref{eq:dH0general}).
This is confirmed numerically in Fig. \ref{fig:addOdagGetPsi0sp}, where for $N=4$, $\Delta=9$ was chosen to make the BCS scar the ground state and the parameter $\theta=\pi/7$ is chosen arbitrarily.
The part of the Hamiltonian governing scars ($H_0$) is
\begin{equation}
\label{eq:spgsH0SpecialEta}
H_\mu + H_{\mathrm{Hub}} + \sum_{j} \left( \Delta e^{i\theta} \CO^{\dagger}_{03,j} + \Delta e^{-i\theta}\CO_{03,j} \right).
\end{equation}

The lowest-energy inter-orbital $\eta$ state (and the global ground state for large enough $\Delta$) is the BCS wavefunction $\ket{z^{03}_0}$ given by Eq. \eqref{eq:spgs} with $\CO^\dagger_j = \CO^{\dagger}_{03,j}$.
All other states in the scar subspace are given by $\ket{z^{03}_n}$, which is the special (inter-orbital $\eta$) case of Eq. \eqref{eq:scarsInZnBasis}.

The state $\ket{z^{03}_0}$ saturates the bound on the absolute value of the one-point function \cite{paperC1}. In our numerical simulations, see Fig. \ref{fig:addOdagGetPsi0sp}, the value of the two-point function (ODLRO) in the BCS state $\ket{z^{03}_0}$ also exceeds the values in all other states and the average ODLRO in the scar subspace (red dashed line in Fig. \ref{fig:addOdagGetPsi0sp}c) is higher than in any state outside the scar subspace (except for five non-scar states). This confirms that the correlations of the type $\CO_{03}$ type are the highest within the scar subspace spanned by $\ket{\phi^{03}_n}$ \eqref{eq:genTowerWf} and $\ket{z_n^{03}}$ \eqref{eq:scarsInZnBasis}, as expected.

In Fig. \ref{fig:addOdagGetPsi0sp} we also show that the projection of all the states $\ket{z_n^{03}}$ in the new scar tower on the subspace spanned by $\ket{\phi^{03}_n}$ (``pr. on $\phi^{03}$'') is exactly equal to 1, which confirms that the two subspaces coincide. The generic, non-scar low-entropy states seen in Fig. \ref{fig:addOdagGetPsi0sp} can again be eliminated by adding to the Hamiltonian eligible $T$ terms from Sec. \ref{sec:specEtaCompHterms}.

The triplet inter-orbital $\eta$ states (magenta triangles) have the same one-point function values for the inter-band pairing as the main tower of scars, as one can see in Fig. \ref{fig:addOdagGetPsi0sp}b). Therefore, they react to the pairing potential in the same way but carry in addition spin-triplet ($\mu=1$) fluctuations, but zero spin-triplet one-point function because only one spin-triplet pair is present in every tower state. 

\begin{figure}[htp!]
	\begin{center}
		\includegraphics[width=0.97\columnwidth]{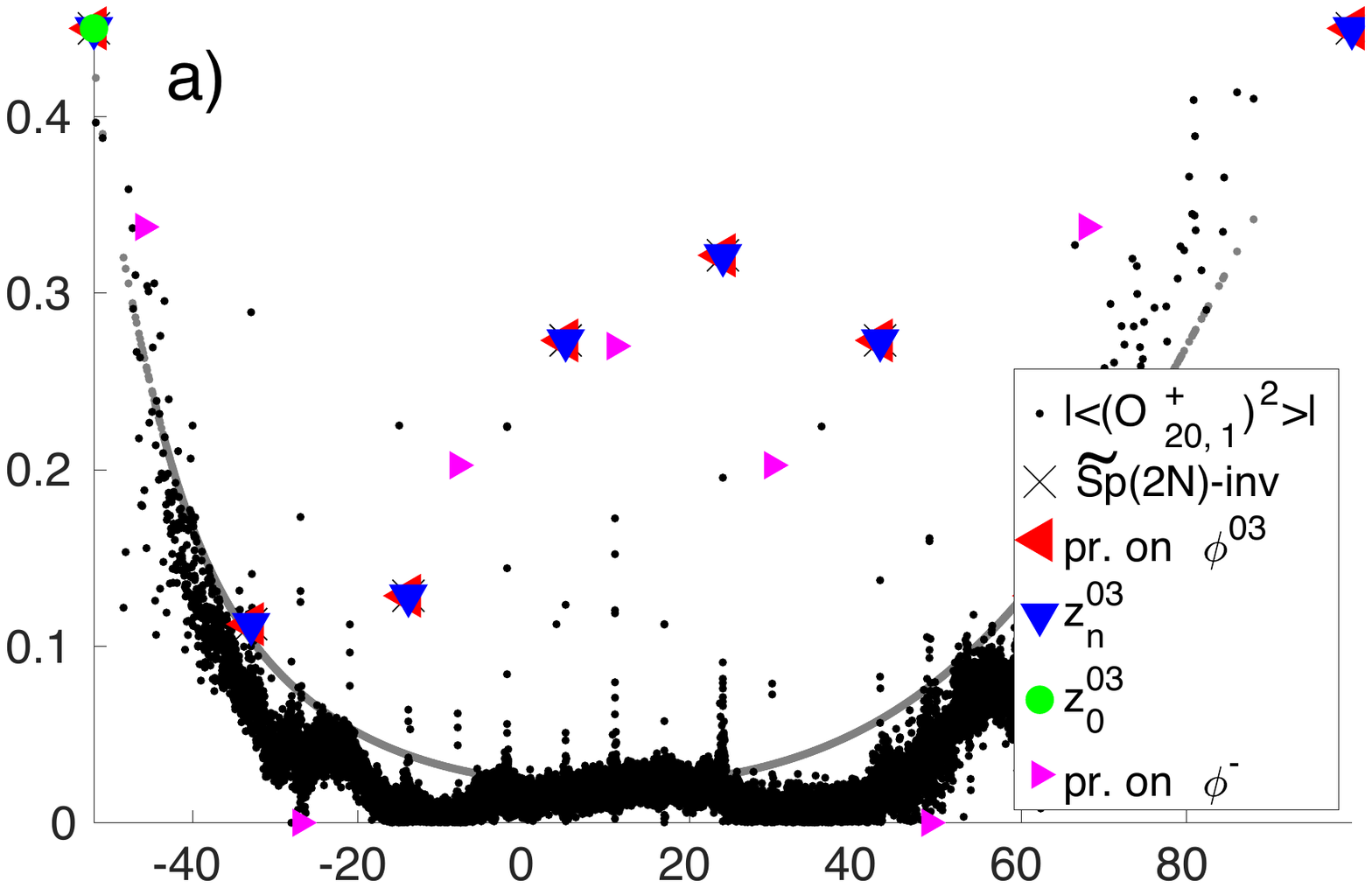}
	\includegraphics[width=\columnwidth]{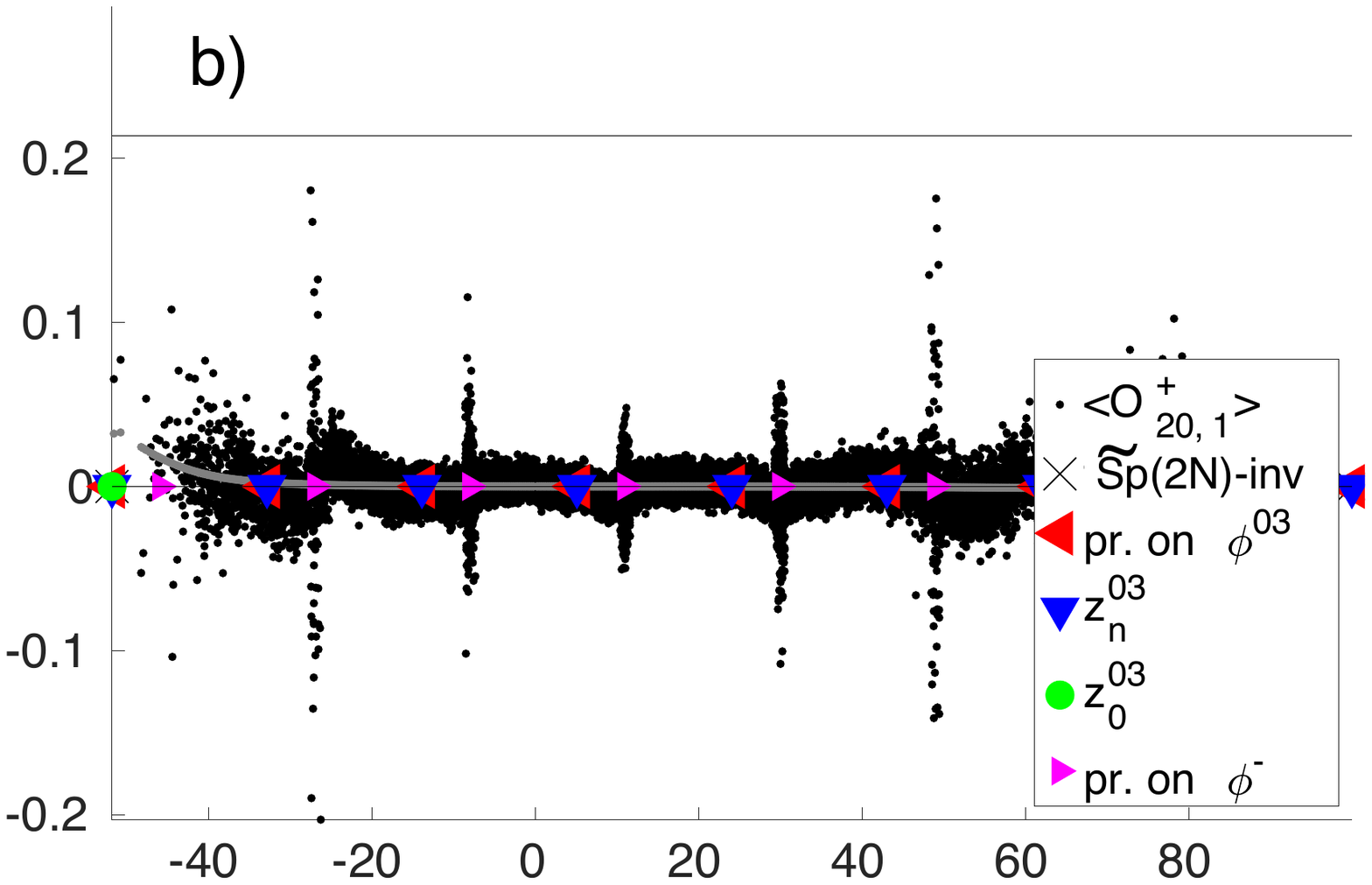}
	\end{center}
\caption{$2e$ vs $4e$ clustering with the inter-orbital pairing potential added. a) The absolute value of the expectation value of the quartet creation operator. b) Real part of the expectation value of the pair creation operator. In both cases, the spin-triplet orbital-singlet pairing $\CO^{\dagger}_{20}$ is considered. }
\label{fig:521g2evs4e}
\end{figure}

In Fig. \ref{fig:521g2evs4e}, we present the evidence that the wavefunctions $\ket{z^{03}_n}$ and their spin-triplet satellite family (``pr. on $\phi^-$'') exhibit $4e$ clustering. As can be seen in Fig. \ref{fig:521g2evs4e}b, both scar families do not have $2e$ pairing, indicated by vanishing expectation values $\braket{\CO^{\dagger}_{20,j=1}}$. However, the $4e$ clustering, Fig. \ref{fig:521g2evs4e}a, is nonzero in the same states, as evidenced by the nonzero expectation value $\braket{(\CO^{\dagger}_{20,j=1})^2}$, which significantly exceeds in magnitude the measurements in the nearby generic states. Measured in the BCS ground state, the absolute value of the $4e$ expectation value is 22 times larger than in an average generic state. It is 12 times larger in an average $\ket{z^{03}_n}$ state than in an average non-scar state. Neither 2e nor 4e expectation values in the scar states depend on the site position due to the O$(N)$ invariance of the scar subspace.

The absolute values of $\braket{(\CO^{\dagger}_{\mu0,j})^2}$ and $\braket{(\CO^{\dagger}_{0\nu,j})^2}$ are all identical and independent of $\mu$ and $\nu$, as follows from Eq. \eqref{eq:squaredRelations}. The $2e$ expectation value is zero in both scar families for all pairing operators except $\CO^{\dagger}_{03,j}$. 
Thus we conclude that for any pairing operator $\CO^{\dagger}_{\mu0,j}$ or $\CO^{\dagger}_{0\nu,j}$, with the exception of $\CO^{\dagger}_{03,j}$, we have exactly zero $2e$ alongside a large $4e$ clustering.

\subsection{Attractive Hubbard interaction}

\begin{figure}[htp!]
	\begin{center}
		\includegraphics[width=\columnwidth]{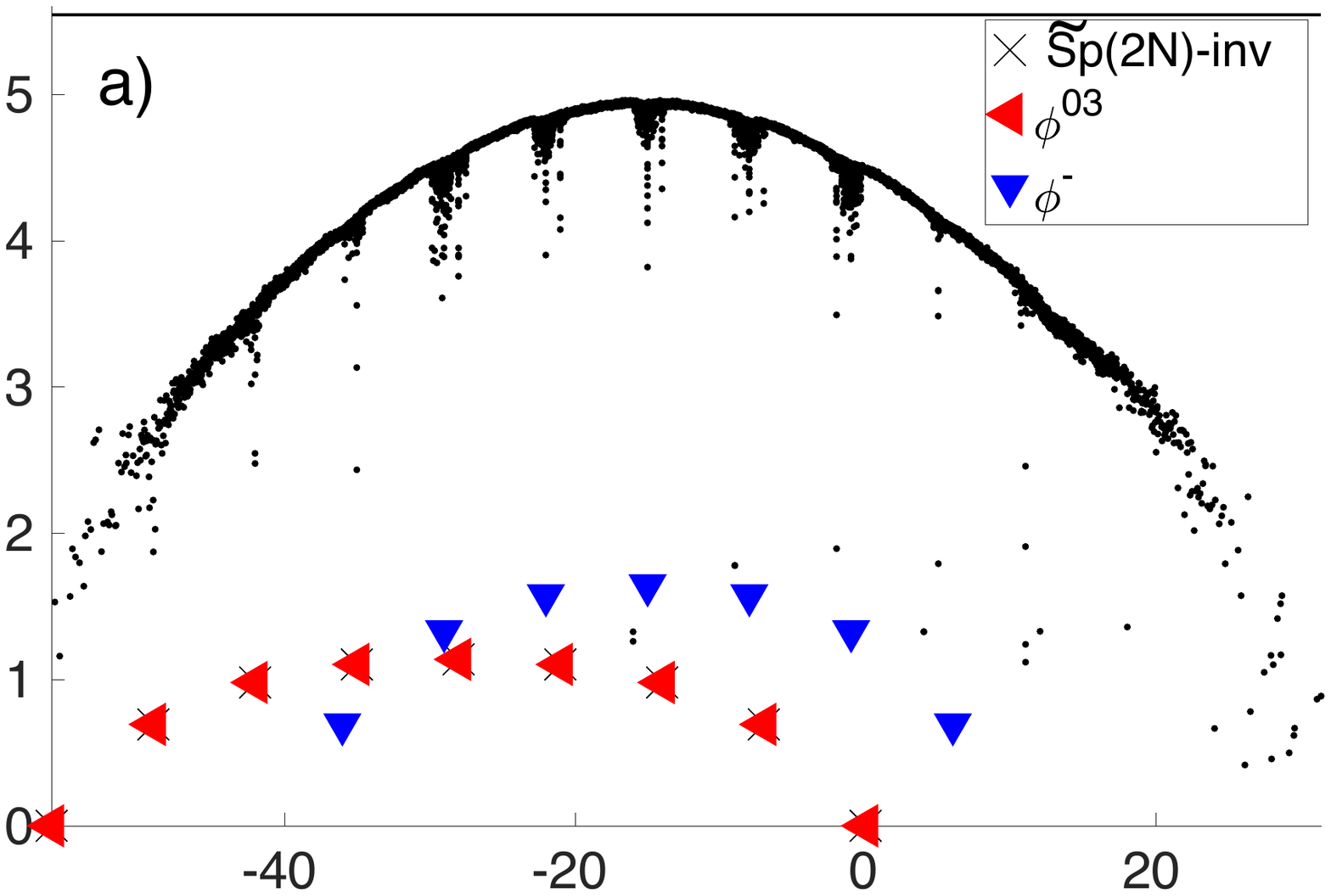}
	\includegraphics[width=\columnwidth]{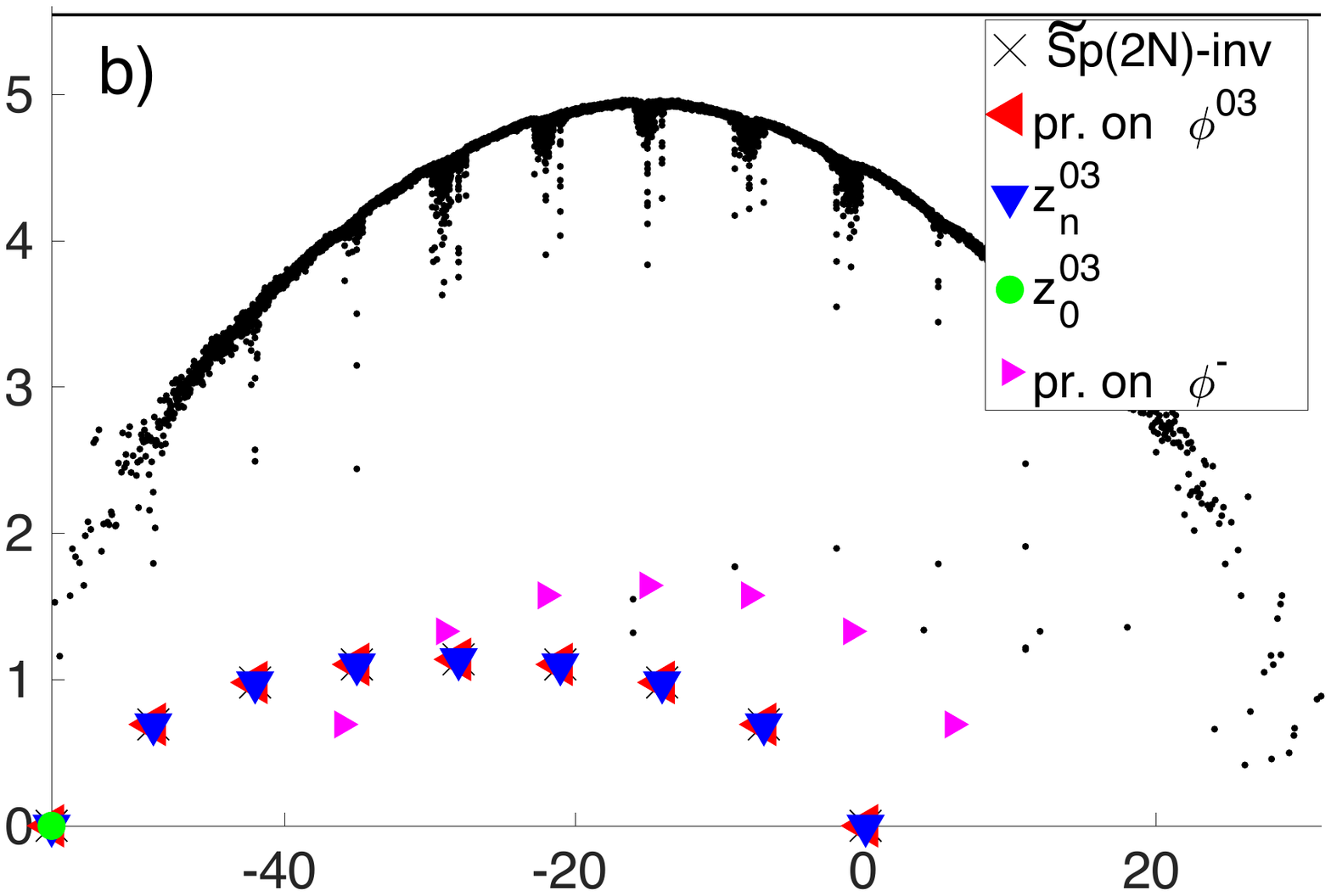}
	\end{center}
\caption{Entanglement entropy for an attractive Hubbard interaction with $U=-6.51$. a) $\Delta=0$; b) $\Delta=0.01$. All other parameters are identical to those used in Fig. \ref{fig:minH1Plots}. }
\label{fig:521AttractiveU}
\end{figure}

In Fig. \ref{fig:521AttractiveU} we show that for an attractive Hubbard interaction, with all other parameters identical to those in Fig. \ref{fig:minH1Plots}, one of the scar states is the ground state even without adding any pairing potential $H_\Delta$. We further demonstrate that if a weak ($\Delta=0.01$) pairing potential is added, it causes the basis rotation within the scar subspace making the BCS inter-orbital scar the ground state without qualitative changes to the non-scar spectrum.

The one-point function in the scars is proportional to $\Delta$ and is accordingly small. However, the two-point function (ODLRO) is large in the scar subspace, regardless of the value of $\Delta$.

\section{Spin-triplet pairing, Type \rom{1}}
\label{sec:triplType1} 

Following the scheme of Ref. \cite{paperC1}, we obtain the scar wavefunctions with unconventional spin-triplet pairing by using the spin-triplet pair creation operator $\CO^{\dagger}_{\mu0,j}$ \eqref{eq:evenBSST} in the general expression \eqref{eq:genTowerWf}:
\begin{align}
\label{eq:BSSTgenTowerWf}
\ket{\phi^{\mu0}_n} = \frac{(\sum_j \CO^{\dagger}_{\mu0,j} )^n \ket{0}}{P_N(n)}, \quad 0\le n\le 2N,
\end{align}
where $\mu=1,2,3$. These scar states are orbital-singlet and therefore correspond to a purely inter-orbital pairing.

The states (\ref{eq:BSSTgenTowerWf}) are endowed with large unconventional superconducting correlations (both one- and two-point) of type $\CO^{\dagger}_{\mu0,j}$. The analytical expressions \eqref{eq:1ptFunInSPGSK1}, \eqref{eq:2ptFunInZInTermsOfMu} and \eqref{eq:2ptFunAvgOverScarSubsp} for the one- and two-point functions, as well as for the average value over the scar subspace, remain valid. The scars $\ket{\phi^{\mu0}_n}$ \eqref{eq:BSSTgenTowerWf} are an example of states that have unitary pairing but non-zero total magnetic moment. 

\subsection{Full symmetry group and its generators \label{sec:triplType1FullSymmerty}}

Both the spin-triplet wavefunctions in this section and the inter-orbital $\eta$ states in the previous section are built using Eq. \eqref{eq:genTowerWf}, with the only difference coming from the different $\CO^\dagger_j$ operators used ($\CO^{\dagger}_{03,j}$ and $\CO^{\dagger}_{\mu 0,j}$, respectively). Comparing these two pair creation operators in Eqs. \eqref{eq:OdagEBTSS} and \eqref{eq:evenBSST}, we notice that one can be converted into the other by swapping the labels of orbitals and spins. 
Because of this simple relation, the $\mu=3$ spin-triplet states in this section also have the full symmetry group Sp$(2N)$. All its generators (and analogues of all the Hamiltonian terms mentioned in \ref{sec:fullSymmGroupForInterOrbital}) can be obtained by swapping the orbital and spin labels in the expressions for the inter-orbital $\eta$ states in Sec. \ref{sec:fullSymmGroupForInterOrbital}.

In this way, we obtain the new version of the $K$ generators
\begin{align}
\label{eq:gensBSSTKSU2}
K'_{A,j} = -\frac{i}{2} \sum_{pq } \tau^A_{pq} \left(c^{p,\dagger}_{j\uparrow}c^q_{j\downarrow}-c^{p,\dagger}_{j\downarrow}c^q_{j\uparrow}\right) ,
\end{align}
cf. Eq. (\ref{eq:gensKSU2main}), which transform as a vector under the orbital $\SU(2)$ group. Note that
one of the generators remains unchanged under the swap: $K'_{2,j} = K_{2,j}$. The corresponding $OT$ term in the Hamiltonian has the form
\begin{align}
\label{eq:gensBSSTKSU2}
K'_A = \sum_j r'_j K'_{A,j},
\end{align}
with real, in general site-dependent, strengths $r'_j$.

Under the transformation exchanging spin and orbital indices, the $w$ and $\bar w$ fermionic operators from Eq. \eqref{uvspinor} become
\begin{align}
\label{eq:uAndvFermionsTriplet}
w'_{jp} \equiv  \frac{c_{j\uparrow}^p-i c^p_{j\downarrow}}{\sqrt{2}}, \quad \bar w'_{jp} \equiv  \frac{c_{j\uparrow}^p+i c^p_{j\downarrow}}{\sqrt{2}}.
\end{align}
These operators annihilate electrons in the spinor states with a definite spin projection onto the $y$ axis.

Applying the spin-orbital swap to the spin-dependent hopping \eqref{eq:soHopping} (which annihilates inter-orbital $\eta$ scars), we obtain the inter-orbital hopping terms
\begin{align}
\label{eq:soHoppingSwapped}
T^A_{\text{orb},ij} = \sum_{pq,\alpha} \tau^A_{pq} c^{p,\dagger}_{i\alpha}c^{q}_{j\alpha}+{\rm H.c.},
\end{align}
which annihilate the $\ket{\phi^{30}_n}$ scar states.

All the generators mentioned above (together with further generators provided in Appendix \ref{sec:AppSpEtaRepTheory}) annihilate the states $\ket{\phi^{30}_n}$ \eqref{eq:BSSTgenTowerWf} and any linear combination of them represents a valid interaction compatible with these states being scars. In addition, these states are annihilated by the generators $K_1$ and $K_2$ \eqref{eq:gensKSU2main}.

We defer the full analytical study of the symmetries of the states $\ket{\phi^{\mu0}_n}$ for $\mu=1,2$ to future studies.
Numerically, we find that some of the generators found for the $\mu=3$ case do annihilate the states with $\mu=1,2$, which suggests that their symmetry groups have a common subgroup.
In particular, the $\ket{\phi^{10}_n}$ states are annihilated by the $K'_A$ generators given in Eq. \eqref{eq:gensBSSTKSU2}, the generators given in Eq. \eqref{eq:tripletSO4Js} in Appendix \ref{sec:AppSpEtaRepTheory}, and the $K_2$ and $K_3$ generators \eqref{eq:gensKSU2main}.
The $\ket{\phi^{20}_n}$ states are annihilated by the generators from Eq. \eqref{eq:tripletSO4Js} and the $K_1$ and $K_3$ generators from Eq. \eqref{eq:gensKSU2main}.
The hopping terms \eqref{eq:soHoppingSwapped} are not compatible with the $\ket{\phi^{10}_n}$ or $\ket{\phi^{20}_n}$ scars.

\subsection{Example Hamiltonian supporting $\ket{\phi^{30}_n}$ scars \label{sec:tr1Hamiltonian}}

The group O$(N)$ is a subgroup of the full symmetry of the states $\ket{\phi^{\mu0}_n}$. Therefore, all the terms from the common Hamiltonian $H_{\mathrm{O}(N)}$, see Eq. \eqref{eq:bareH}, including the Hubbard interaction \eqref{eq:HhubShort} and the pairing potential, can be used.
Additional valid $H_0$ terms include the O$(N)$-invariant Hubbard interaction \eqref{eq:ONHubbardU} and for the $\ket{\phi^{30}_n}$ states the generators $K_{\zeta,1}$ and $K_{\zeta,2}$ that can be obtained from the $K$ generators \eqref{eq:gensKSU2main} by the Shiba transformation \cite{hubbard1DbookShiba}.

\begin{figure}[htp!]
	\begin{center}
				\includegraphics[width=0.9\columnwidth]{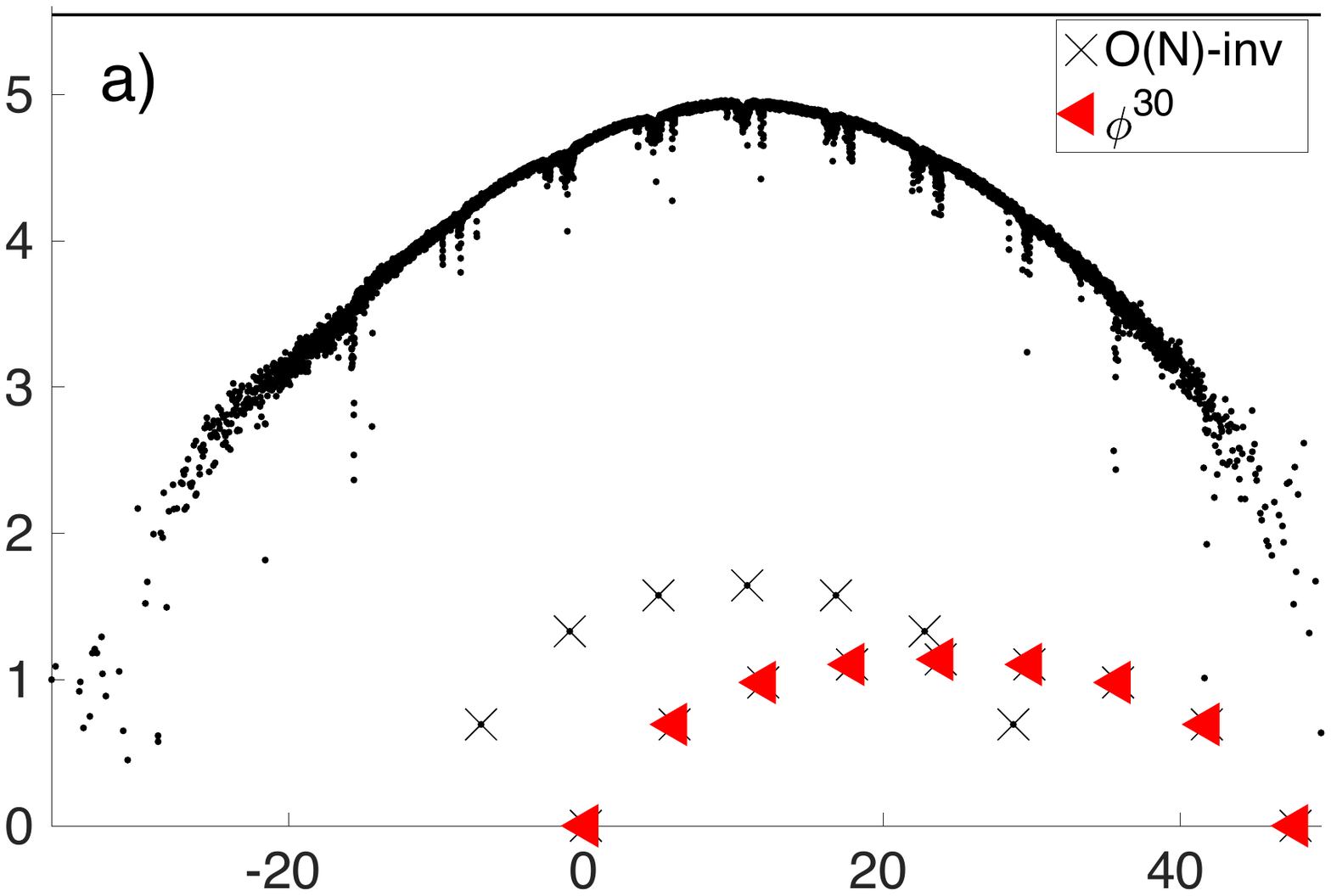}		
				\includegraphics[width=0.97\columnwidth]{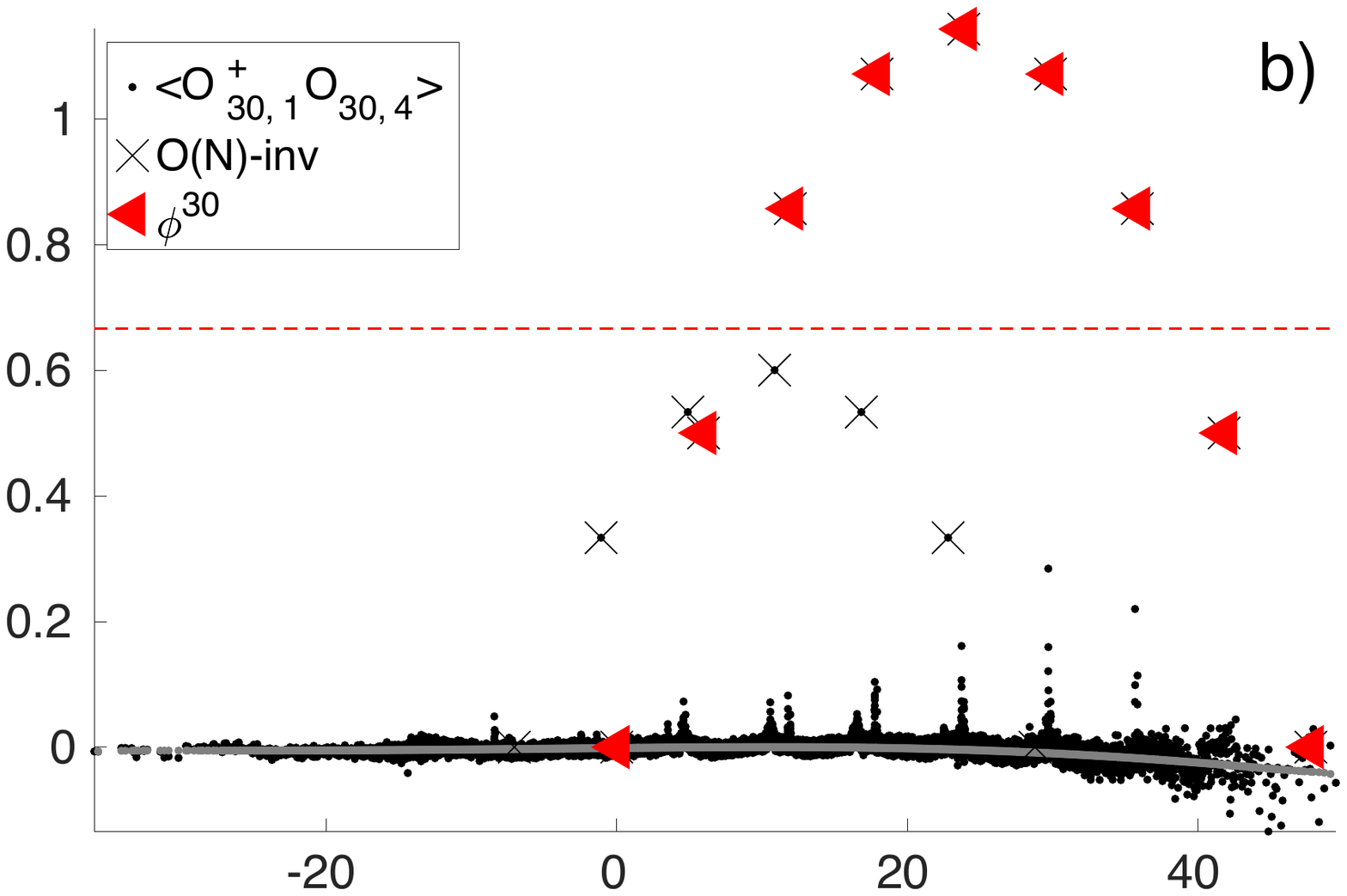}
				\includegraphics[width=\columnwidth]{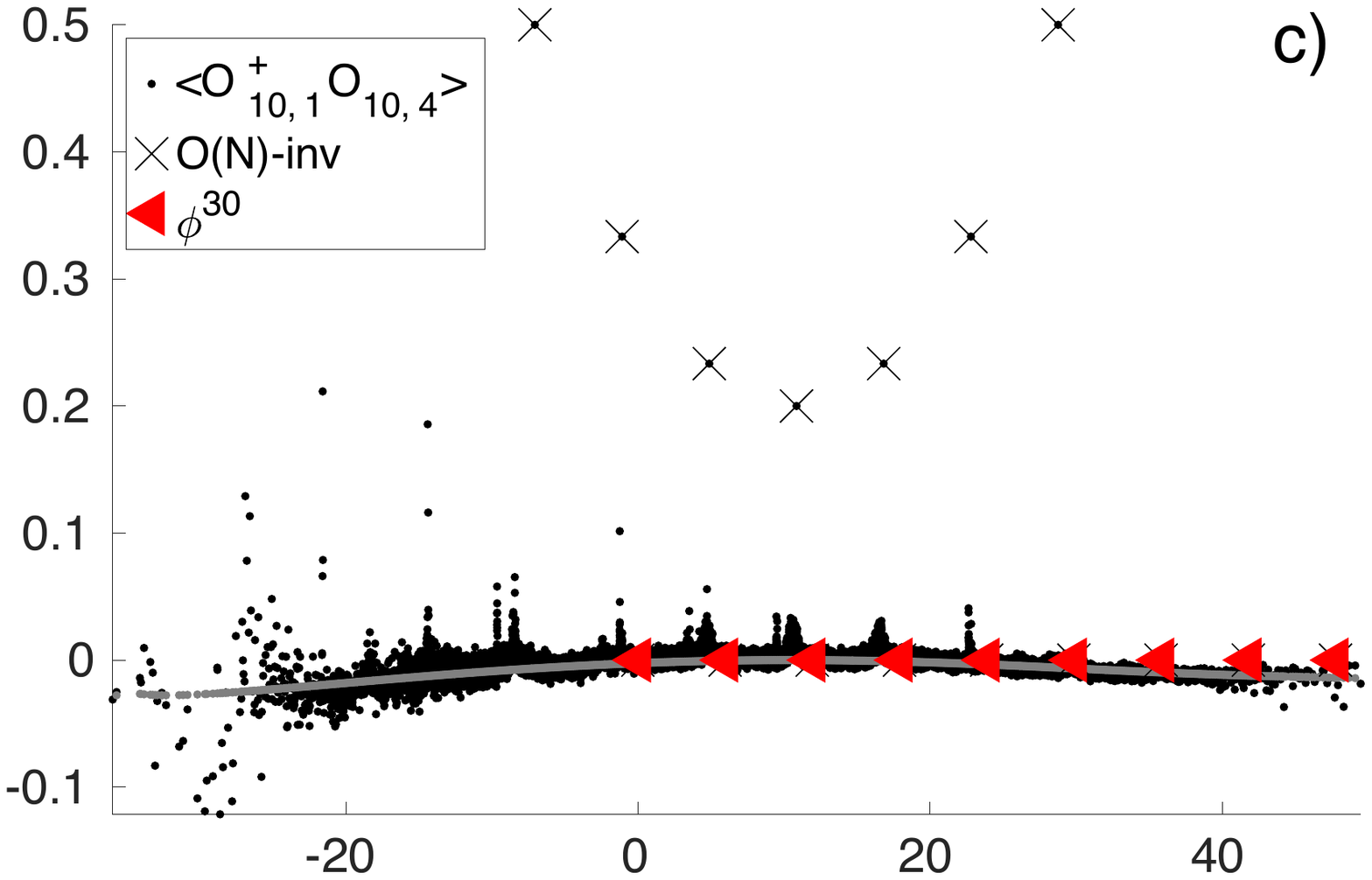}		
	\end{center}
\caption{Numerical results for the model (\ref{eq:tripletGamma0MediumH}) with $N=4$. a) Entanglement entropy. b) Spin-triplet pairing ODLRO in the $\mu=3$ channel $\braket{\CO^{\dagger}_{30,1}\CO_{30,N} }$. Dashed line is the average two-point function over the $\ket{\phi^{30}_n}$ scar subspace. c) Spin-triplet pairing ODLRO in the $\mu=1$ channel $\braket{\CO^{\dagger}_{10,1}\CO_{10,N} }$.
}
\label{fig:BSSTgamma0}
\end{figure}

For numerical analysis, we use the following example Hamiltonian: 
\begin{align}
\label{eq:tripletGamma0MediumH}
    H^{A}_{\rom{1}} = H_{\mathrm{O}(N)}  +  K'_A,
\end{align}
with $A=1$.
The results are shown in Fig. \ref{fig:BSSTgamma0}, where the symmetry-breaking term $\sum_l O_l T_l$ defined in \eqref{eq:TOT2} is added together with its extension $0.5008\sum_A K'_A O K'_A$ specific for the spin-triplet scars. In $H_{\mathrm{O}(N)}$ we use the parameters $\mu_x=\mu_y=-0.27$, $U=6.51$ and a site-independent $r_j'=0.2$ (results for site-dependent strengths are not shown but are not noticeably different). Results obtained for other values of $A$ in Eq. (\ref{eq:tripletGamma0MediumH}) are similar and discussed along with other example Hamiltonians in Appendix \ref{sec:tr1HamiltonianAddNumerAppendix}.
 
As expected, the spin-triplet scars $\ket{\phi^{\mu0}_n}$ form an equally spaced tower with the energy separation equal to $2\mu_{\rm eff}=U+\mu_x+\mu_y$, as follows from Eqs. \eqref{eq:uvXdef} and \eqref{eq:bareH0effectiveMu}. The absolute value of the spin-triplet pairing ODLRO in the $\mu=3$ channel is on average 91 times larger in the scars $\ket{\phi^{30}_n}$ than in generic thermal states. In addition, another tower of scars with the same energy separation is observed. It features a spin-triplet ODLRO in both $\mu=1$ and $\mu=3$ channels that is significantly stronger than in generic states. It is annihilated by $T^A_{\text{orb}}$ and the $T$ terms in the Hamiltonian  \eqref{eq:tripletGamma0MediumH}, namely, by $K'_A$ and the hopping (\ref{eq:bareOT}).

\subsection{Spin-triplet BCS scar is the ground state \label{sec:tr1BCSgsHamiltonian}}

\begin{figure}[htp!]
	\begin{center}
				\includegraphics[width=0.8\columnwidth]{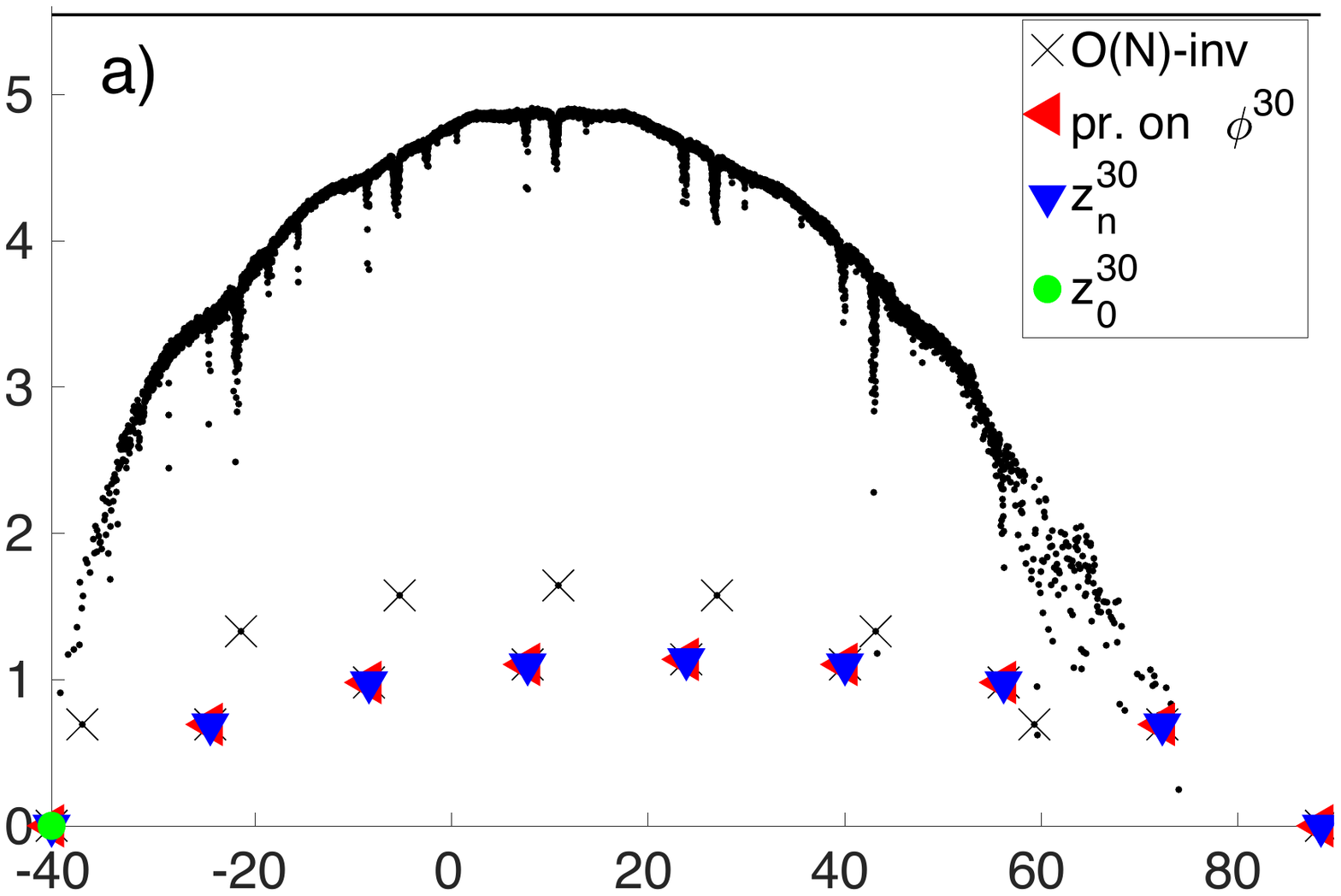}
			\includegraphics[width=0.88\columnwidth]{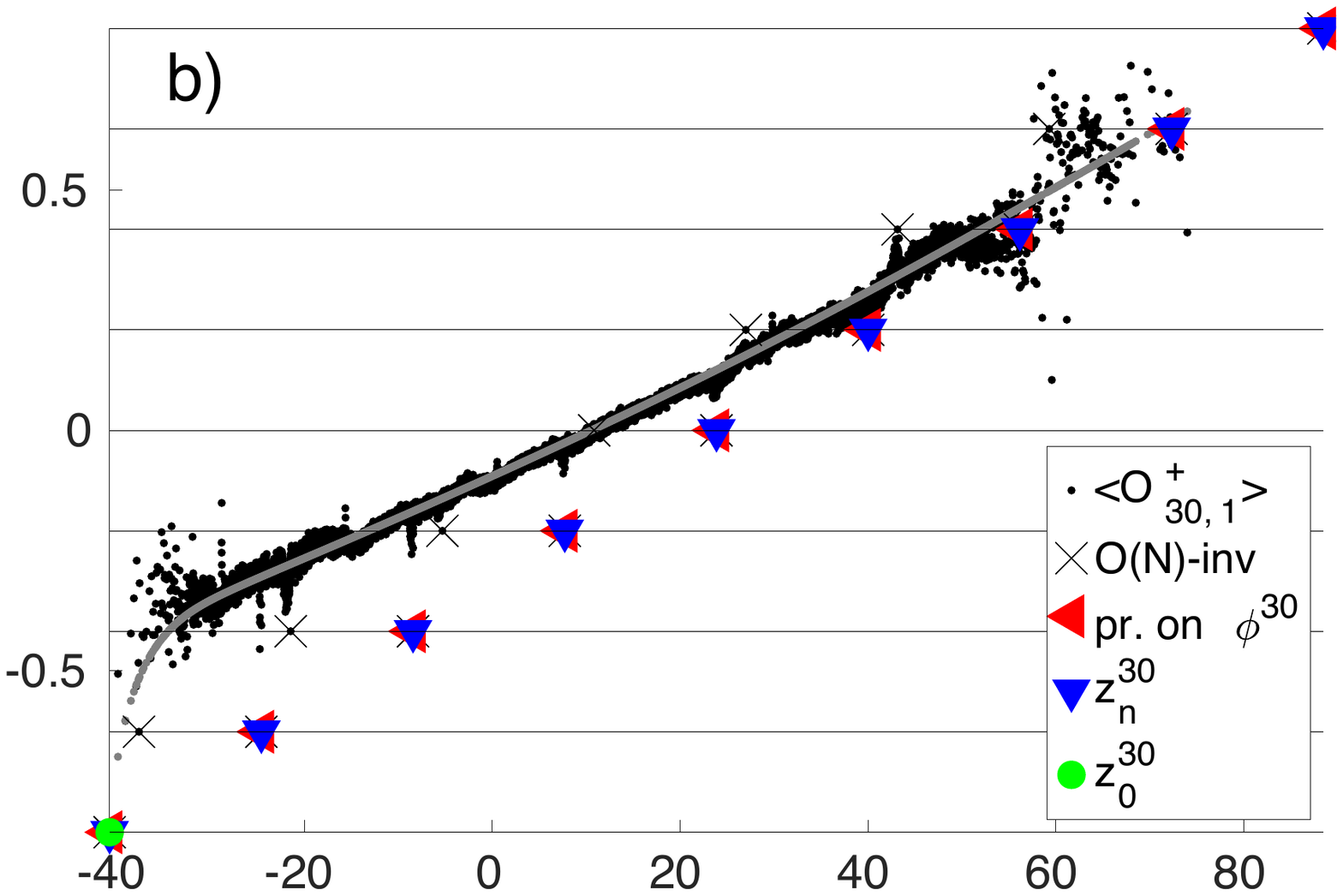}
				\includegraphics[width=0.88\columnwidth]{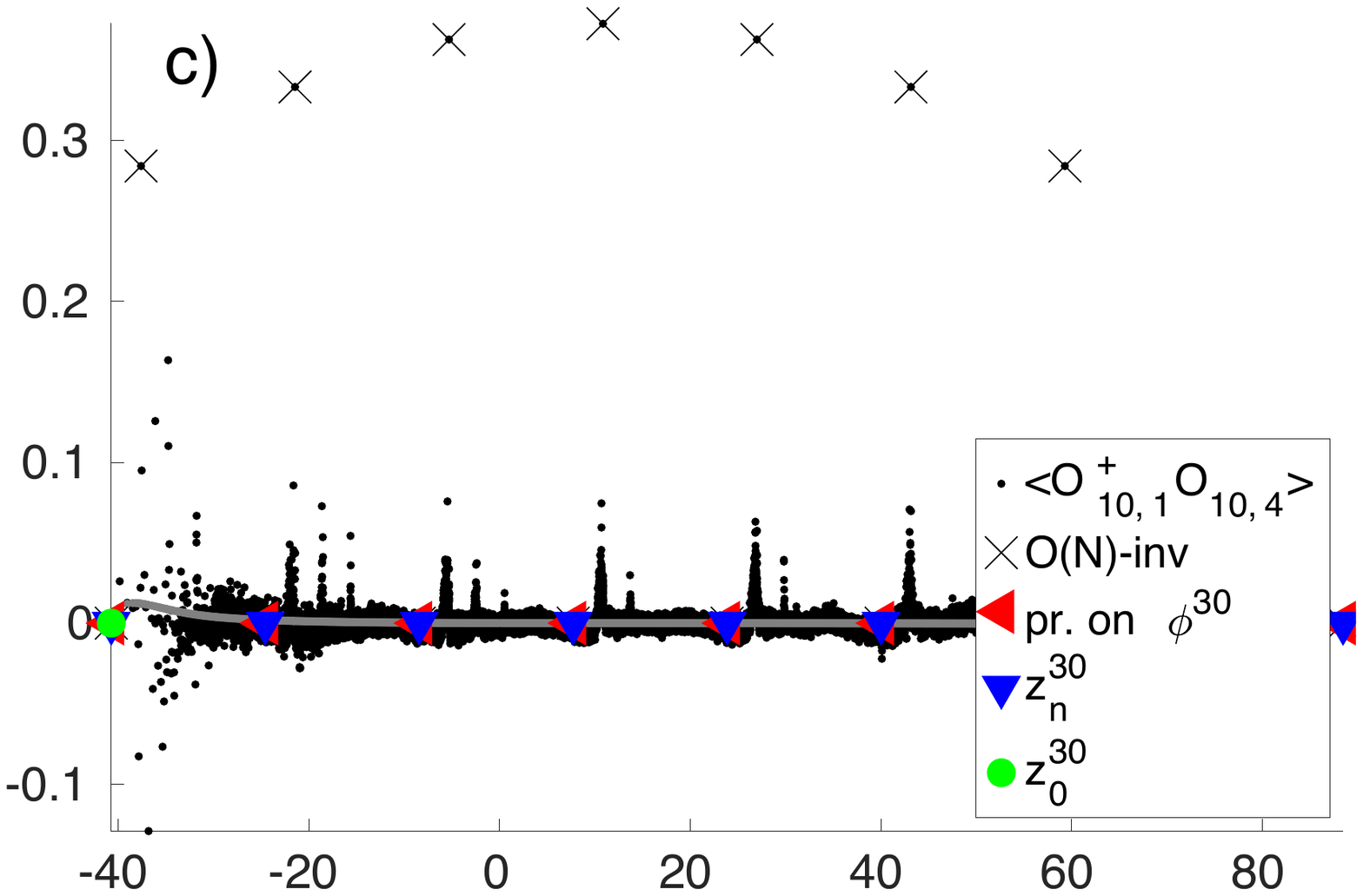}														
				\includegraphics[width=0.85\columnwidth]{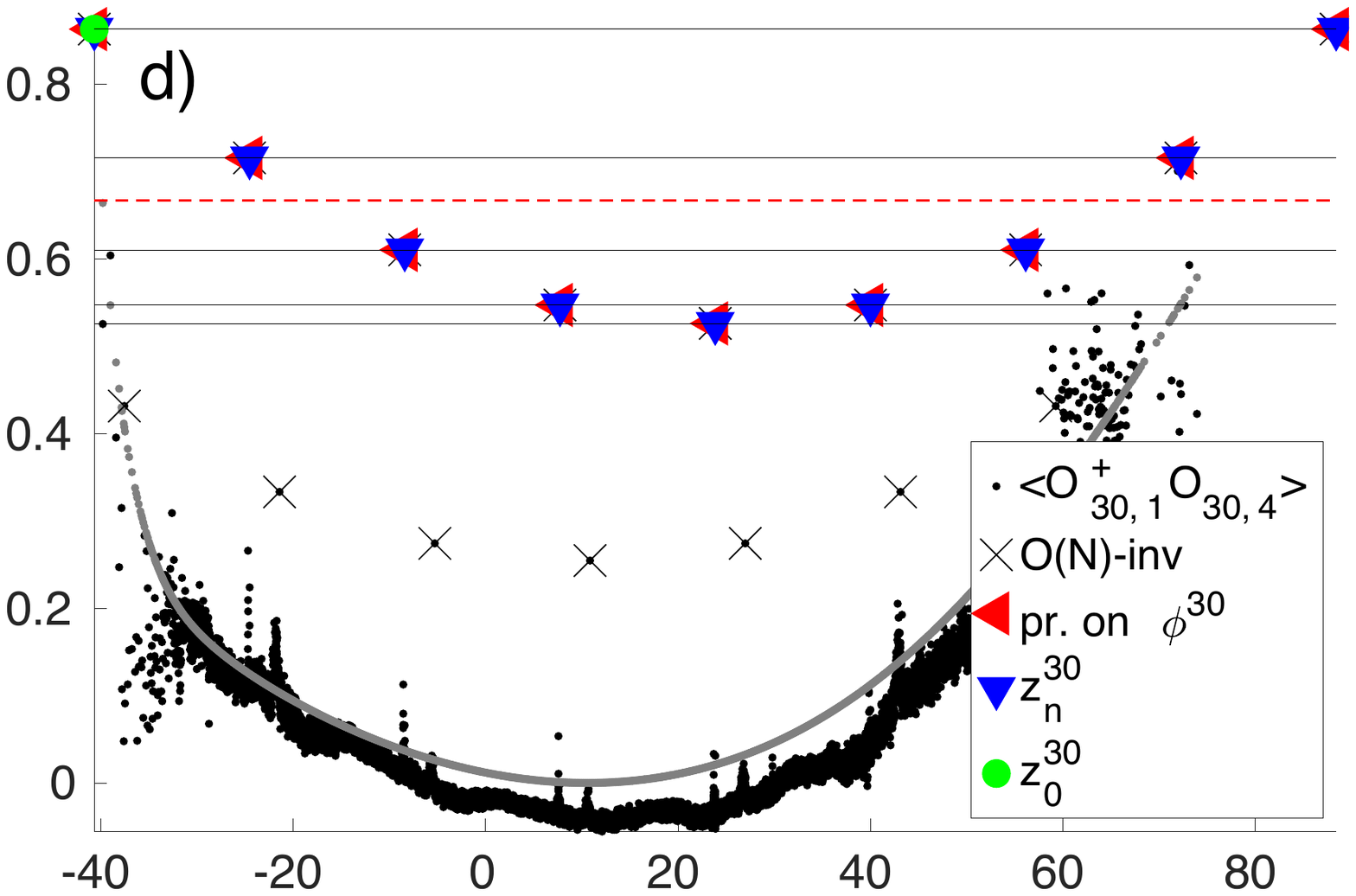}
	\end{center}
\caption{Numerical results for the Hamiltonian \eqref{eq:tripletGamma0MediumH} with the spin-triplet pairing potential \eqref{eq:dH0general} added to it. a) Entanglement entropy. b) Real part of the one-point function $\braket{\CO^{\dagger}_{30,j} }$. c) Two-point function $\braket{\CO^{\dagger}_{10,1}\CO_{10,N} }$. d) Two-point function $\braket{\CO^{\dagger}_{30,1}\CO_{30,N} }$. Solid horizontal lines indicate analytical values. Dashed line is the average two-point function over the $\ket{z^{30}_n}$  scar subspace. \\
}
\label{fig:BSSTmediumHwithGamma}
\end{figure}

In order to make a spin-triplet scar the ground state, we add the pairing potential $H_\Delta$ defined in Eq. \eqref{eq:dH0general} with $O_j = \CO_{\mu0,j}$ to the Hamiltonian  \eqref{eq:tripletGamma0MediumH}. The results for $N=4$, $\Delta=7.5$, and $\theta=\pi/7$ are shown in Fig. \ref{fig:BSSTmediumHwithGamma}.

The BCS scar state $\ket{z_0}$ \eqref{eq:spgs} with $O_j = \CO_{30,j}$ becomes the ground state. It saturates the upper bound on the absolute value of the expectation value $\braket{\CO_{30,j}}$ (Fig. \ref{fig:BSSTmediumHwithGamma}b) and in our simulation also has highest spin-triplet ODLRO $\braket{ \CO^\dagger_{30,i} \CO_{30,j}}$ (Fig. \ref{fig:BSSTmediumHwithGamma}d). The values of the one- and two-point function in the excitations $\ket{z^{30}_n}$ \eqref{eq:scarsInZnBasis} above the BCS ground state agree with the analytical expressions. The $\mu=1,2$ one-point functions (not shown) are zero. Their average spin-triplet pairing ODLRO (dashed red line in the panel d) is larger than that in any generic, non-scar state, thus confirming that this subspace, as intended, features the largest pairing correlations of the $\CO_{30,j}$ type. 

As we observe in Fig. \ref{fig:BSSTmediumHwithGamma}b, in the additional tower featuring both $\mu=1$ and $\mu=3$ spin-triplet ODLRO (indicated by black crosses) the value of the one-point function $\braket{\CO_{30,j}}$ coincides with that in the $\ket{z^{30}_n}$ states with $1\leq n\leq N-1$. Therefore, these states also strongly react to the pairing potential, which rotates them with respect to the $\Delta=0$ case. In the rotated basis, they are still annihilated by all the $T$ terms of the Hamiltonian, remain equally-spaced in energy (with the same spacing as in the main tower), and preserve strong spin-triplet ODLRO in the $\mu=1$ channel (Fig. \ref{fig:BSSTmediumHwithGamma}c). The corresponding $\mu=1$ one-point function remains zero. 

The spin-triplet pairing expectation value $\braket{\CO^{\dagger}_{30,j} }$ shown in Fig. \ref{fig:BSSTmediumHwithGamma}b is the only nonzero pairing both in the scar states $\ket{z^{30}_n}$, which are obtained by a rotation of the basis \eqref{eq:BSSTgenTowerWf}, and in the satellite tower. 
On the other hand, as we explained in Sec. \ref{sec:scIntro}, the square of any of the pair creation operators produces the same operator and it is therefore sufficient to consider only one of them.

For any pairing type $\CO^{\dagger}_{\mu0,j}$ or $\CO^{\dagger}_{0\nu,j}$ except for $\CO^{\dagger}_{30,j}$ we observe (numerical results for $\CO^{\dagger}_{03,j}$ shown in Fig. \ref{fig:BSST2evs4e} in Appendix) that the $4e$ clustering in the absence of the $2e$ pairing of the same type is large in the scar states and is significantly stronger in the scar states compared to an average non-scar state. As in the case of the inter-orbital $\eta$ states one should keep in mind eq. \eqref{eq:squaredRelations} when interpreting these results.


\section{Spin-triplet pairing, type \rom{2} \label{sec:triplType2}}

In this section, in contrast to the inter-orbital $\eta$ states in Sec. \ref{sec:ioPairingSpecialEta} and type-\rom{1} spin-triplet states in Sec. \ref{sec:triplType1}, we consider another family of spin-triplet O(N)-invariant states that are {\it not} of the general form \eqref{eq:genTowerWf} studied in Ref. \cite{paperC1}: 
\begin{align}
\label{eq:triplType2Wfs}
\ket{\phi_{m,n}}  \equiv \frac{  (\eta^{\dagger}_{\downarrow\downarrow})^m (\eta^{\dagger}_{\uparrow\uparrow})^n }{P_{n,m}} |0\rangle ,
\end{align}
where $0\le m+n\le 2N$, $\eta^{\dagger}_{\alpha\beta}  = \sum_j c^{x,\dagger}_{j\alpha}c^{y,\dagger}_{j\beta}$ and $P_{n,0} = P_{0,n}= \sqrt{N! n!/(N-n)!}$.

In these states, we have for the one-point function
\begin{align}
\braket{ \phi_{n+1,0} | \CO_{1 0, j} |  \phi_{n,0} } = \frac{-2\sqrt{n(N-n+1)}}{N}.
\end{align}
and for the two-point function
\begin{align}
\label{eq:tripletType2ODLRO}
\langle  \phi_{m,n}|&\CO^{\dagger}_{1 0, i} \CO_{1 0, j} |\phi_{m,n}\rangle = 4 \frac{(N-n)n+(N-m)m}{N(N-1)} .
\end{align}
Neither of these expressions depend on the sites where the expectation value is measured. For the two-point function \eqref{eq:tripletType2ODLRO} this indicates the presence of spin-triplet pairing ODLRO of the $\CO^{\dagger}_{1 0, j}$ type. The expressions above also hold if we replace $\CO^{\dagger}_{1 0, j}$ with the $\mu=2$ pairing channel $\CO^{\dagger}_{20, j}$.

\subsection{ Hamiltonian \label{sec:type2tripletH}}

The Hamiltonian $H_{\mathrm{O}(N)}$ is compatible with the $\ket{\phi_{m,n}}$ scars. However, for these states one can use instead of Eq. (\ref{eq:H0eta}) a more general, spin- and orbital-dependent, chemical potential term 
\begin{align}
\label{eq:HmuON}
\tilde H_\mu = \sum_{j,p,\alpha} \mu_{p\alpha} n_{jp\alpha},
\end{align}
which lifts the degeneracies between the states $\ket{\phi_{m,n}}$ and other O$(N)$ singlets that might otherwise mask them.

Besides the Hubbard interaction \eqref{eq:HhubShort} already used in $H_{\mathrm{O}(N)}$, its two further generalizations are valid $H_0$ terms: the O($N$)-symmetric Hubbard interaction 
\begin{align}
\label{eq:ONHubbardU}
H^{\text{O}(N)}_{\text{Hub}}= U\sum_{j}\prod_{p,\alpha} \left(\frac{1}{2}\!-\!n_{jp\alpha}\right),
\end{align}
which was originally proposed in Ref. \cite{paper3Case1Majorana}, and also 
\begin{align}
\label{eq:zetaHub}
\tilde{H}_{\mathrm{Hub}} = \frac{U}{2}\sum_j n_j (n_j - 1).
\end{align}

Valid $OT$ terms have some similarities to the ones used for the type-\rom{1} spin-triplet scars.
They include the spin-blind imaginary hopping, $K_3$ defined Eq. \eqref{eq:gensKSU2main}, and all components of the orbital-dependent hopping $T^A_{\text{orb}}$ \eqref{eq:soHoppingSwapped}.

\subsubsection{Numerical test}

The example Hamiltonian
\begin{align}
\label{eq:HfullTripletType2min}
H_{\rom{2}} =H_{\mathrm{O}(N)} + H_{\text{Z}}^{3} + T_{\mathrm{SO}}^3
\end{align}
includes $H_{\mathrm{O}(N)}$ \eqref{eq:bareH}, the SO coupling $T_{\mathrm{SO}}^3$ \eqref{eq:Hsooriginal}, and the Zeeman term (\ref{eq:siteDepZeeman}). It is very similar to the Hamiltonian \eqref{eq:Hminimal} we used for the inter-orbital $\eta$ states, with two differences. Here we use the same chemical potential for both orbitals and the $A=3$ component in the Zeeman term, which splits the $\ket{\phi_{m,n}}$ states.
Similar to the inter-orbital $\eta$ states, the term essential for singling out the scars $\ket{\phi_{m,n}}$ is the SO term $T_{\mathrm{SO}}^3$. It can be replaced by $T^A_{\text{orb}}$ \eqref{eq:soHoppingSwapped} with $A=1$ or $2$. 

In the numerical simulation, we add the symmetry-breaking $OT$ term $1.252\sum_l O_l T_l$ from Eq. \eqref{eq:TOT2} and also $0.5008\sum_j T_{\mathrm{SO},j}^3 O_j T_{\mathrm{SO},j}^3$ with the generators \eqref{eq:Hsooriginal} to the Hamiltonian \eqref{eq:HfullTripletType2min}.
We use $N=4$ and set $\mu_x=\mu_y=-0.27$, a uniform Zeeman strength $r_j^Z=0.1$, $U=6.51$, and choose site-independent amplitudes of the SO term: $r_j^{\mathrm{SO}} = 0.2$ (results for site-dependent SO amplitudes are qualitatively the same).

\begin{figure}[htp!]
	\begin{center}
				\includegraphics[width=0.89\columnwidth]{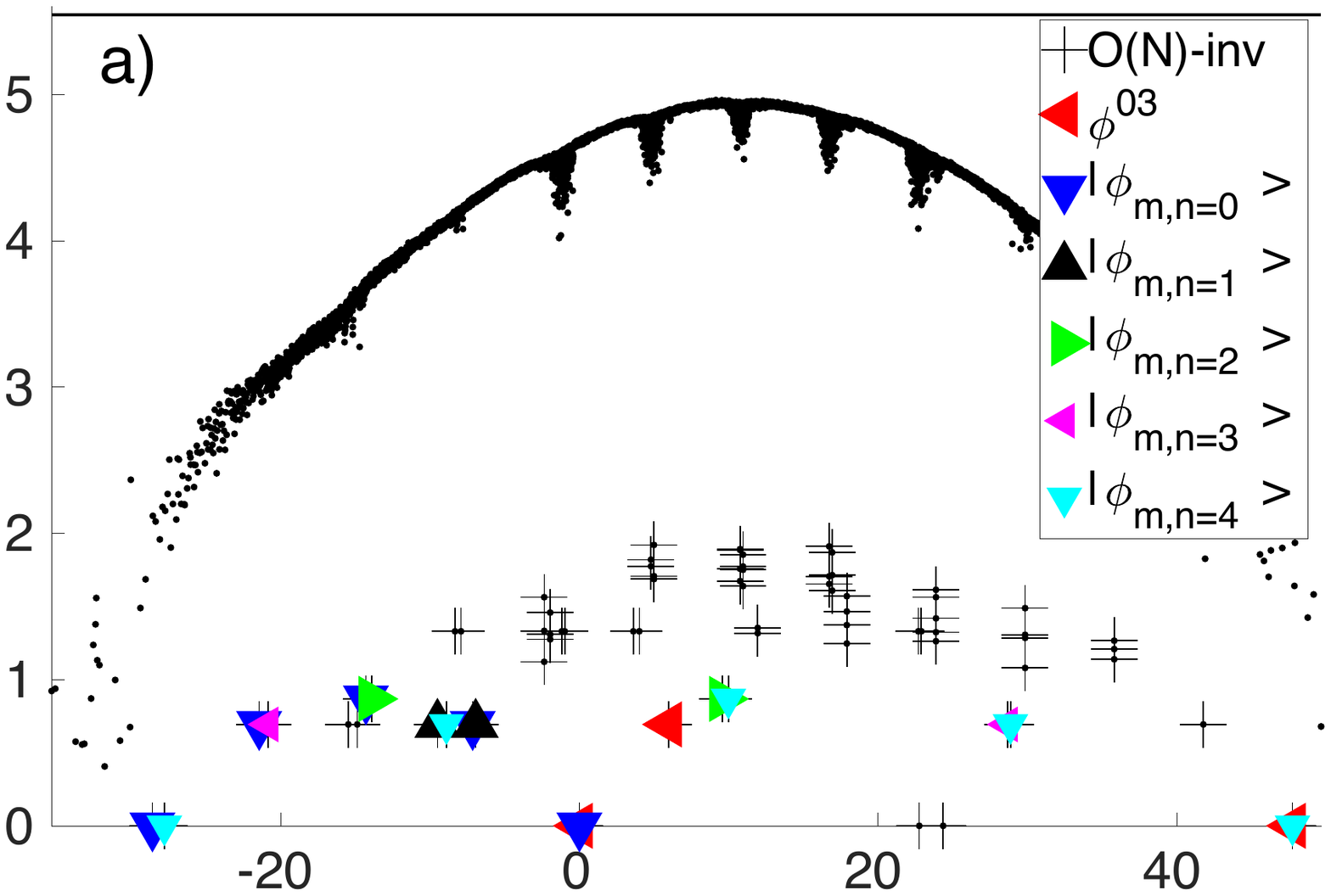}				
				\includegraphics[width=\columnwidth]{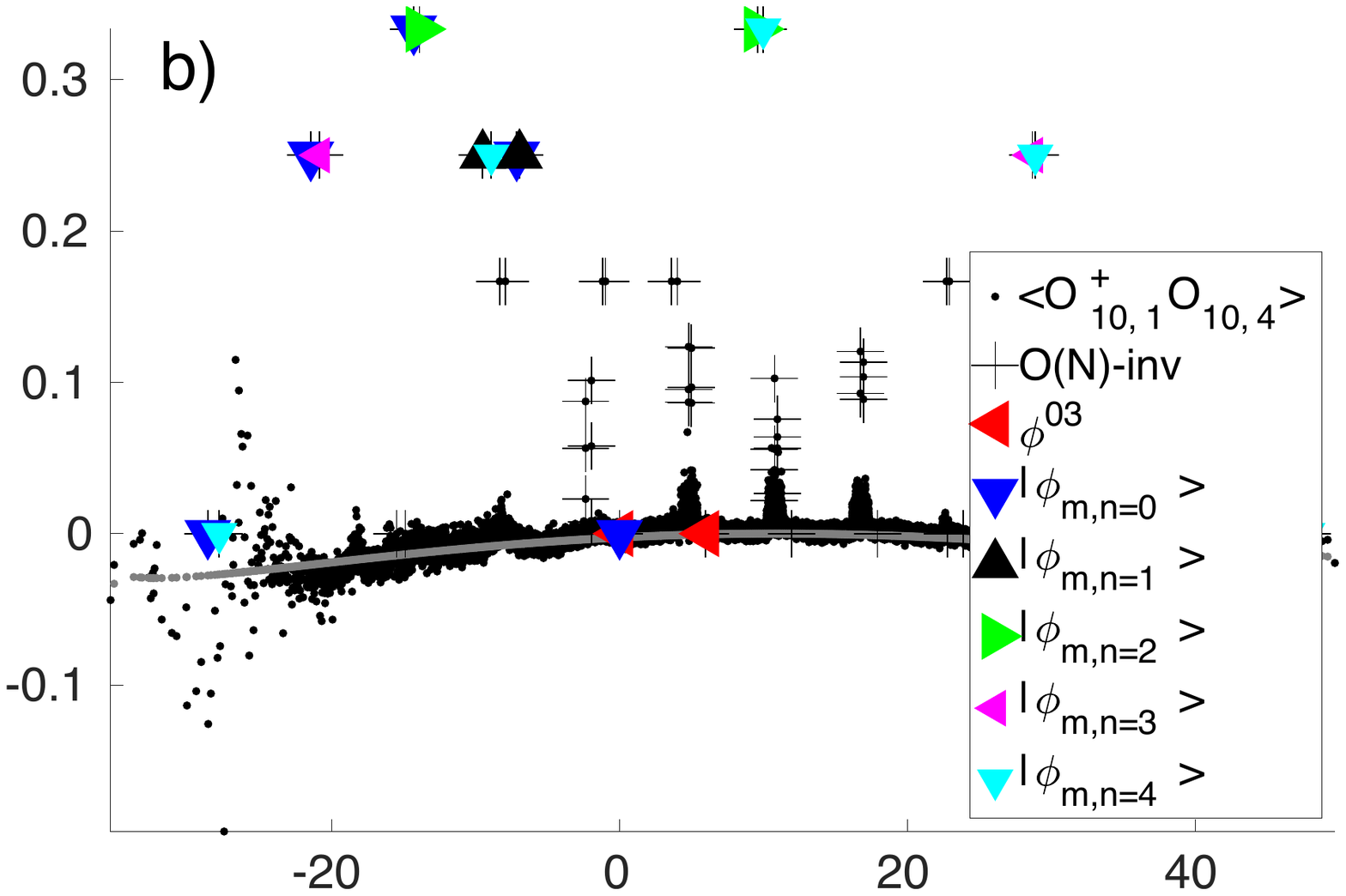}
	\end{center}
\caption{Numerical results for the Hamiltonian $H_{\rom{2}}$ \eqref{eq:HfullTripletType2min}. a) Entanglement entropy. b) Two-point function $\braket{\CO^{\dagger}_{10,1}\CO_{10,N} }$.
}
\label{fig:byHandWfTripletN4}
\end{figure}

The results are presented in Fig. \ref{fig:byHandWfTripletN4}, where we observe the scars series $\ket{\phi_{m,n}}$ with $n=0,m=\{0,1,2,3,4\}$; $n=1,m=\{0,4\}$; $n=2,m=\{0,4\}$;  $n=3,m=\{0,4\}$; and $n=4,m=\{0,1,2,3,4\}$. The scar states possess strong two-point spin-triplet correlations of equal magnitude in the $\mu=1$ (right panel) and $\mu=2$ (not shown) channels. The two-point function in the $\mu=3$ channel is zero in all the scar states. The corresponding one-point functions vanish, because there are no terms in the $H_0$ part of the Hamiltonian governing the scar subspace that would change the particle number.  

The number of other, generic O$(N)$-invariant states at low entropy can be reduced by making the chemical potentials orbital-dependent. This would also make all the inter-orbital $\eta$ scars $\ket{\phi^{03}_n}$ visible, eliminating one of the two differences to the Hamiltonian \eqref{eq:Hminimal}. Only some of these states are visible with orbital-independent chemical potentials in Fig. \ref{fig:byHandWfTripletN4}. Scars $\ket{\phi^{03}_n}$ are compatible with all the components of the Zeeman term.

\section{Discussion and Outlook}

We have presented four families of scars that occur in two-orbital spinful fermionic lattice models. Each family is a subspace where the states with strongest unconventional pairing (inter-orbital, spin-triplet, or both) are concentrated. The off-diagonal long-range order in all four scar families can be attributed to the O$(N)$ invariance of these subspaces, which leads to the two-point correlation function $\braket{\CO^\dagger_i\CO_j}$ being independent of the distance between points $i$ and $j$. At large distances this property is well established for superconductors.

All example Hamiltonians \eqref{eq:Hminimal}, \eqref{eq:tripletGamma0MediumH}, and \eqref{eq:HfullTripletType2min} necessarily include chemical potential term for making scars non-degenerate. The term that singles out scar states depending on the family is the on-site SO coupling $T_{\mathrm{SO}}^3$ \eqref{eq:Hsooriginal} or its counterpart $K'_A$ \eqref{eq:gensBSSTKSU2} that arises from the switching of the spin and orbital degrees of freedom. The two-orbital Hubbard term $H_{\mathrm{Hub}}$ \eqref{eq:HhubShort} is compatible with all our scar families and although optional is present in all our minimal Hamiltonians to make them strongly interacting. 

The Hamiltonians we study numerically are merely examples of the many Hamiltonians supporting each scar family. Other compatible Hamiltonians can be constructed following $H_0+OT$ structure and we list the suitable building blocks of types $H_0$ and $T$ for each scar family. This way we expect that a connection to specific materials will eventually be made.

The BCS ground state wavefunction \eqref{eq:spgs} builds in $4e$-clustering by construction. We do observe exactly zero $2e$-pairing and simultaneous large $4e$-clustering measured in the scar states for several pairing types, as discussed in the end of Sec. \ref{sec:ws521g} for inter-orbital $\eta$ states and in the end of Sec. \ref{sec:tr1BCSgsHamiltonian} for the type-\rom{1} spin-triplet states. The interpretation of these results is complicated by the fact that the absolute value of $4e$-clustering operator in our two-orbital system is independent of the pairing type, as explained in the text leading to Eq. \eqref{eq:squaredRelations}.

Scar subspaces constructed in this work exclusively possess large unconventional superconducting pairing correlations. While we guarantee that the BCS scar state can always be made the ground state by adding a sufficiently strong mean-field pairing potential, one of the scars may become the ground state even without an additional potential. For example, this happens for the inter-orbital $\eta$ scars if we flip the sign of the Hubbard interaction to make it attractive, as demonstrated in Fig. \ref{fig:521AttractiveU}. Further research could identify the Hamiltonian parameter regimes when the superconducting scar subspace is located in the low-energy part of the spectrum. It would also be interesting to analyze the consequences of such a configuration for the superconducting and thermodynamic properties at a finite temperature.

The two-orbital system we consider is the minimal model allowing to build unconventionally-paired scars with local Hamiltonians. We expect that MBS with unconventional pairing and even richer structure exist for fermionic systems with more than two flavors, and their study would be an interesting direction for future investigations.

In this proof-of-principle work we intentionally restrict ourselves to the same-site pairing making our results lattice-independent. It would however be natural to work out the full classification relating pairing types beyond same-site to the compatible scar families and specific lattices. Combined with the present study it would reveal the candidate materials for observing MBS with unconventional pairing.

All the scars we considered are annihilated by the imaginary hopping as a result of O$(N)$ being a common subgroup of $G$---the full symmetry group of the MBS. Considering other groups $G$ would produce other MBS families that are likely to include states with unconventional pairing. In particular, we expect that MBS annihilated by real hopping exist and can be obtained from the states in this work by a simple mapping, similar to the two $\eta$ scar families that exist in the single-orbital Hubbard model \cite{Pakrouski:2021jon}.

Although we do not aim to make quantitative predictions for any particular material, the models studied in this work do reflect, at least qualitatively, many important aspects of the physics of real multi-orbital superconductors. In particular, Eq. \eqref{eq:HhubShort} provides a reasonable minimal model of a local, Hubbard-like, interaction, while Eq. \eqref{eq:Hsooriginal} can be used to describe the SO coupling in various multi-orbital systems, e.g. Sr$_2$RuO$_4$.
Sufficiently small perturbations added to the Hamiltonians may mix the MBS subspace with the rest of the Hilbert space, but the salient features associated with MBS will not be completely suppressed, e.g. revivals will remain but acquire a finite lifetime as was demonstrated in the single-orbital case \cite{KolbStabilityPRXQuantum2023}.


\section{Acknowledgments}

We are grateful to Zimo Sun for the help in identifying the full symmetry group of the inter-orbital $\eta$ states and the wavefunctions for the type-\rom{2} spin-triplet MBS, as well as for useful comments on the manuscript. We thank Manfred Sigrist and Ilaria Maccari for multiple insightful discussions, and Igor Klebanov and Fedor Popov for collaboration on related projects.

This work was supported by a Discovery Grant 2021-03705 from the Natural Sciences and Engineering Research Council of Canada (KS). KS is grateful to the Institute for Theoretical Physics, ETH Zurich for hospitality and the Pauli Center for Theoretical Studies for financial support. KP is supported by the Swiss National Science Foundation, grant CRSK-$2\_237767$.
The simulations presented in this work were performed on computational resources managed and supported by Princeton's Institute for Computational Science $\&$ Engineering and OIT Research Computing.


\appendix

\section{Motivation for the pair creation operators definition \label{sec:motivatePairing}}

The general mean-field pairing Hamiltonian in a two-orbital superconductor has the form
\begin{equation}
\label{H-pair-2-band}
  \hat H_{pair}=\frac{1}{2}\sum_{ij,pq,\alpha\beta}\Delta^{pq}_{\alpha\beta}(i,j)c^{p,\dagger}_{i\alpha}c^{q,\dagger}_{j\beta} + \mathrm{H.c.},
\end{equation}
where the gap function $\hat\Delta$ is an antisymmetric $4N\times 4N$ matrix in the site, orbital, and spin spaces ($N$ is the number of lattice sites). We consider 1D here to avoid unnecessary complications that would come from a specific lattice.
The gap function can be represented in the following form:
\begin{equation}
  \Delta^{pq}_{\alpha\beta}(i,j)=\sum_{\mu,\nu=0}^3 \varphi_{\mu\nu}(i,j) (i\hat\sigma_\mu\hat\sigma_2)_{\alpha\beta} (i\hat\tau_\nu\hat\tau_2)_{pq} ,
\end{equation}
where $\hat\sigma_0$ and $\hat{\bm{\sigma}}=(\hat\sigma_1,\hat\sigma_2,\hat\sigma_3)$ are the unit matrix and the Pauli matrices in the spin space, while $\hat\tau_0$ and $\hat{\bm{\tau}}=(\hat\tau_1,\hat\tau_2,\hat\tau_3)$ are the unit matrix and the Pauli matrices in the orbital space. 

The sixteen functions $\varphi_{\mu\nu}(i,j)$ correspond to different pairing channels. Similar to the spin structure, we refer to the channels with $\nu=0$ and $\nu=1,2,3$ as orbital-singlet and orbital-triplet, respectively. 
The anti-commutation of the fermionic operators imposes certain constraints on the functions $\varphi_{\mu\nu}(i,j)$, which are summarized in Table \ref{table: phis}. 
For example, the orbital-singlet and spin-singlet pairing has to be odd in the spatial indices: 
$\varphi_{00}(i,j)=-\varphi_{00}(j,i)$, whereas the orbital-singlet and spin-triplet pairing has to be even: $\varphi_{\mu 0}(i,j)=\varphi_{\mu 0}(j,i)$ ($\mu=1,2,3$), etc. 
Note that the purely intra-orbital pairing is described by $\varphi_{\mu 1}$ and $\varphi_{\mu 2}$, whereas the inter-orbital pairing is described by $\varphi_{\mu 0}$ and $\varphi_{\mu 3}$.

We use the following simplest expression for the even-pairing gap functions:
\begin{equation}
\label{phi-2-band-even}
  \varphi_{\mu\nu}^{(even)}(i,j)=\eta_{\mu\nu}\delta_{ij},
\end{equation}
which corresponds to a uniform superconducting state with same-site pairing. The coefficient $\eta_{\mu\nu}$ plays the role of the superconducting order parameter. One can now introduce the pair creation operators for each local pairing channel:
\begin{equation}
\label{O-even-2-band}
  \hat O_{\mu\nu,j}^{\dagger}=\frac{1}{2} (i\hat\sigma_\mu\hat\sigma_2)_{\alpha\beta} (i\hat\tau_\nu\hat\tau_2)_{pq} c^{p,\dagger}_{j\alpha}c^{q,\dagger}_{j\beta}.
\end{equation}
For the relevant combinations of $\mu$ and $\nu$ here, see Table \ref{table: O-even}. The factor $1/2$ is a matter of convention and can be dropped. 

For the odd gap functions in a 1D lattice, the simplest expression, corresponding to a uniform superconducting state with the nearest-neighbor pairing, has the following form:
\begin{equation}
\label{phi-2-band-odd}
  \varphi_{\mu\nu}^{(odd)}(i,j)=\eta_{\mu\nu}(\delta_{i,j-1}-\delta_{i-1,j}).
\end{equation}
One can consider different pairing channels separately and introduce the corresponding pair creation operators in the odd channels:
 \begin{eqnarray}
  \hat O_{\mu\nu,j}^{\dagger} &=& \frac{1}{2} (i\hat\sigma_\mu\hat\sigma_2)_{\alpha\beta} (i\hat\tau_\nu\hat\tau_2)_{pq}\nonumber \\
  &&\times (c^{p,\dagger}_{j\alpha}c^{q,\dagger}_{j+1,\beta}-c^{p,\dagger}_{j+1,\alpha}c^{q,\dagger}_{j\beta}) \nonumber\\
  &=& (i\hat\sigma_\mu\hat\sigma_2)_{\alpha\beta}(i\hat\tau_\nu\hat\tau_2)_{pq} c^{p,\dagger}_{j\alpha}c^{q,\dagger}_{j+1,\beta}.
 \end{eqnarray}
For the relevant combinations of $\mu$ and $\nu$ here, see Table \ref{table: O-odd}. 

Explicit examples of the pair creation operators we use are given by Eqs. \eqref{eq:OdagEBTSS} and \eqref{eq:evenBSST} for the local pairing.
Inter-orbital pairs are described by the operators $\hat O_{\mu 0,j}^\dagger$ and $\hat O_{\mu 3,j}^\dagger$, 
where $\mu=0$ corresponds to inter-orbital spin-singlet pairing and $\mu=1,2,3$ correspond to inter-orbital spin-triplet pairing.

\begin{table}[t]
\caption{Spatial parity of the pairing functions in a two-orbital superconductor; $s$ ($t$) stands for the singlet (triplet) pairing.}
\begin{tabular}{|c|c|c|}
    \hline
    orbital & spin & $\varphi_{\mu\nu}(i,j)$ \\ \hline
    $s$ & $s$ & $\varphi_{00}(i,j)=-\varphi_{00}(j,i)$  \\
    $t$ & $s$ & $\varphi_{0\nu}(i,j)=\varphi_{0\nu}(j,i)$,\quad $\nu=1,2,3$  \\ 
    $s$ & $t$ & $\varphi_{\mu 0}(i,j)=\varphi_{\mu 0}(j,i)$,\quad $\mu=1,2,3$ \\ 
    $t$ & $t$ & $\varphi_{\mu\nu}(i,j)=-\varphi_{\mu\nu}(j,i)$,\quad $\mu,\nu=1,2,3$ \\ \hline
\end{tabular}
\label{table: phis}
\end{table}

\begin{table}[t]
\caption{Pair creation operators for two-orbital local pairing in a 1D lattice.}
\begin{tabular}{|c|c|}
    \hline
    $(\mu,\nu)$ & $\hat O_{\mu\nu,j}^\dagger$ \\ \hline
    $(0,\nu)$, $\nu=1,2,3$ & $(i\hat\sigma_2)_{\alpha\beta}(i\hat\tau_\nu\hat\tau_2)_{pq} c^{p,\dagger}_{j\alpha}c^{q,\dagger}_{j\beta}$ \\
    $(\mu,0)$, $\mu=1,2,3$ & $(i\hat\sigma_\mu\hat\sigma_2)_{\alpha\beta}(i\hat\tau_2)_{pq} c^{p,\dagger}_{j\alpha}c^{q,\dagger}_{j\beta}$ \\
    \hline		
\end{tabular}
\label{table: O-even}
\end{table}

\begin{table}[t]
\caption{Pair creation operators for two-orbital odd pairing in a 1D lattice.}
\begin{tabular}{|c|c|c|}
    \hline
    $(\mu,\nu)$ & $\hat O_{\mu\nu,j}^\dagger$ \\ \hline
    $(0,0)$ & $(i\hat\sigma_2)_{\alpha\beta}(i\hat\tau_2)_{pq} c^{p,\dagger}_{j\alpha}c^{q,\dagger}_{j+1,\beta}$ \\
    $(\mu,\nu)$, $\mu,\nu=1,2,3$ & $(i\hat\sigma_\mu\hat\sigma_2)_{\alpha\beta} (i\hat\tau_\nu\hat\tau_2)_{pq}c^{p,\dagger}_{j\alpha}c^{q,\dagger}_{j+1,\beta}$ \\
    \hline		
\end{tabular}
\label{table: O-odd}
\end{table}


\section{Inter-orbital $\eta$ states}
In this section we discuss certain technical details that complement the discussion of the inter-orbital $\eta$ states $\ket{\phi^{03}_n}$ in Sec. \ref{sec:ioPairingSpecialEta}.

\subsection{Additional generators\label{sec:AppSpEtaRepTheory}}

Full symmetry group of the inter-orbital $\eta$ states is discussed in Sec. \ref{sec:fullSymmGroupForInterOrbital} along with some of its generators. Here we list the generators that didn't find use in our example Hamiltonians but are otherwise valid building blocks for a Hamiltonian supporting the $\ket{\phi^{03}_n}$ MBS. They include
\begin{align}\label{CMdefMain}
\CM_{ij}^{w \bar w} = w_{i\uparrow}^\dagger {\bar w}_{j\uparrow} + {\bar w}^\dagger_{j\downarrow} w_{i\downarrow} , \quad \CM_{ij}^{{\bar w} w } = \left(\CM^{w{\bar w}}_{ji}\right)^\dagger, 
\end{align}
and
\begin{align}\label{CNdefMain}
\CN^{w{\bar w}}_{ij}=  w^\dagger_{i\downarrow}  {\bar w}_{j\uparrow}+w_{i\uparrow} {\bar w}^\dagger_{j\downarrow}, \quad \CN^{{\bar w}w}_{ij}= \left(\CN^{w{\bar w}}_{ji}\right)^\dagger,
\end{align}
as well as
\begin{align}
\label{eq:spEtaSO4JsDef}
J^A_+ = \frac{1}{2}\left(S_A + K_A\right) = \frac{1}{2}\sum_{j,\alpha\beta}w_{j\alpha}^\dagger \sigma^A_{\alpha\beta}w_{j\beta}, \nonumber \\
J^A_- = \frac{1}{2}\left(S_A - K_A\right) = \frac{1}{2}\sum_{j,\alpha\beta}{\bar w}_{j\alpha}^\dagger \sigma^A_{\alpha\beta}{\bar w}_{j\beta}.
\end{align}

\subsection{Additional Hamiltonians \label{sec:appMoreGeneralHForIoEta}}

Similar to the example \eqref{eq:Hminimal} in the main text, numerous further Hamiltonians supporting the inter-orbital $\eta$ scars may be constructed following the $H_0+OT$ structure using the building blocks provided. For instance, one could consider a Hamiltonian which includes all the components of $T^A_{\text{s}}$ \eqref{eq:soHopping} and $K_A$ \eqref{eq:gensKSU2main}:
\begin{align}
\label{eq:Hfull}
H =H_{\mathrm{O}(N)} + \sum_{A,j} r^{\text{Z}}_{jA} H^A_{\text{Z},j}   + \sum_{A,ij} T^A_{\text{s},ij} + \sum_A \lambda_A K_A,
\end{align}
where $A=1$, $2$, or $3$, with $H_{\mathrm{O}(N)}$ defined in Eq. \eqref{eq:bareH}.

In the numerical simulations for the Hamiltonian (\ref{eq:Hfull}) (not shown) we observe that the results are broadly identical to those presented in Fig. \ref{fig:minH1Plots}, except that the triplet inter-orbital scars family is absent because not all the operator components $A$ correspond to valid generators of its full symmetry group.

\subsection{Pairing correlations in a degenerate scar subspace}

For $H^{c}_0$ \eqref{eq:bareH0} the states $\ket{\phi_n^{03}}$ and $\ket{\phi^{\pm}_n}$ become degenerate when $U= - \sum_{p=1}^2 \mu_p$. Consider an equal-weight superposition of all the scar states within one of these families
\begin{align}
\label{eq:psiSE}
|\phi_s\rangle = \frac{1}{2N_m+1}\sum_{n=0}^{2N_m}|\phi_n\rangle.
\end{align}
where $N_m$ is the number of states in the family. Denoting this superposition $\ket{\phi_{\text{se}}}$ for the family $\ket{\phi_n^{03}}$ and $|\phi_{\text{tsem}}\rangle $ for the family $\ket{\phi^{\pm}_n}$, we obtain the following expectation values for the relevant pairing correlations:
\begin{align}
\label{O3se}
&\left\langle\phi_{\text{se}}|\CO^{\dagger}_{03,j}|\phi_{\text{se}}\right\rangle  = \frac{\sum_n \sqrt{n(2N+1-n)}}{N(2N+1)}
\end{align}
and
\begin{align} 
\label{O3tsem}
\left\langle\phi_{\text{tsem}}|\CO^{\dagger}_{03,j}|\phi_{\text{tsem}}\right\rangle &= \frac{\sum_n \sqrt{n(2N-1-n)}}{N(2N-1)}.
\end{align}

\subsection{Small-$U$ regime}

\begin{figure}[htp!]
	\begin{center}
	\includegraphics[width=0.76\columnwidth]{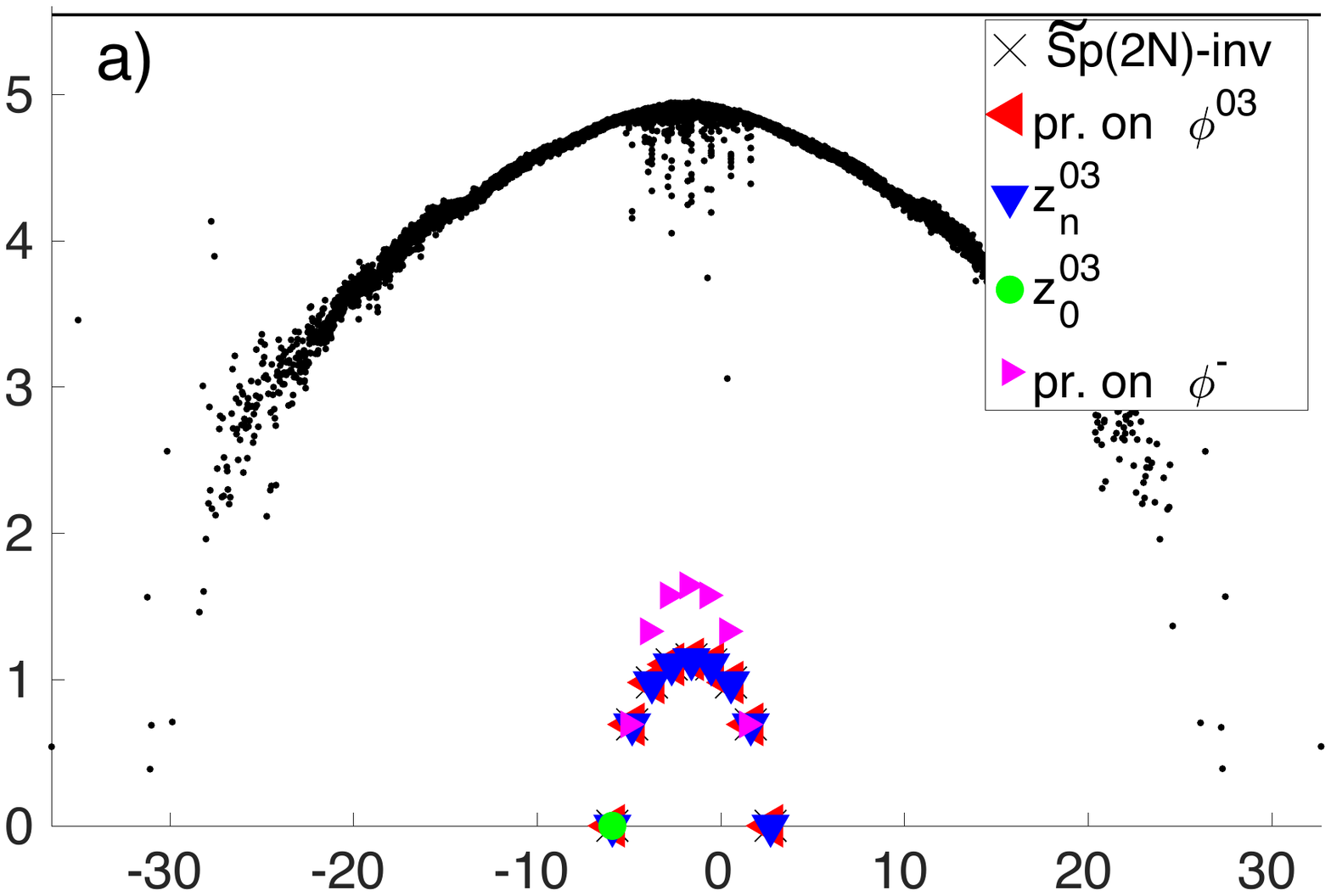}
	\includegraphics[width=0.83\columnwidth]{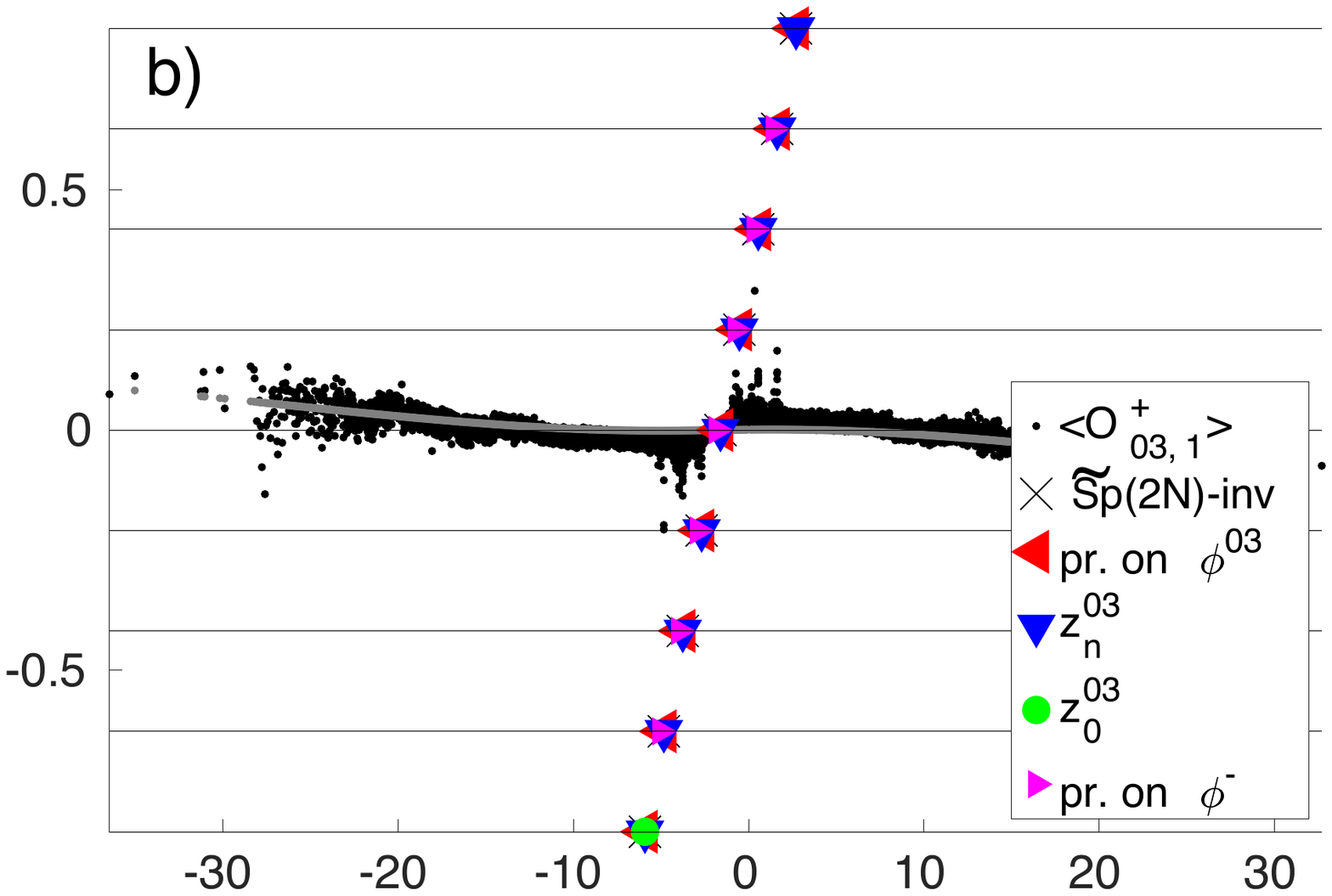}	
		\includegraphics[width=0.82\columnwidth]{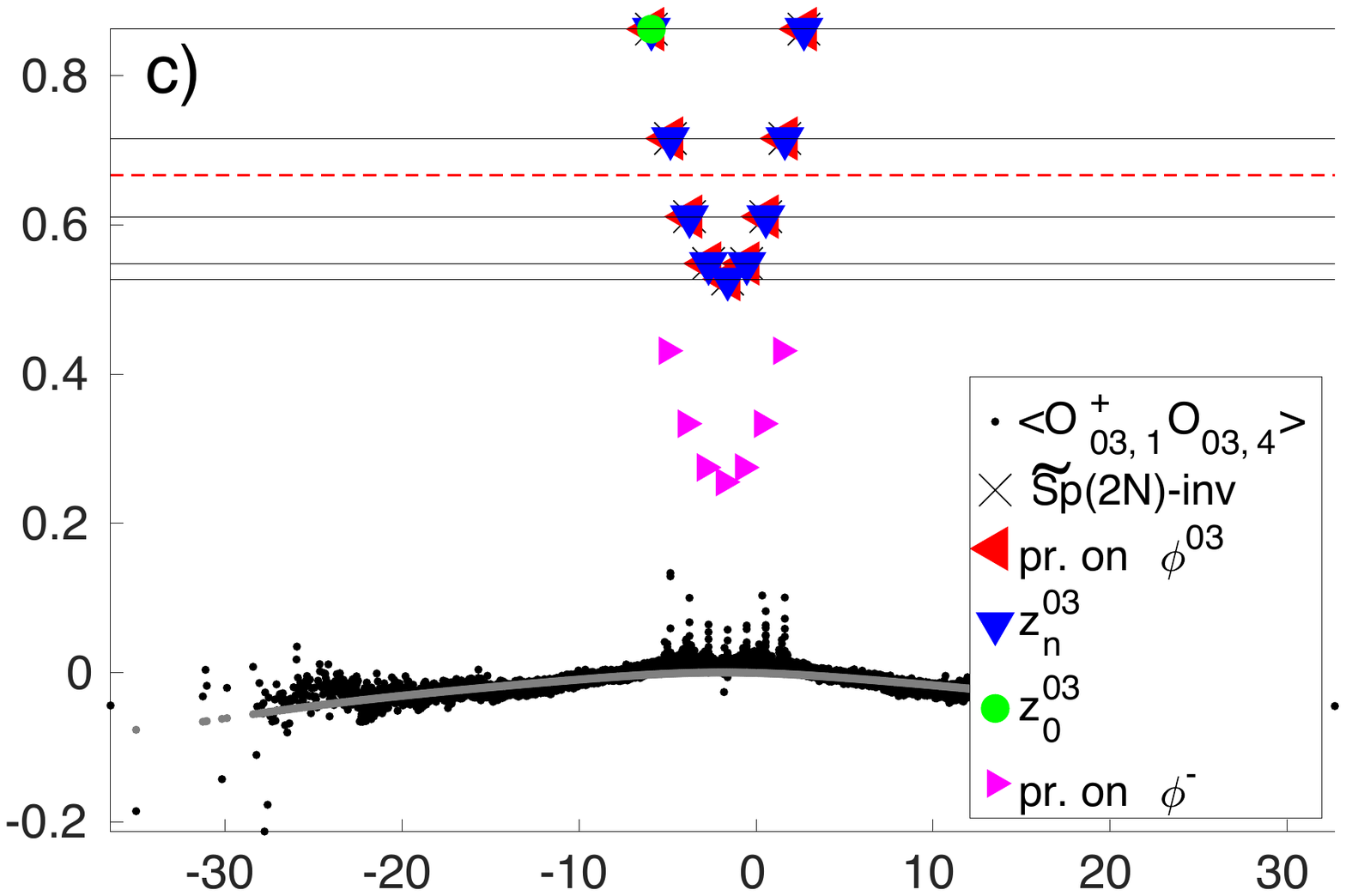}	
			\includegraphics[width=0.82\columnwidth]{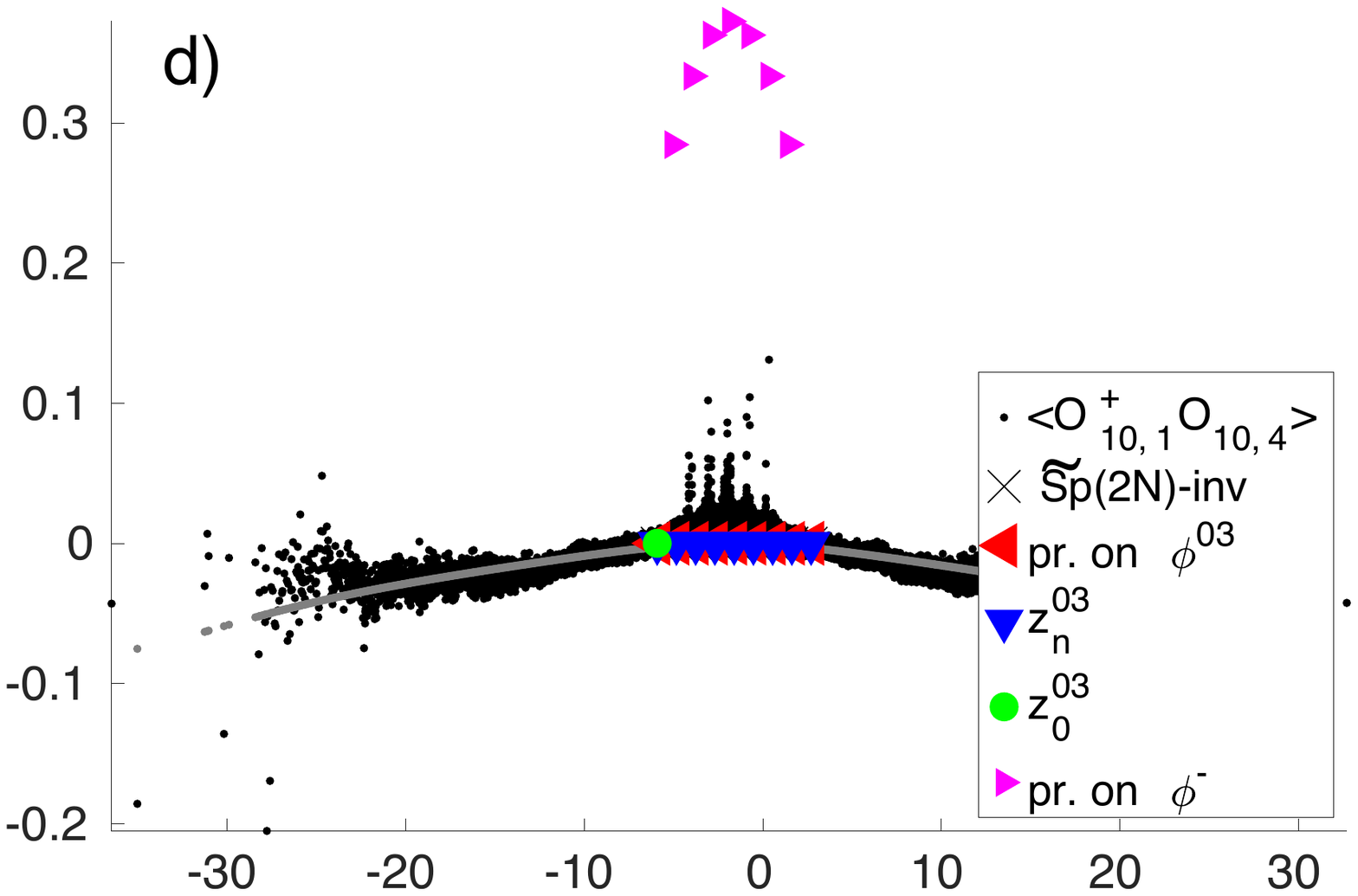}	
	\end{center}
\caption{Inter-orbital pairing potential \eqref{eq:dH0general} added to the Hamiltonian \eqref{eq:Hminimal}. Parameters are the same as in Fig. \ref{fig:addOdagGetPsi0sp}, except that $\Delta=9$ and $U=0.1$. a) Entanglement entropy. b) Real part of the one-point function $\braket{\CO^{\dagger}_{03,j}}$ for inter-orbital pairing; horizontal lines indicate analytical values. c) Two-point function $\braket{\CO^{\dagger}_{03,1} \CO_{03,N}}$ for inter-orbital pairing; the dashed line indicates average over the scar subspace $\ket{z_n^{03}}$. d) Two-point function $\braket{\CO^{\dagger}_{10,1} \CO_{10,N}}$ for spin-triplet pairing; the corresponding one-point function is zero because only one spin-triplet pair is present in every of the tower states.}
\label{fig:addOdagGetPsi0spSmallU}
\end{figure}

In Fig. \ref{fig:addOdagGetPsi0spSmallU} we show the numerical results for a finite $\Delta=9$ and a small Hubbard interaction $U=0.1$. This corresponds to the regime where the solution of the mean-field Hamiltonian is deformed perturbatively by weak interactions.

As expected, we observe the full family of the $\ket{z_n}$ scars (including the BCS scar $\ket{z_0}$) together with the satellite family which results from the rotation of the triplet inter-orbital states. Both families have strong superconducting correlations.
This demonstrates the stability of the mean-field solution (the inter-orbital $\eta$ MBS) to the interactions.

\section{Type-\rom{1} spin-triplet MBS}
This section contains additional generators and Hamiltonians for the $\ket{\phi_n^{30}}$ MBS, along with the numerical test exhibiting the $4e$ clustering in these states.

\subsection{Additional generators}
Using the spin-orbital swap mapping mentioned in the main text, we can obtain the additional generators that annihilate the triplet states $\ket{\phi^{30}_n}$ \eqref{eq:BSSTgenTowerWf}.
Substituting the definition of $w'_{jp}$ and $\bar w'_{jp}$ \eqref{eq:uAndvFermionsTriplet} into Eqs. \eqref{CMdefMain} and \eqref{CNdefMain}, we obtain the analogous mapped generators:
\begin{align}
\label{tripletCMdef}
\CM_{ij}^{w' {\bar w}'} = \frac{1}{2} [ (c^{x,\dagger}_{i\uparrow} + i c^{x,\dagger}_{i\downarrow}) ( c^{x}_{j\uparrow} + i c^{x}_{j\downarrow}) \\ \nonumber
 +(c^{y,\dagger}_{j\uparrow} - i c^{y,\dagger}_{j\downarrow}) ( c^{y}_{i\uparrow} - ic^{y}_{i\downarrow}) ]
\end{align}
and 
\begin{align}
\label{tripletCNdef}
\CN_{ij}^{w' {\bar w}'} = \frac{1}{2}[ (c^{x,\dagger}_{i\downarrow} + i c^{y,\dagger}_{i\downarrow})( c^x_{j\uparrow} + i c^y_{j\uparrow}) \\ \nonumber
 + (c^x_{i\uparrow}- ic^y_{i\uparrow})( c^{x,\dagger}_{j\downarrow} - i c^{y,\dagger}_{j\downarrow}) ].
\end{align}
Further, using \eqref{eq:uAndvFermionsTriplet} and \eqref{eq:spEtaSO4JsDef}, we can write down the mapped version of the generators of the SO(4) group acting on both spin and orbital degrees of freedom:
\begin{align}
\label{eq:tripletSO4Js}
J^{\prime,A}_+ = \frac{1}{2}\sum_{j,pq}w_{jp}^{\prime,\dagger} \tau^A_{pq}w'_{jq}, \quad 
J^{\prime,A}_-  = \frac{1}{2}\sum_{j,pq}{\bar w}_{jp}^{\prime,\dagger} \tau^A_{pq}{\bar w}'_{jq}.
\end{align}

\subsection{Additional Hamiltonians \label{sec:tr1HamiltonianAddNumerAppendix}}

In the main text we demonstrated a simple example Hamiltonian \eqref{eq:tripletGamma0MediumH} that supports the $\ket{\phi^{30}_m}$ scars.
Another example Hamiltonian can be constructed by including all the generators of the full symmetry group:
\begin{align}
\label{eq:tripletGamma0MaximalH}
 & H'_{I} =   H_{\mathrm{O}(N)}  + \sum_A K'_A  \nonumber\\ 
 & +\sum_{ij} \left[ \CM_{ij}^{w' {\bar w}'} + \CM_{ji}^{{\bar w}' w'} + \CN_{ij}^{w' {\bar w}'} + \CN_{ji}^{{\bar w}' w'} \right] \nonumber\\
  & + \sum_A(J^{\prime,A}_+ + J^{\prime,A}_- ).
\end{align}
The operators $\CM_{ij}^{w' {\bar w}'}$, which are defined in Eq. \eqref{tripletCMdef}, are summed over nearest-neighbour and same-site combinations of $i,j$ which ensures Hermiticity. 
The same remark applies also to the term including generators $\CN_{ij}^{w' {\bar w}'}$ defined in Eq. \eqref{tripletCNdef}. The operators $J^{\prime,A}_\pm$ are defined in \eqref{eq:tripletSO4Js}.
The numerical results for the Hamiltonian \eqref{eq:tripletGamma0MaximalH} (not shown) differ from those in Fig. \ref{fig:BSSTgamma0} in that only the tower of scars $\ket{\phi^{30}_n}$ \eqref{eq:BSSTgenTowerWf} remains at low entanglement, while the satellite family is absent.

To make a scar with spin-triplet pairing the ground state, we can add the pairing potential $H_\Delta$ \eqref{eq:dH0general} with $O_j = \CO_{30,j}$ to the Hamiltonian \eqref{eq:tripletGamma0MaximalH}. 
The numerical results (not shown) are different from the example in the main text in that the additional scar family seen in Fig. \ref{fig:BSSTmediumHwithGamma} is absent and as a result there are no low-entanglement states with significant $\mu=1$ spin-triplet ODLRO. The tower of scars  $\ket{\phi^{30}_n}$ (in the rotated basis) looks the same as in Fig. \ref{fig:BSSTmediumHwithGamma}.

\subsection{$4e$ clustering}

\begin{figure}[htp!]
	\begin{center}
				\includegraphics[width=0.92\columnwidth]{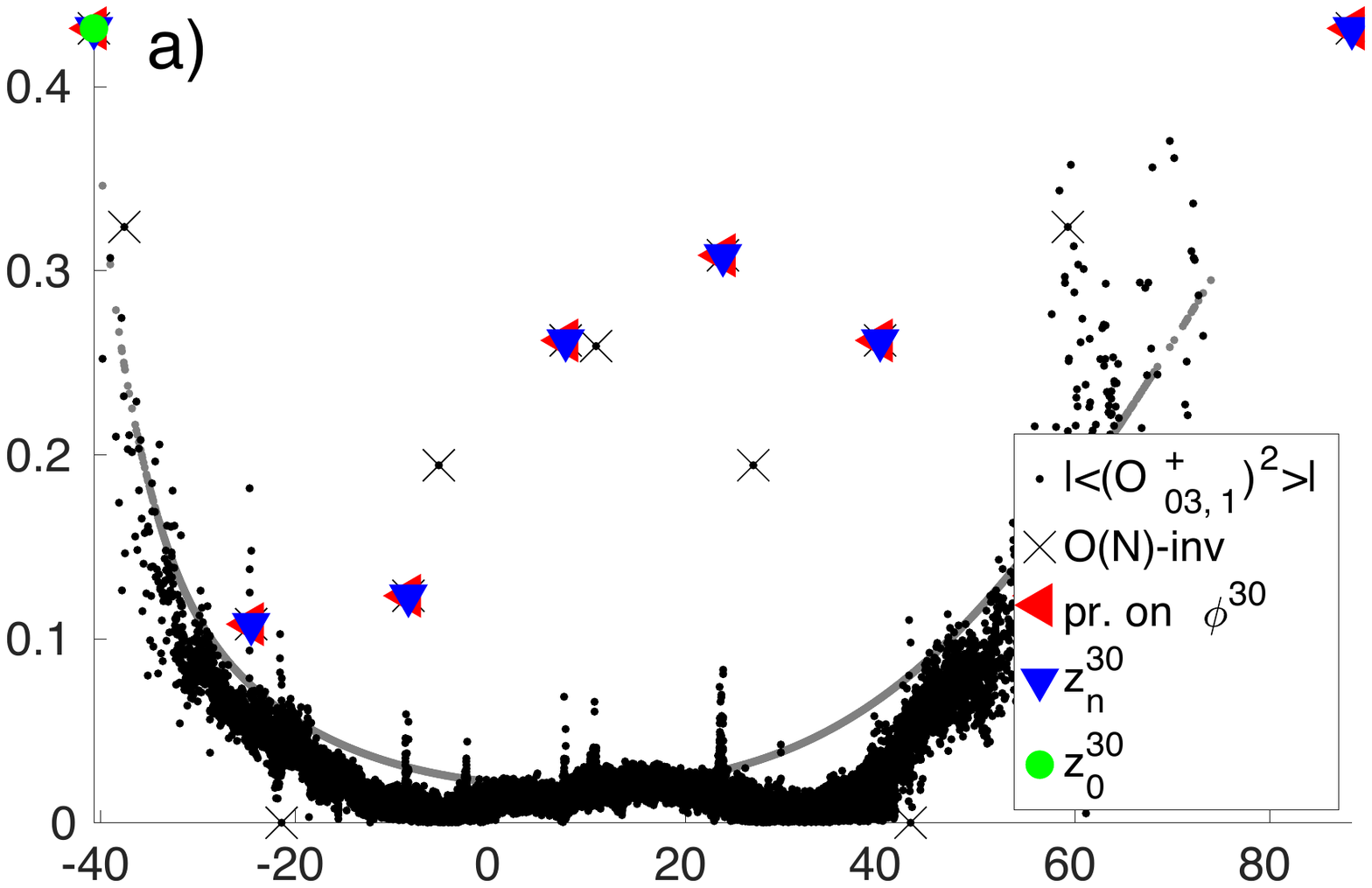}
                \includegraphics[width=\columnwidth]{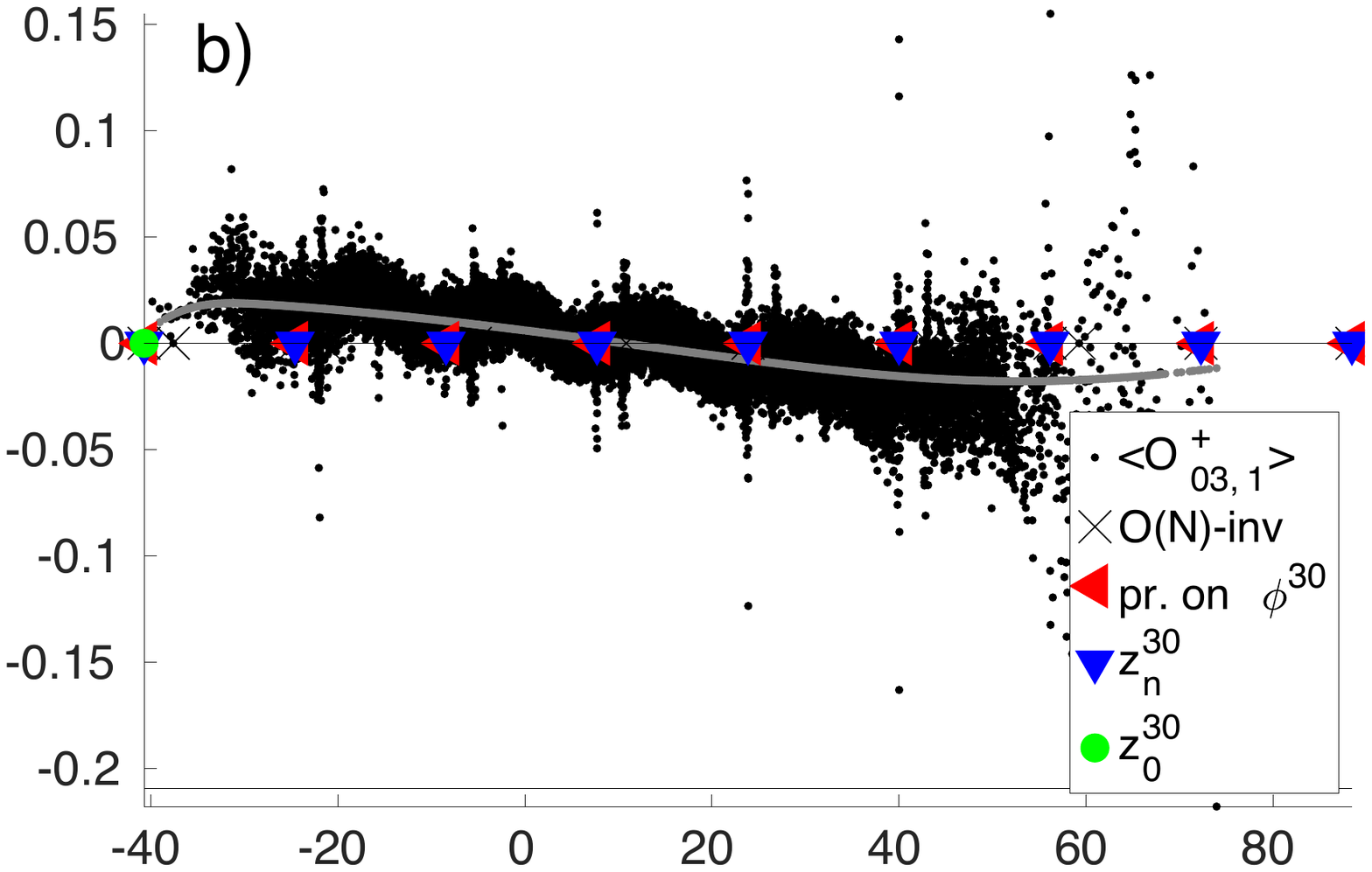}
	\end{center}
\caption{$2e$ (right panel) and $4e$ (left panel) clustering in the spin-triplet type-\rom{1} scar states $\ket{\phi^{30}_n}$ \eqref{eq:BSSTgenTowerWf}. Absolute value is shown for the $4e$ case $\braket{(\CO^{\dagger}_{03,j})^2}$, while the real part is shown for the $2e$ case  $\braket{\CO^{\dagger}_{03,j}}$ (right panel).}
\label{fig:BSST2evs4e}
\end{figure}

In Fig. \ref{fig:BSST2evs4e} we show the $4e$ (left panel) and $2e$ (right panel) clustering of type $\CO^{\dagger}_{03,j}$ measured in every eigenstate. The Hamiltonian is identical to the one used in the main text (Sec. \ref{sec:tr1BCSgsHamiltonian} and Fig. \ref{fig:BSSTmediumHwithGamma}).

The $4e$ clustering (in the absence of the $2e$ pairing of the same type) is a factor of 22 larger in the BCS ground state and a factor of 12 in an average scar than in an average non-scar state.   
The same statement can be made for any other pairing type $\CO^{\dagger}_{\mu0,j}$ or $\CO^{\dagger}_{0\nu,j}$, except for $\CO^{\dagger}_{30,j}$.

\section{ Spin-triplet type-\rom{2} states. Correlations in a superposition state.}

The states $\ket{\phi_{n,m}}$ \eqref{eq:triplType2Wfs} are not a special case of the scheme from Ref. \cite{paperC1}. It is possible to create a particle number-indefinite superposition of them by making the whole subspace $\ket{\phi_{n,m}}$ \eqref{eq:triplType2Wfs} degenerate leading to eigenstates in arbitrary superposition
\begin{align}
\label{eq:superposition}
|\lambda\rangle = \sum_{n=0}^{N} \lambda_n |\phi_{n,0}\rangle, \quad \sum_n \lambda_n^2  = 1 .
\end{align}
States $\ket{\phi_{0,n}}$ become degenerate when $\mu^x_{\uparrow} = - \mu^y_{\uparrow}$ and states $\ket{\phi_{m,0}}$ when $\mu^x_{\downarrow} = - \mu^y_{\downarrow}$.

In the state \eqref{eq:superposition}, the one-point function is given by
\begin{align}
\langle  \lambda|&\CO^{\dagger}_{1 0, j} |\lambda\rangle  = -\frac{2}{N}\sum_n \lambda_n \lambda_{n-1}\sqrt{n(N-n+1)} ,
\end{align}
and the two-point function remains distance-independent:
\begin{align}
\langle  \lambda|&\CO^{\dagger}_{1 0, i} \CO_{1 0, j} |\lambda\rangle =  4\sum_{n}\lambda_n^2 \frac{(N-n)n}{N(N-1)} .
\end{align}
Both expectation values can be made finite in the limit of large $N$ by a choice of $\lambda_n$ peaking at $n=N/2$. For example, for 
\begin{align}
\label{eq:lambdaStar}
    \lambda^\star_n = \sqrt{\frac{1}{2^N}\frac{N!}{n!(N-n)!}},
\end{align}
we have $ \langle  \lambda^\star|\CO^{\dagger}_{1 0, i} \CO_{1 0, j} |\lambda^\star\rangle = 1$.
The one-point function for the same weight choice is equal to $-1$.

Should the wavefunction manipulation become feasible on some kind of quantum simulator, these results provide a guidance how a wavefunction with a strong unconventional superconducting pairing may be created by combining the $\ket{\phi_{n,m}}$ MBS.


\input{paperC2UnconventionalSCinScars_v2.bbl}

\end{document}

%% file: paperC2UnconventionalSCinScars_v2.bbl
%